\documentclass[twocolumn]{aastex701}

\usepackage{xcolor}

\begin{document}

\title{X-ray Intraday Variability of the Blazar OJ 287 Observed with XMM-Newton}

\author[0000-0000-0000-0001]{Tao Huang}
\affiliation{State Key Laboratory of Radio Astronomy and Technology, Xinjiang Astronomical Observatory, CAS, 150 Science 1-Street, Urumqi 830011, China}
\affiliation{School of Astronomy and Space Science, University of Chinese Academy of Sciences, No.1 Yanqihu East Road, Beijing 101408, China}
\email[]{tao31882@google.com}

\author[0000-0002-9331-4388]{Alok C. Gupta}
\affiliation{State Key Laboratory of Radio Astronomy and Technology, Xinjiang Astronomical Observatory, CAS, 150 Science 1-Street, Urumqi 830011, China}
\affiliation{Aryabhatta Research Institute of Observational Sciences (ARIES), Manora Peak, Nainital 263001, India}
\email[show]{acgupta30@gmail.com} 

\author[0000-0003-0721-5509]{Lang Cui}
\affiliation{State Key Laboratory of Radio Astronomy and Technology, Xinjiang Astronomical Observatory, CAS, 150 Science 1-Street, Urumqi 830011, China}
\affiliation{School of Astronomy and Space Science, University of Chinese Academy of Sciences, No.1 Yanqihu East Road, Beijing 101408, China}
\affiliation{Xinjiang Key Laboratory of Radio Astrophysics, 150 Science 1-Street, Urumqi 830011, China}
\email[show]{cuilang@xao.ac.cn}

\author[0000-0002-0786-7307]{Ashutosh Tripathi}
\affiliation{State Key Laboratory of Radio Astronomy and Technology, Xinjiang Astronomical Observatory, CAS, 150 Science 1-Street, Urumqi 830011, China}
\email[]{ashutoshtripathi@xao.ac.cn}

\author[0000-0001-7199-2906]{Yongfeng Huang}
\affiliation{School of Astronomy and Space Science, Nanjing University, Nanjing 210023, China}
\affiliation{Key Laboratory of Modern Astronomy and Astrophysics (Nanjing University), Ministry of Education, Nanjing 210023, China}
\email[]{hyf@nju.edu.cn}

\author[0000-0003-3337-4861]{P.\ U.\ Devanand}
\affiliation{Aryabhatta Research Institute of Observational Sciences (ARIES), Manora Peak, Nainital 263001, India}
\email[]{devanandullas@gmail.com}

\author[0000-0001-9815-2579]{Xiang Liu}
\affiliation{State Key Laboratory of Radio Astronomy and Technology, Xinjiang Astronomical Observatory, CAS, 150 Science 1-Street, Urumqi 830011, China}
\affiliation{Xinjiang Key Laboratory of Radio Astrophysics, 150 Science 1-Street, Urumqi 830011, China}
\email[]{liux@xao.ac.cn}
\correspondingauthor{ACG, LC}

\begin{abstract}
\noindent
We present X-ray intraday variability, cross-correlated variability, and power spectrum density analysis of the binary black hole blazar candidate OJ 287. The X-ray pointed observations of the source were carried out on eight occasions by the EPIC-pn camera on board the XMM-Newton satellite from November 2005 to November 2022. These good time intervals range between 3.6 hours and 24.1 hours. Three energy bands -- 0.2-2 keV (soft), 2-10 keV (hard), and 0.2-10 keV (total) -- have been used to estimate variability. Low amplitude variations are observed in 4, 5, and 6 light curves in soft, hard, and total energy bands, respectively. Only two observation IDs has shown variation in the all energy bands. The discrete correlation function of the light curves in soft and hard energy bands peaks at zero lag, suggesting that the emission in both bands was cospatial and came from the same population of leptons. Red noise dominates the power spectral densities of variable light curves. According to our flux and spectrum investigations, both particle acceleration and synchrotron cooling mechanisms contribute significantly to the emission from this blazar.
\end{abstract}


\section{Introduction} \label{sec:introduction}
\noindent
Blazars, which include flat-spectrum radio quasars (FSRQs) and BL Lacertae objects (BLLs), are a subclass of radio-loud active galactic nuclei (AGNs). According to \citet{1995PASP..107..803U}, they are distinguished by highly collimated, relativistic charged particle jets orientated at tiny angles ($\leq \rm{15}^{\circ}$) to the observer's line of sight. These jets launched as tightly focused plasma outflows driven by powerful magnetic and gravitational fields near the central super-massive black hole (SMBH) \citep{2002ApJ...579..530W}, which has a mass between $10^6-10^{10}M_{\odot}$ \citep{1984ARA&A..22..471R}. Blazars demonstrate extreme variability in their flux, spectrum and polarization across the entire electromagnetic (EM) spectrum. Blazars variability across the whole EM spectrum, influencing various observational parameters such as flux, spectral index, degree of polarization, etc. Due to their predominantly nonthermal emission and the current observational limitations in spatial resolution, which hinder direct examination of the fine structure within the emission region, time-domain variability becomes a crucial diagnostic tool for constraining the spatial scale of the emitting region. Blazars recognized for their variability across nearly all accessible timescales, ranging from minutes to years. Variability is typically categorized based on timescale into intraday variability (IDV), which occurs within less than a day \citep{1995ARA&A..33..163W}; short-term variability (STV), which spans days to weeks; and long-term variability (LTV), which covers months to years \citep{2004A&A...422..505G}. \\
\\
Blazar's broadband spectral energy distributions (SEDs) exhibit a distinctive double-humped profile \citep{1998MNRAS.299..433F}: the lower-energy hump, which peaks between the infrared (IR) and soft X-ray regions, is attributed to synchrotron radiation from relativistic electrons within the jet; the higher-energy hump, peaking in the GeV - TeV $\gamma-$ray energies, is typically attributed either to inverse-Compton (IC) scattering, which includes synchrotron self-Compton (SSC) and external Compton (EC) processes, or to hadronic cascade processes \citep[e.g.][and references therein]{1983ApJ...264..296M,2020MNRAS.498.5424R}. \\
\\
One of the BL Lac objects that has been investigated the most is OJ 287 (z = 0.306). Its optical light curve (LC), which dates back to the late 19th century, was discovered in 1967 \citep{1967AJ.....72..757D}. \citet{1988ApJ...325..628S} used a century-long LC and find a quasi-periodic outburst pattern that occurred approximately every 12 years, indicating the existence of a unique dynamical mechanism at its centre. They put up a binary SMBH model and predicted that the next significant outburst will take place between 1994 and 1995. The international OJ-94 monitoring campaign was inspired by this prediction and was successful in detecting double-peaked optical outbursts. The major outburst was observed in 1994, and the next outbursts were separated by $\sim$1.2 years \citep{1996A&A...305L..17S}. \citet{1996A&A...305L..17S} developed the disk-impact model based on these findings, which postulates that the observed double-peaked flares are caused by periodic impact of the secondary SMBH on the primary's accretion disc. This hypothesis was partially confirmed by further observations. OJ 287 once more showed two significant outbursts between 2005 and 2007, with a gap of almost two years \citep{2009ApJ...698..781V}. During the 2015-2019 cycle, orbital precession is thought to be the cause of the first outburst, which happened in December 2015 \citep{2016ApJ...819L..37V}, and the second, which happened in July 2019 after an interval of $\sim$3.5 years  was predicted by BH impact-induced bremsstrahlung flare \citep{2020ApJ...894L...1L}. OJ 287 is one of the blazars that has been shown to display the most quasi-periodic oscillation (QPO) activity, appearing across several EM wavelengths and covering a wide range of periods from several minutes to decades, in addition to this commonly acknowledged 12-year optical outburst occurrence \citep[e.g.,][and references therein]{1985Natur.314..146C,1985Natur.314..148V,1988ApJ...325..628S,2020MNRAS.499..653K,2025MNRAS.541.3008Z}.\\
\\
Because of its $\sim$12-year optical outbursts, OJ 287 is one of the most dynamically active blazars. As a result, different groups are studying OJ 287 using single-wavelength and quasi-simultaneous multiwavelength combined observations to examine its different observational properties over a wide range of timescales \citep[e.g.][and references therein]{2018MNRAS.473.1145K,2018ApJ...863..175G,2019ApJ...877..151W,2020ApJ...894L...1L,2020MNRAS.498L..35K,2021ApJ...923...51K,2021A&A...654A..38P,2022ApJ...924..122G,2022ApJS..260...39G,2023ApJ...957L..11G,2025A&A...700A..16T}. OJ 287 has been frequently observed by various X-ray missions after its first detection by the Einstein satellite in 1979 -- 1980 \citep{1988ApJ...330..776M}. By using EXOSAT data, \citet{1994ApJS...95..371S} showed that OJ 287 had a comparatively high X-ray flux and a photon index of $\sim$2.3, suggesting that it was in a very bright state at the time. In the analysis of Swift/XRT observations from 2015 and 2018, \citet{2020ApJ...890...47P} found a time delay of ultralviolate bands with respect to X-ray emission the X-ray. XMM-Newton and Suzaku observations were used to perform X-ray IDV studies of OJ 287 \citep{2007MmSAI..78..741C,2024MNRAS.532.3285Z}. \\
\\
Variabilities on the intra-day timescales are the most puzzling of all the variability phenomena seen in AGNs, and they might be connected to activity in the most core regions close to the SMBH. Blazars, on the other hand, show flux variation throughout a variety of observable timescales, from a few minutes to over a decade \citep[e.g.,][and references therein]{2017ApJ...841..123P,2022ApJ...939...80D}. In order to constrain the size of the emitting region and estimate the mass of the central SMBH, as well as to gain a better understanding of the radiation mechanisms close to the central engine of blazars, it is crucial to study intraday variability \citep{1995ARA&A..33..163W}. \\
\\
To better understand X-ray IDV flux and spectral properties and search for QPOs in blazars, our group started a pilot project using various X-ray missions, e.g., XMM-Newton, Chandra, Suzaku, NuStar, etc., and published a series of papers \citep[e.g.][]{2009A&A...506L..17L,2010ApJ...718..279G,2014MNRAS.444.3647B,2016NewA...44...21B,2016MNRAS.462.1508G,2017MNRAS.469.3824K,2017ApJ...841..123P,2018ApJ...859...49P,2019ApJ...884..125Z,2021ApJ...909..103Z,2021MNRAS.506.1198D,2022MNRAS.511.3101P,2022ApJS..262....4N,2022ApJ...939...80D,2025ApJS..278...20D,2024MNRAS.532.3285Z}. The main motivation for this study is to understand the X-ray variability of the binary black hole blazar OJ 287 on the IDV timescale and also search for QPOs in the blazar. Despite decades of observation, OJ 287 has been extensively studied in the optical and radio bands, but in other energy bands, study remains relatively scarce. Using publicly available XMM-Newton archive data, we examined eight pointed observations of the blazar OJ 287. We analyzed the IDV of flux and spectra, as well as power spectral density analysis, to search for possible QPO on the IDV timescale. This study contributes to a deeper understanding of the X-ray IDV properties of this blazar. \\
\\
The structure of this paper is as follows. Section \ref{sec:Reduction} discusses the archival EPIC-pn data and its analysis. Section \ref{sec:Methods} briefly describes the analysis techniques used in this study. Section \ref{sec:results} presents the results of data analysis, followed by a discussion in Section \ref{sec:discussion} and the conclusions in Section \ref{sec:conclusion}.

\begin{deluxetable*}{cccccccccc}
\tablenum{1}
\tablecaption{Analysis Record of EPIC-PN Observations of OJ 287 from the XMM-Newton }
\tablewidth{0pt}
\tablehead{
\colhead{} & 
\colhead{} & 
\colhead{} & 
\colhead{} & 
\colhead{} & 
\multicolumn{3}{c}{$\mu$ (counts s$^{-1}$)\tablenotemark{b}}& \nocolhead{} \\
\colhead{ObsID} & 
\colhead{Date of Obs.} & 
\colhead{Rev} & 
\colhead{GTI\tablenotemark{a}} & 
\colhead{region} & 
\colhead{Soft} & 
\colhead{Hard} & 
\colhead{Total} & 
\colhead{Mean} &
\colhead{{$\Gamma_{eff}^c$}}\\
\colhead{} & 
\colhead{yyyy-mm-dd} & 
\colhead{} & 
\colhead{(ks)} & 
\colhead{(pixels)} & 
\colhead{(0.2--2 keV)} & 
\colhead{(2--10 keV)} & 
\colhead{(0.2--10 keV)} & 
\colhead{HR}&\nocolhead{}
}
\decimalcolnumbers
\startdata
0300480301 & 2005-11-03 & 1081 & 28.0 & r $\leqslant$ 580 &  1.12 $\pm$ 0.12 & 0.21 $\pm$ 0.06 & 1.35 $\pm$ 0.13 & $-0.64 \pm 0.01$& 1.86 \\
0401060201 & 2006-11-17 & 1271 & 44.9 & r $\leqslant$ 650 &  0.86 $\pm$ 0.10 & 0.22 $\pm$ 0.06 & 1.07 $\pm$ 0.12 & $-0.60 \pm 0.01$ & 1.76\\
0502630201 & 2008-04-22 & 1533 & 53.5 & r $\leqslant$ 550 &  0.85 $\pm$ 0.10 & 0.22 $\pm$ 0.07 & 1.05 $\pm$ 0.12 & $-0.59 \pm 0.01$ & 1.71\\
0679380701 & 2011-10-15 & 2170 & 21.6 & r $\leqslant$ 650 &  2.83 $\pm$ 0.18 & 0.71 $\pm$ 0.10 & 3.52 $\pm$ 0.21 & $-0.60 \pm 0.01$ & 1.76\\
0761500201 & 2015-05-07 & 2822 & 86.7 & r $\leqslant$ 680 &  2.51 $\pm$ 0.18 & 0.35 $\pm$ 0.07 & 2.85 $\pm$ 0.20 & $-0.75 \pm 0.01$ & 2.10\\
0830190501 & 2018-04-18 & 3362 & 22.2 & r $\leqslant$ 500 &  2.21 $\pm$ 0.81 & 0.34 $\pm$ 0.08 & 2.55 $\pm$ 0.19 & $-0.73 \pm 0.01$ & 2.04\\
0854591201 & 2020-04-24 & 3732 & 13.1 & r $\leqslant$ 550 &  16.04 $\pm$ 0.52 & 0.81 $\pm$ 0.12 & 16.78 $\pm$ 0.53 & $-0.90 \pm 0.01$& 2.70 \\
0913992101 & 2022-11-22 & 4204 & 29.1 & r $\leqslant$ 800 &  2.26 $\pm$ 0.20 & 0.44 $\pm$ 0.10 & 2.68 $\pm$ 0.31 & $-0.67 \pm 0.01$ & 1.86\\
\enddata
\tablenotetext{a}{GTI = good time interval}
\tablenotetext{b}{$\mu$ = mean count rate}
\tablenotetext{c}{$\Gamma_{eff}$ corresponding to each observation indicates the effective photon index obtained from the respective observed mean HR by assuming an absorbed power-law model folded through the {\it EPIC-PN} instrumental response.}
\label{tab:oj287_obslog}
\end{deluxetable*}

\section{Data Selection and Reduction} \label{sec:Reduction}
\subsection{Data Selection}
\noindent
The European Space Agency (ESA) launched the high-sensitivity X-ray astronomy satellite XMM-Newton in 1999. With an orbital period of about 48 hours, a perigee of $\sim$7,000 kilometres, an apogee of $\sim$114,000 kilometres, and an inclination of 40$^{\circ}$, the satellite is in a very eccentric elliptical orbit. It has several cutting-edge scientific instruments, such as an optical monitor (OM), a reflector grating spectrometer (RGS), and an EPIC (European Photon Imaging Camera) detector. Three CCD (charge-coupled device) cameras, one EPIC-pn camera, and two EPIC-MOS cameras make up the EPIC detector. XMM–Newton is a great tool for analysing high-energy radiation from astronomical sources because it has extraordinary spectral coverage and sensitivity in the (0.15 - 15.0) keV energy range when compared to other X-ray observatories.\\
\\
The effective area of EPIC-MOS is substantially lower than that of EPIC-pn because the RGS diverts some of the incident radiation because the EPIC-pn camera is positioned behind the telescope that houses the RGS grating. This is especially noticeable in the hard X-ray band ($\textgreater$ 4 keV), where EPIC-MOS's effective area decreases at a substantially faster rate than EPIC-pn's. A major advantage of the EPIC-pn camera is its exceptionally high time resolution, which allows it to identify and investigate rapid variability in the X-ray radiation of astronomical objects. Furthermore, the photon accumulation effect from powerful sources is typically less pronounced with the EPIC-pn camera \citep{2001A&A...365L..27T}. We used observational data from the XMM–Newton public archive of the blazar OJ 287 for this study. The archival records show that between 2005 and 2022, XMM–Newton made 10 pointed observations of OJ 287, each lasting at least three hours. Eight observation IDs were kept for analysis after we successively eliminated one observation with missing EPIC-pn data and one observation with a Good Time Interval (GTI) less than 10 ks. The observation log for OJ 287 is given in Table \ref{tab:oj287_obslog} which includes the following information: average flux, average hardness ratio, region size we utilised to produce the light curves (LCs), observation ID, observation time, and GTI. 

\subsection{Data Reduction}
\noindent
The EPIC-pn camera outperforms the MOS camera in the 0.15 -- 15 keV energy range in terms of count rate and high-energy response capability. However, soft proton flares frequently have a significant impact on the results in the energy range over 10 keV. We have limited the energy range to 0.2 -- 10 keV for analysis in order to guarantee data quality and make the most of the EPIC-pn camera's best effective area response.\\
\\
Using the most recent Calibration Files\footnote{http://www.cosmos.esa.int/web/xmm-newton/sas-threads}, we created an event list for the PN detector using the \emph{epproc} program after reprocessing the raw data in accordance with the suggested procedure of XMM-Newton SAS v21.0.0. We utilised the \emph{tabgtigen} program to build a GTI file that satisfied RATE $\leq$ 0.4 after initially identifying soft proton flare contamination in the 10 -- 12 keV LCs in order to acquire cleansed data. The GTI files were supplied as input to the \emph{evselect} program along with the initial event list in order to carry out event filtering. Then, using the selection criteria of (PATTERN $\leq$ 4) and (FLAG $=$ 0), the event data were separated into three energy intervals: total (0.2 -- 10 keV), soft (0.2 -- 2 keV), and hard (2 -- 10 keV). In order to include the majority of the photons, We obtained the source counts from defined circular regions, usually with a radius of 25$^{\prime\prime }$ to 40$^{\prime\prime }$ (see Table \ref{tab:oj287_obslog} for details). These regions were centred on the object source. Background regions, whose radius is frequently matched in radius to the source region, are chosen from the same event maps by excluding areas with any source emission. The \emph{epatplot} tool was used to analyse the pile-up effects. In the event that pile-up was identified, it was mitigated by extracting source events from an annular region and excluding the source's central region. In our search, none of the observations showed evidence of pile-up. Due to limitations in the signal-to-noise ratio, a time bin size of 200 seconds was adopted for Observation IDs 0401060210, 0679380701, 0761500201, and 0854591201, while a 100 seconds binning was applied to all other observations. We used the \emph{epiclccorr} task to create light curves that were background-corrected. To increase the accuracy of the final LCs, high-background intervals at the end of each observation were removed. \\
\\
We analysed eight pointed XMM-Newton observations of the blazar OJ 287 drawn from the public archive. From the first observation on April 12, 2005, to the most current one on November 22, 2022, the dataset covers a total of $\sim$17 years, with GTIs ranging from 13.1 to 86.7 ks. A thorough examination of the flux and spectral variations of OJ 287 on IDV and LTV timescales is made possible by the high cadence pointed observations, and long-term coverage. As seen in Figs. \ref{total-LCs} and \ref{LCs:soft and hard}, the LCs for the three bands were extracted and plotted. Significant variation on IDV timescales in the GTI from 3.6 to 24.1 hours was only observed in a small number of LCs.

\begin{deluxetable*}{lccccccccccccc}
\tablenum{2}
\tablecaption{X-Ray variability parameters in soft, hard and total bands of OJ 287 }
\tablewidth{0pt}
\tablehead{
\colhead{Observation ID} & 
\multicolumn{3}{c}{Soft} & 
\multicolumn{3}{c}{Hard} & 
\multicolumn{3}{c}{Total} & 
\colhead{$|\tau|$(ks)} & 
\colhead{$|\tau|_{\rm corr}$(ks)} & 
\colhead{$|\tau|$(ks)} & 
\colhead{$|\tau|$(ks)} \\
\colhead{} & 
\multicolumn{3}{c}{(0.2 -- 2 keV)} & 
\multicolumn{3}{c}{(2 -- 10 keV)} & 
\multicolumn{3}{c}{(0.2 -- 10 keV)} & 
\colhead{Total} & 
\colhead{Total} & 
\colhead{Soft} & 
\colhead{Hard} \\
\colhead{} & 
\colhead{$F_{\rm var}$ (percent)} & 
\colhead{Sig} & 
\colhead{{{Var?}}} & 
\colhead{$F_{\rm var}$ (percent)} & 
\colhead{Sig} & 
\colhead{{{Var?}}} & 
\colhead{$F_{\rm var}$ (percent)} & 
\colhead{Sig} & 
\colhead{{{Var?}}} & 
\colhead{(0.2 -- 10 keV)} & 
\colhead{(0.2 -- 10 keV)} & 
\colhead{(0.2 -- 2 keV)} & 
\colhead{(2 -- 10 keV)}
}
\decimalcolnumbers
\startdata
0300480301 & 3.08 $\pm$ 0.67 & 4.60 & yes & 4.04 $\pm$ 1.50 & 2.69 & no & $1.95\pm0.56$ & 3.49 & yes & $0.25\pm0.09$ & $0.19\pm0.07$ & $0.22\pm0.07$ & -- \\
0401060201 & 5.93 $\pm$ 1.63 & 3.64 & yes & 4.66 $\pm$ 1.17 & 3.99 & yes & 5.89 $\pm$ 1.58 & 3.73 & yes & 0.64 $\pm$ 0.25 & 0.49 $\pm$ 0.19 & 0.65 $\pm$ 0.25 & 0.29 $\pm$ 0.12 \\
0502630201 & $2.66\pm0.16$ & 4.39 & yes & $5.03\pm1.23$ & 4.06 & yes & $2.84\pm0.54$ & 5.23 & yes & $0.14\pm0.03$ & $0.11\pm0.03$ & $0.17\pm0.06$ & $0.06\pm0.03$ \\
0679380701 & 1.00 $\pm$ 0.46  & 2.18 & no & 4.75 $\pm$ 0.91 & 5.21 & yes & 1.91 $\pm$ 0.40 & 4.72 & yes & 1.27 $\pm$ 0.27 & 0.97 $\pm$ 0.36 & -- & 0.50 $\pm$ 0.17 \\
0761500201 & 1.77 $\pm$ 0.29 & 6.19 & yes & 1.57 $\pm$ 0.80 & 1.95 & no & 1.98 $\pm$ 0.27 & 7.33 & yes & 1.05 $\pm$ 0.37 & 0.81 $\pm$ 0.29 & 0.93 $\pm$ 0.31 & -- \\
0830190501 & 1.63 $\pm$ 0.56 & 2.88 & no & 9.05 $\pm$ 1.44 & 6.29 & yes & 1.25 $\pm$ 0.53 & 2.37 & no & -- & -- & -- & 0.08 $\pm$ 0.02 \\
0854591201 & 0.22 $\pm$ 0.35 & 0.63 & no & 5.23 $\pm$ 1.34 & 3.91 & yes & 0.15 $\pm$ 0.28 & 0.54 & no & -- & -- & -- & 0.17 $\pm$ 0.06 \\
0913992101 & 1.41 $\pm$ 0.52 & 2.70 & no & 0.45 $\pm$ 1.36 & 0.33 & no & 1.59 $\pm$ 0.49 & 3.25 & yes & 0.54 $\pm$ 0.12 & 0.41 $\pm$ 0.09 & -- & -- \\
\enddata
\vspace{0.2cm}
{{Note.}}{{ Columns (2), (5), and (8) represent fractional variance ($F_{\rm var}$). Columns (3), (6), and (9) provide ``Sig", which represents significance and is calculated by $F_{\rm var}$/$({F_{\rm var}})_{\rm err}$. Columns (4), (7), and (10) are denoted by ``Var", which represents variability and is ``yes" for variable LC and ``no" for non-variable LC. Columns (11), (13), and (14) provide $|\tau|$, which is the variability timescale. Column (12): $|\tau|_{corr}$ is the red shift-corrected variability timescale, calculated as $|\tau|_{\rm corr} = |\tau|(1 + z)^{-1}$.}}
\label{tab:variability}
\end{deluxetable*}

\section{Analytical Methods} \label{sec:Methods}
\noindent
In this section, we present the methodology employed to analyze the X-ray data from the blazar OJ 287.

\subsection{Excess Variance and Fractional Variability}\label{subsec:excess variance}
\noindent
The intrinsic variability in the astronomical source's LC, which is what remains after the variations brought on by observational errors are removed, is measured using excess variance. It does not reflect variations caused by noise or measurement errors, but rather the intrinsic changes of the source. The X-ray variability intensity of blazars is usually measured using the excess variance $\sigma_{XS}^{2}$ and fractional rms variability amplitude $F_{var}$ \citep{2002ApJ...568..610E}. An LC with $N$ data points $X_{i}$, measured at times $t_{k}$, and associated measurement errors $\sigma_{err,k}$ is considered. The following formula can be used to determine the excess variance:

\begin{equation}
\sigma_{XS}^{2}=S^{2}-\bar \sigma_{err}^{2} ~,
\end{equation}

\noindent
where $S^{2}$ refers to the sample variance of the LC, by follow
\begin{equation}
    S^2 = \frac{1}{N-1}\sum^N_{k=1}(x_k - \bar x)^2 ~,
\end{equation}

\noindent
and $\bar \sigma_{err}^{2}$ represents the average squared error of the uncertainties
\begin{equation}
    \bar \sigma_{err}^{2} = \frac{\sum^N_{i=1}\sigma^2_{err,k}}{N}.
\end{equation}

\noindent
The fractional variance is defined as
\begin{equation}
F_{var}=\sqrt{\frac{S^{2}-\bar \sigma_{err}^{2}}{\bar x^{2}}},
\end{equation} 

\noindent
and the error associated with the fractional variance is calculated by \citep[e.g.][]{2003MNRAS.345.1271V}
\begin{equation}
(F_{var})_{err}=\sqrt{\left[\sqrt \frac{1}{2n}\frac{\bar \sigma_{err}^{2}}{F_{var}\bar x^{2}}\right]^2 +\left[\sqrt{\frac{\bar \sigma _{err}^{2}}{n}}\frac{1}{\bar x}\right]^{2}}.
\end{equation}
The values of excess variance and fractional variance for the three energy bands are summarized in Table \ref{tab:variability}. 

\subsection{Flux Variability Timescale}\label{subsec:timescale}
\noindent
To determine the flux variability timeline, we employed the technique first proposed by \citet{1974ApJ...193...43B}, which was later applied by \cite{2019ApJ...884..125Z} to study X-ray variability in Mrk 421
\begin{equation}
\tau_{var}=\left |\frac{\Delta t}{\Delta \ln F} \right |,
\end{equation}
here, $\Delta t$ represents the time interval between flux measurements $F_1$ and $F_2$ with $F_1> F_2$ such that the flux difference is $\Delta \ln F = \ln \frac{F_1}{F_2}$. Following the approach of \citet{2008ApJ...672...40H}, we selected all flux values that satisfy the condition {{$|F_j-F_k|> \sigma_{F_j}+ \sigma_{F_k}$}} for calculation, where {{$\sigma_{F_j}$ and $\sigma_{F_k}$}} represent the respective measurement uncertainties. This criterion ensures that the detected flux difference is statistically significant and not dominated by observational noise. The timescale associated with each valid pair was calculated as $\tau_{jk} = \left | t_j - t_k \right |$, and the shortest variability timescale was taken to be the minimum among them, i.e., $\tau=min \{ \tau_{jk} \}$, where $j=1,...M-1,k=j+1,....M$, and $M$ denotes the total count of flux measurements. The uncertainty in the derived timescale $\tau_{var}$ was estimated following the method by \citet{2019ApJ...884..125Z}
\begin{equation}
\Delta \tau_{var}\simeq \sqrt{\frac{F_{1}^{2}{\sigma_{F_{2}^{2}}}+F_{2}^{2}{\sigma_{F_{1}^{2}}}}{F_{1}^{2}F_{2}^{2}(ln[F_{1}/F_{2}])^{4}}} \quad \Delta t .
\end{equation}
The variability timescales are summarized in Table \ref{tab:variability}. 

\subsection{Power Spectral Density}
\noindent
The distribution of variability power across different frequencies in a time series is commonly characterized by the power spectral density (PSD), which provides a fundamental diagnostic framework for identifying QPO signatures in LCs. In the context of AGNs, PSDs typically exhibit a red noise pattern at low frequencies, indicating stronger long-timescale variability. This pattern gradually transitions into white noise at higher frequencies, where instrumental noise and measurement uncertainties dominate. To evaluate the PSD, we begin by computing the \emph{periodogram} following the procedure described in \citet{2003MNRAS.345.1271V}. This is carried out using functions available in the SciPy library, and the resulting power is normalized in units of (rms/mean)$^{2}$/Hz which facilitates direct comparison between different datasets. For model fitting, we adopt the Bayesian framework combined with maximum likelihood estimation proposed by \citet{2010MNRAS.402..307V}. In this method, we minimize the following fit statistic
\begin{equation}
S=2\sum_{j=1}^{N/2}\frac{I_{j}}{P_{j}}+\ln{P_j}.
\end{equation}
Here, the fit statistic $S$ is defined as twice the negative log-likelihood function. For each Fourier frequency $f_j$ the values $P_j$ and $I_j$ denote the predicted spectral power of the modele model and the observed periodogram value, respectively. A QPO is considered statistically significant if the power at any given frequency exceeds the red-noise continuum model by at least $3\sigma$, corresponding to a $99.73\%$ confidence level. To characterize the PSD of LCs, we adopt the model widely used in the studies by \citep{2012A&A...544A..80G, 2022ApJ...939...80D, 2015ApJ...805...91M}:
\begin{equation}
P(f)=N f^{-\alpha}+C.
\end{equation}
\noindent
This model has three key parameters: N is a normalization factor that scales the overall amplitude of the spectrum; the spectral index $\alpha$ that describes how the power varies with frequency, and $C$ is a constant added to account for the influence of Poisson noise (white), particularly at higher frequencies. 

\subsection{Discrete Correlation Function}
\noindent
To analyze the correlation between LCs in different energy bands, we employed the discrete correlation function (DCF) proposed by \citet{1988ApJ...333..646E} and investigated possible time lags. The analysis begins with the calculation of the unbinned discrete correlation values, $UDCF_{ik}$, for all possible pairs of soft and hard band data points. Each $UDCF_{ik}$ is defined by

\begin{equation}
UDCF_{ik}=\frac{(a_i-\bar a)(b_k-\bar b)}{\sqrt{\sigma_a^2 \sigma_b^2}} ~,
\end{equation}
where $a_i$ and $b_k$ denote individual flux measurements in the soft and hard bands, with $\bar a$, $\bar b$, $\sigma_a$ and $\sigma_b$ denoting their corresponding means and standard deviations. Each $UDCF_{ik}$ is assigned a time lag $\Delta t_{ik} = t_k - t_i$. All values are then grouped into bins of width $\Delta\tau$, and the mean DCF is computed for each lag $\tau$, defined within the range $\tau - \frac{\Delta\tau}{2} \leq \Delta t_{ik} \leq \tau+\frac{\Delta\tau}{2}$:
\begin{equation}
DCF \left (\tau \right )=\frac{1}{M} \sum UDCF_{ik}, 
\end{equation}
with $M$ being the number of UDCF pairs within that lag bin. The corresponding statistical uncertainty for each DCF bin is estimated using:
\begin{equation}
\sigma_{DCF} \left (\tau \right )=\frac{\sqrt{\sum [UDCF_{ik}-DCF(\tau)]}}{M-1}.
\end{equation}
\noindent
The DCF curve with a value $>$ 0 indicates that the two signals are correlated at the given lag, while a DCF value $<$ 0 suggests an anti-correlation and DCF = 0 shows no correlation between the two datasets. Furthermore, when the soft and hard signals are identical (where $a = b$), the DCF can be simplified to an autocorrelation function (ACF), which can be used as a diagnostic tool to identify periodic features in unevenly sampled astronomical LCs. Obvious oscillations in the ACF spectrum usually indicate the presence of periodic components. Section \ref{cross-correlated} presents a discussion of the correlation results.

\subsection {Hardness Ratio}
\noindent
To describe the spectral shape of X-ray source, we employ the hardness ratio (HR), which compares the normalized differential counts in two energy bands. The most common way to calculate a hardness ratio as
\begin{equation}\label{eq13}
HR=\frac {H-S}{H+S}
\end {equation}
where $H$ and $S$ denote background-subtracted source counts in the soft and hard bands, respectively. The uncertainty in a HR measurement, denoted as $\sigma_{HR}$, is given by \citep[e.g.][]{2022ApJ...939...80D}
\begin{equation}
\sigma_{HR}=\frac{2\sqrt{(H^2\sigma_{S}^{2}+S^2\sigma_{H}^{2})}}{(H+S)^2}.
\end {equation}
Here, $\sigma_S$ and $\sigma_H$ denote the measurement uncertainties associated with the hard and soft bands, respectively. \\
\\
In order to provide a physical interpretation for HR values we obtain, we associate an effective photon index $\Gamma_{eff}$ with the observed HR. For each observation, we initially generated source and background spectra along with corresponding response (RMF) and ancillary (ARF) files using SAS, as described in SAS data reduction thread\footnote{\url{https://www.cosmos.esa.int/web/xmm-newton/sas-thread-pn-spectrum}}. The source and background regions are the same as those we used for LC extraction. We then folded an absorbed power-law model (with the galactic column $n_H$ fixed to $2.37\times10^{20}\, cm^{-2}, \citet{HI4PI16}$) through observation-specific response files of each observation using Xspec \citep{Arn96}. For each observation, we then calculated the model-predicted count rate in the soft (0.2-2.0 keV) and hard (2.0-10.0) keV for a range of photon indices from $\Gamma=1-3$. This was done without fitting the spectra, and the normalization constant of the model was set to an arbitrary value of $10^{-3}$ since HR value is independent of normalization.  Then, the response-dependent HR was calculated using \autoref{eq13} for these photon indices. This response-dependent HR and $\Gamma$ relation can be used to find the effective photon index $\Gamma_{eff}$ corresponding to the mean HR value obtained from LC by simple interpolation.

\begin{figure*}
    \centering
    \includegraphics[scale=0.35]{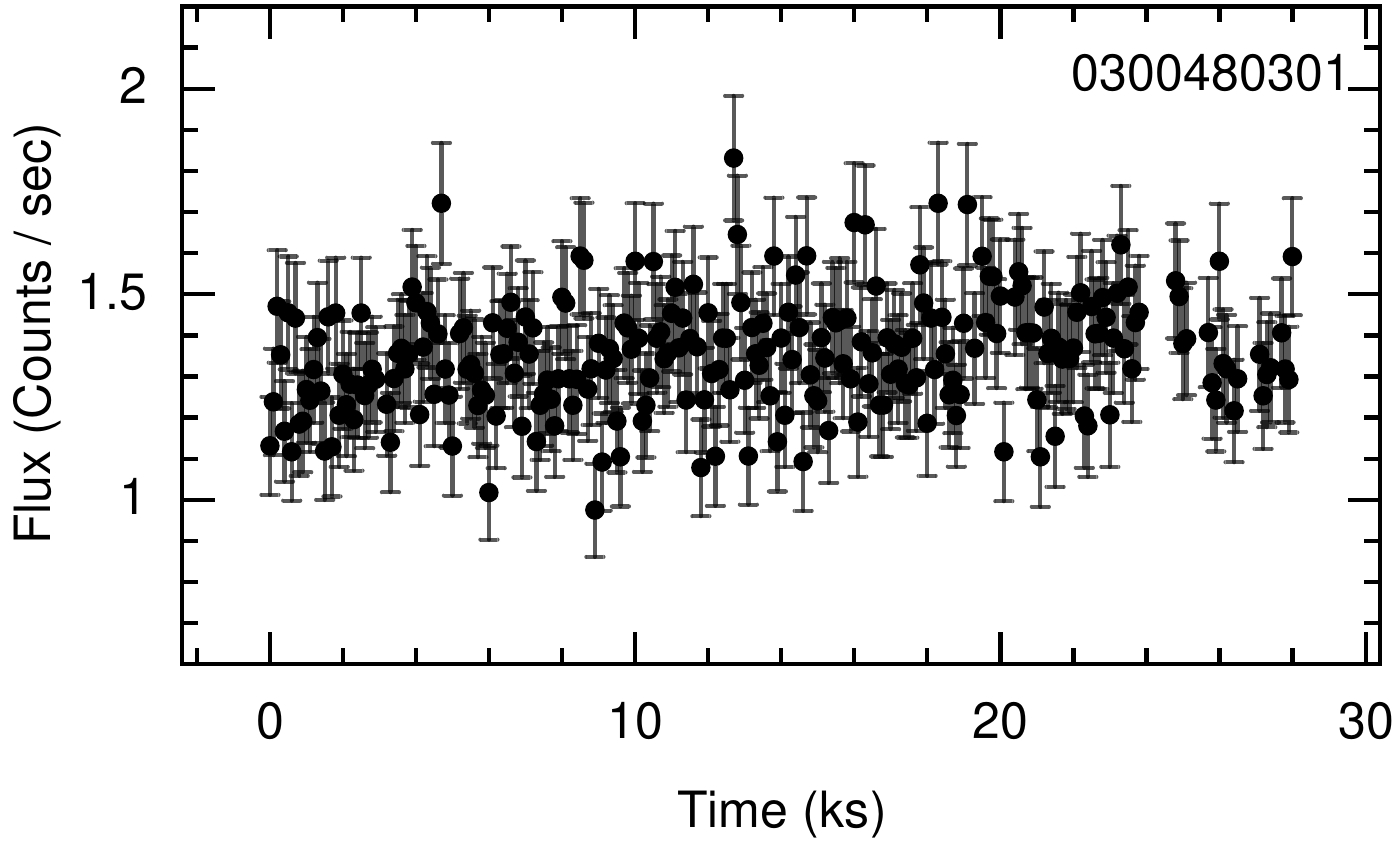}
    \includegraphics[scale=0.35]{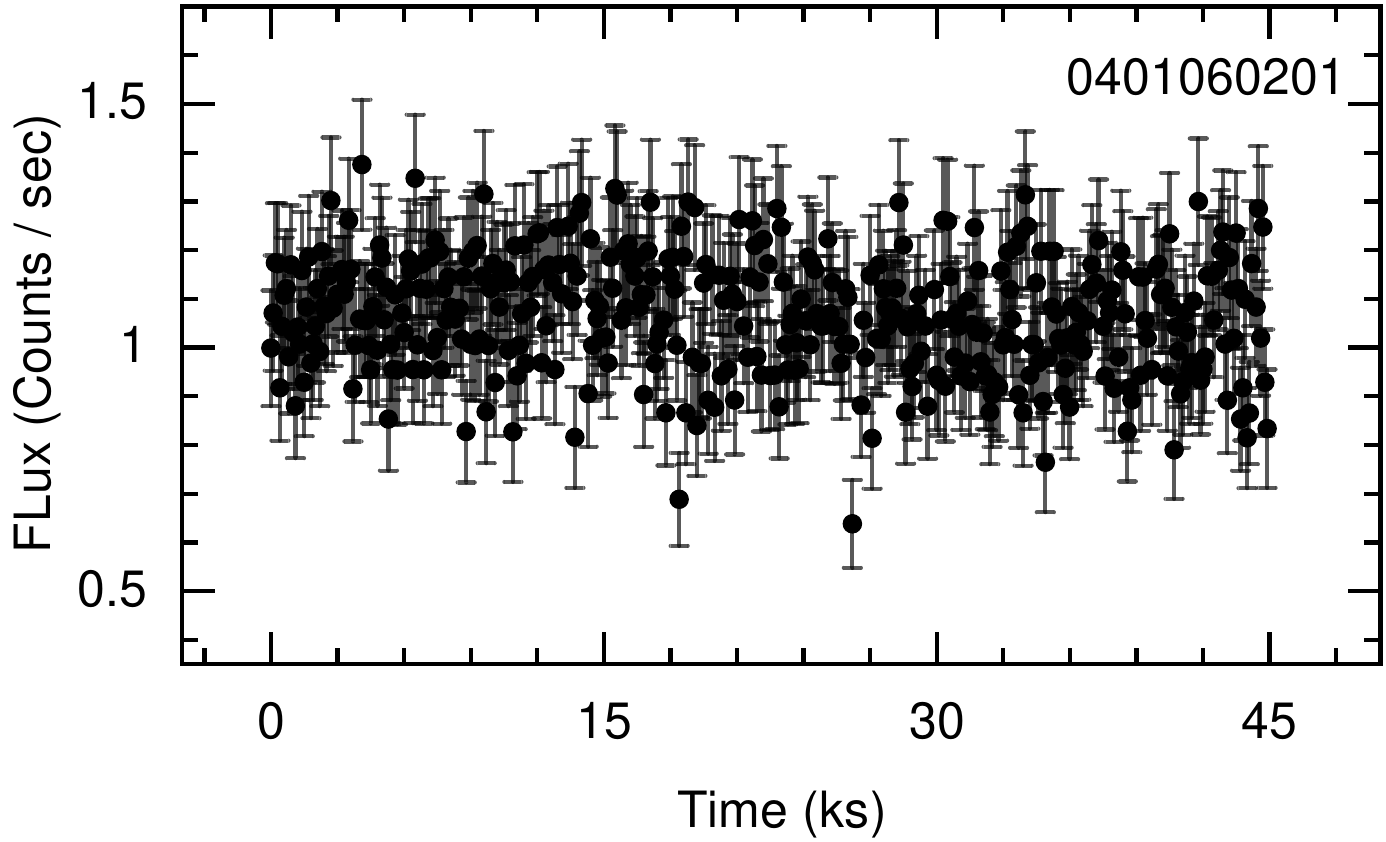}

    \vspace{0.1in}
    \includegraphics[scale=0.35]{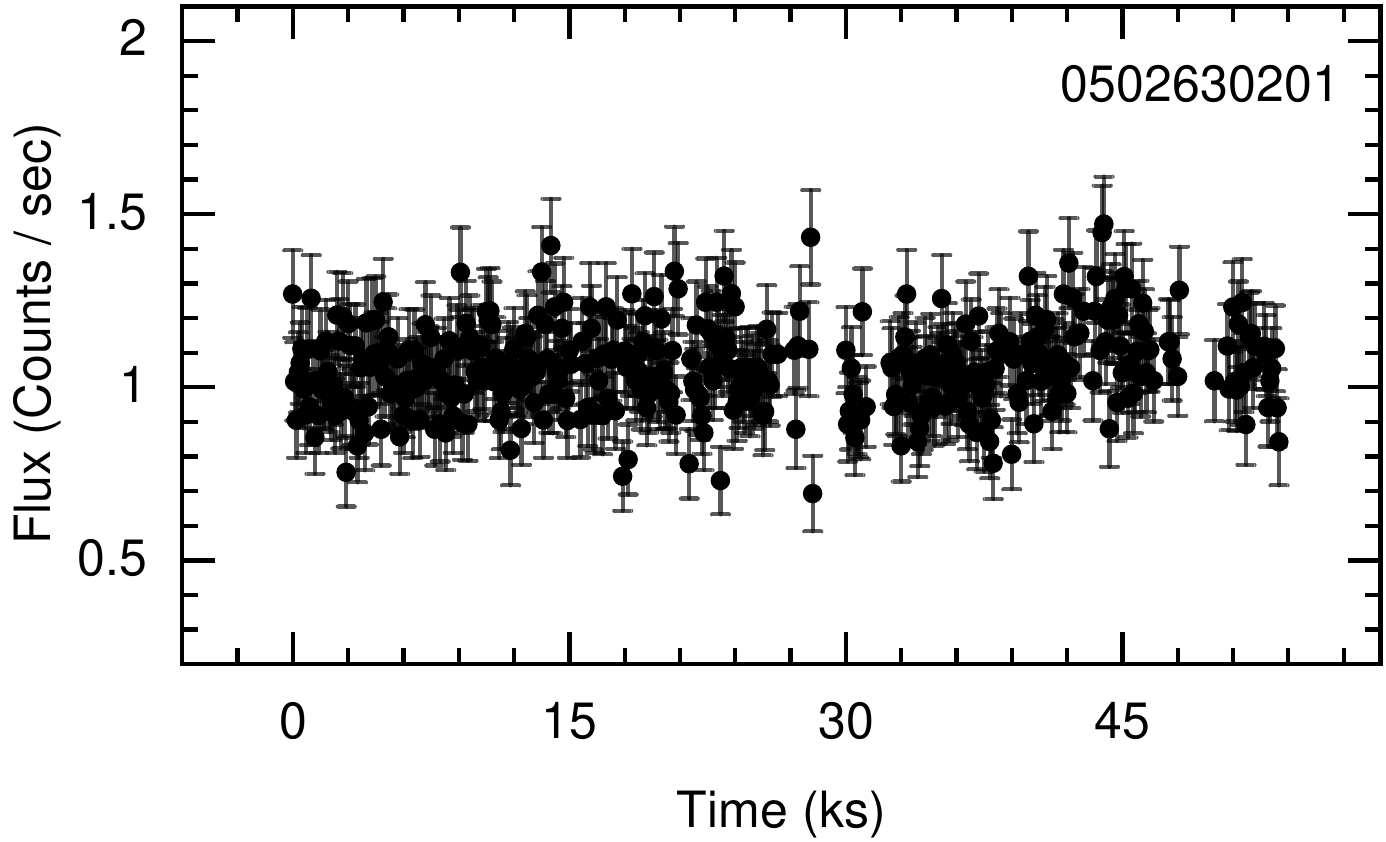}
    \includegraphics[scale=0.35]{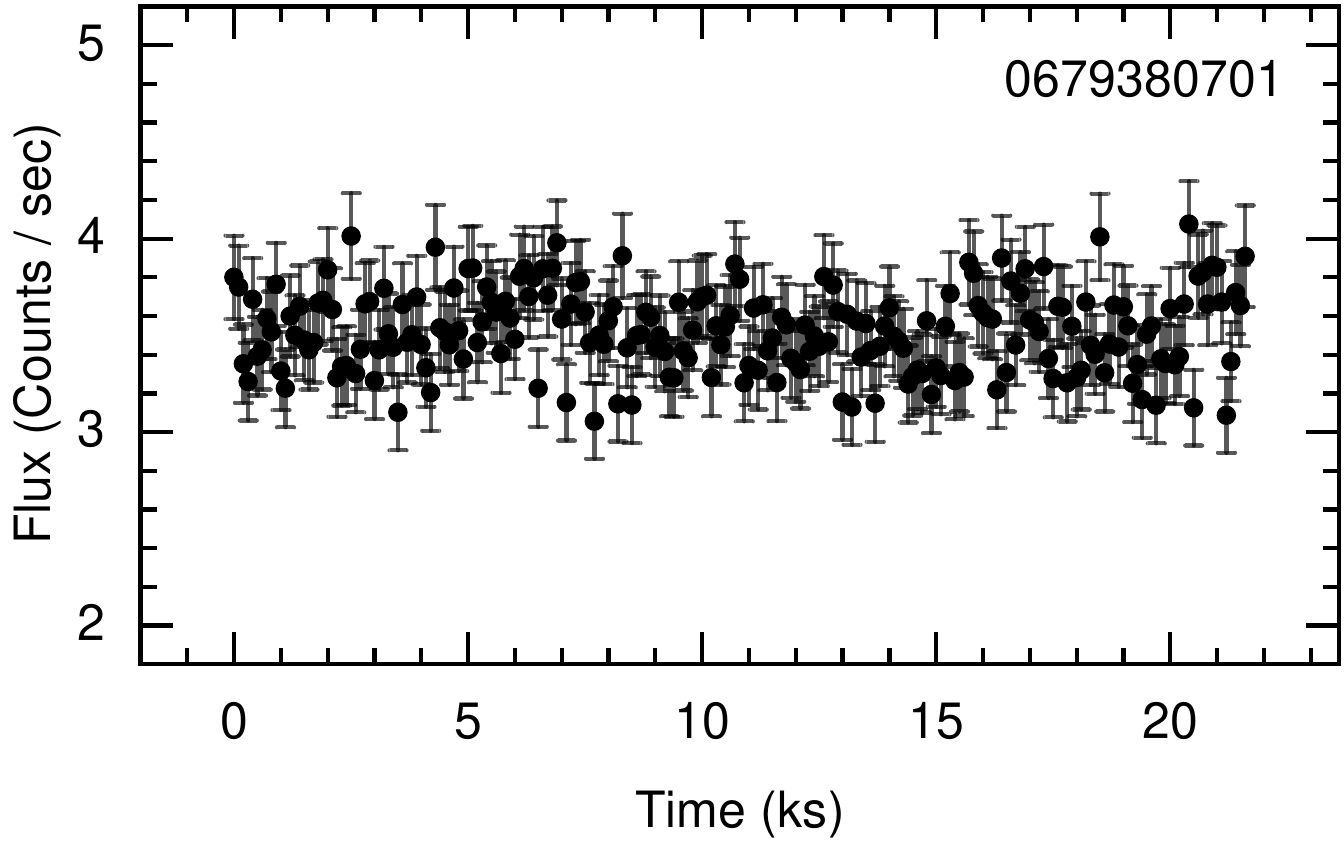}

    \vspace{0.1in}
    \includegraphics[scale=0.35]{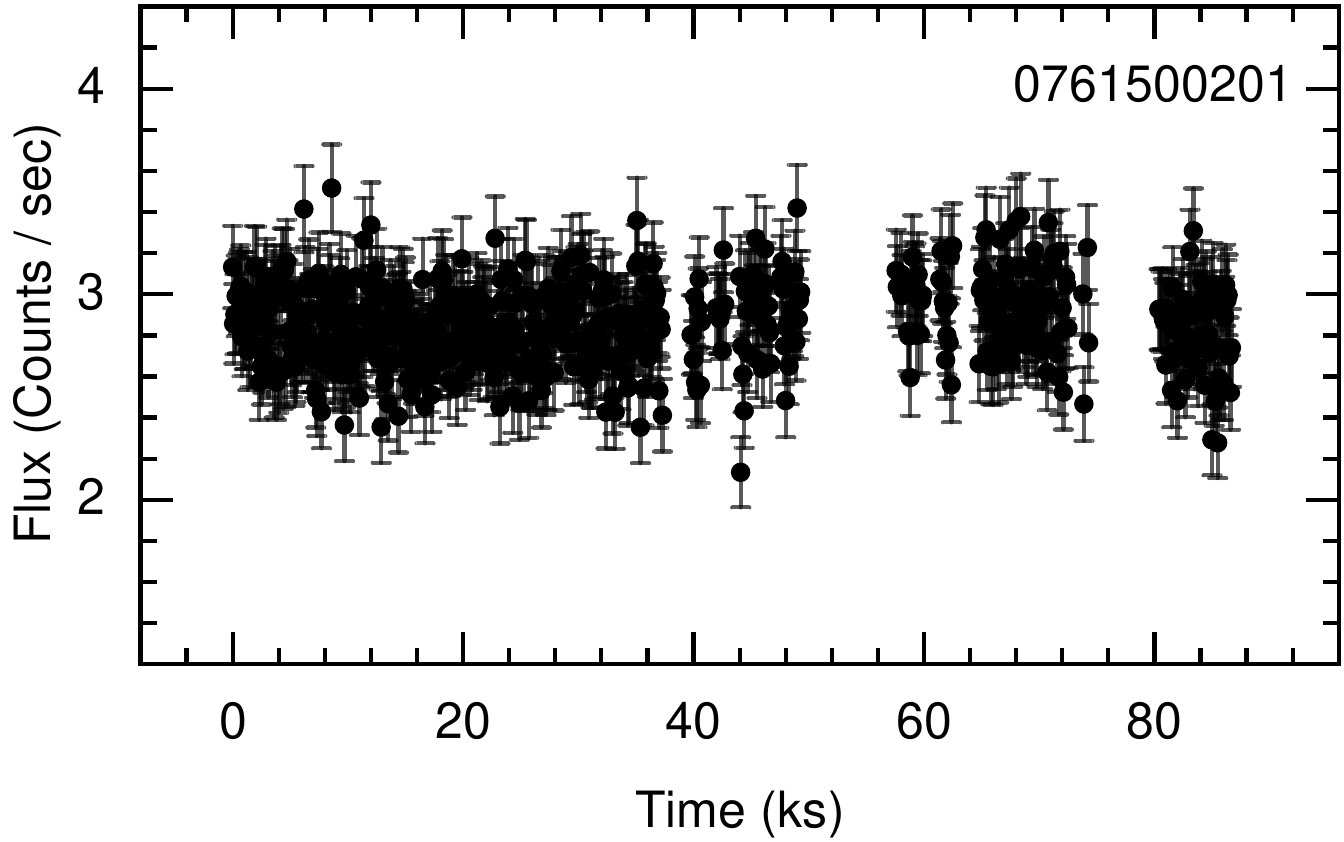}
    \includegraphics[scale=0.35]{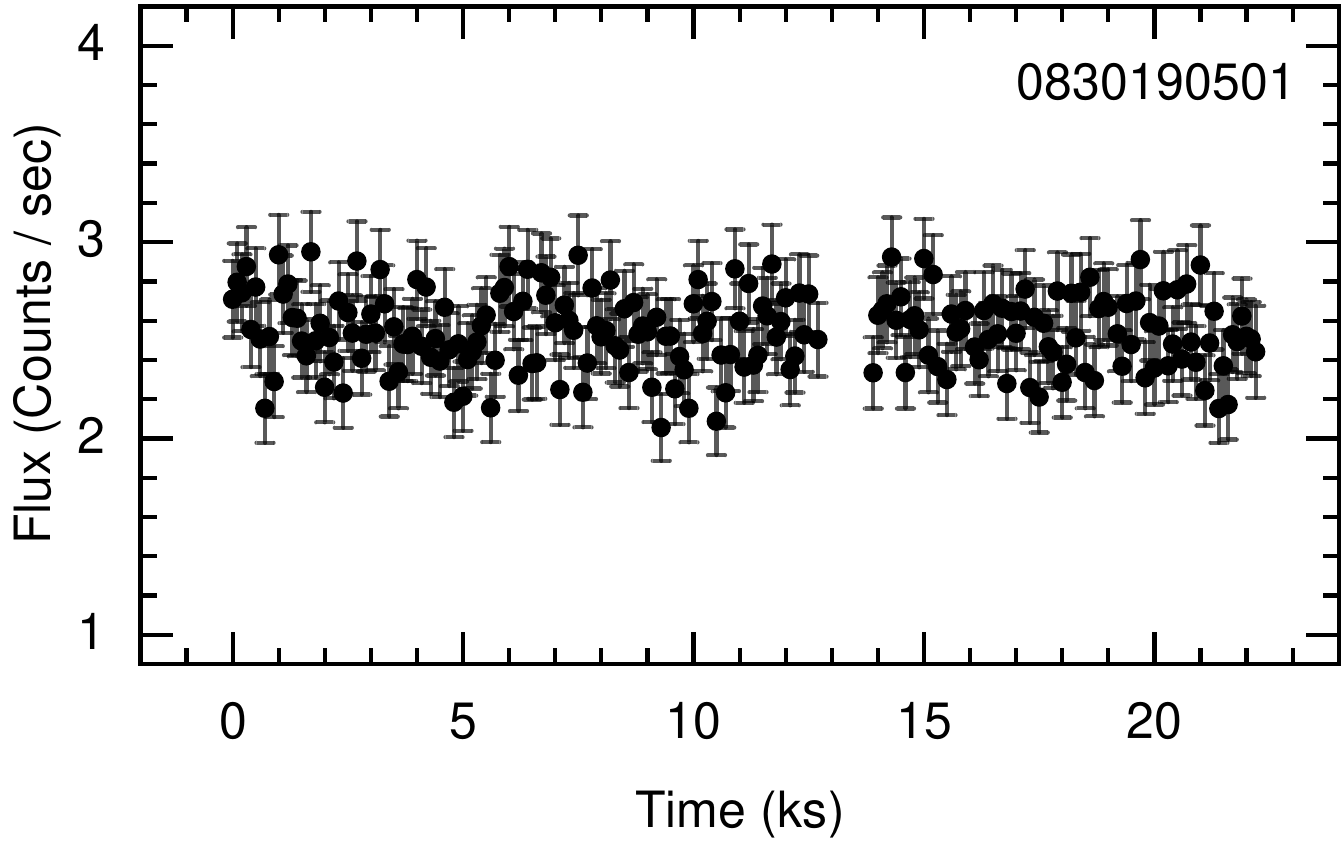}
    
    \vspace{0.1in}
    \includegraphics[scale=0.35]{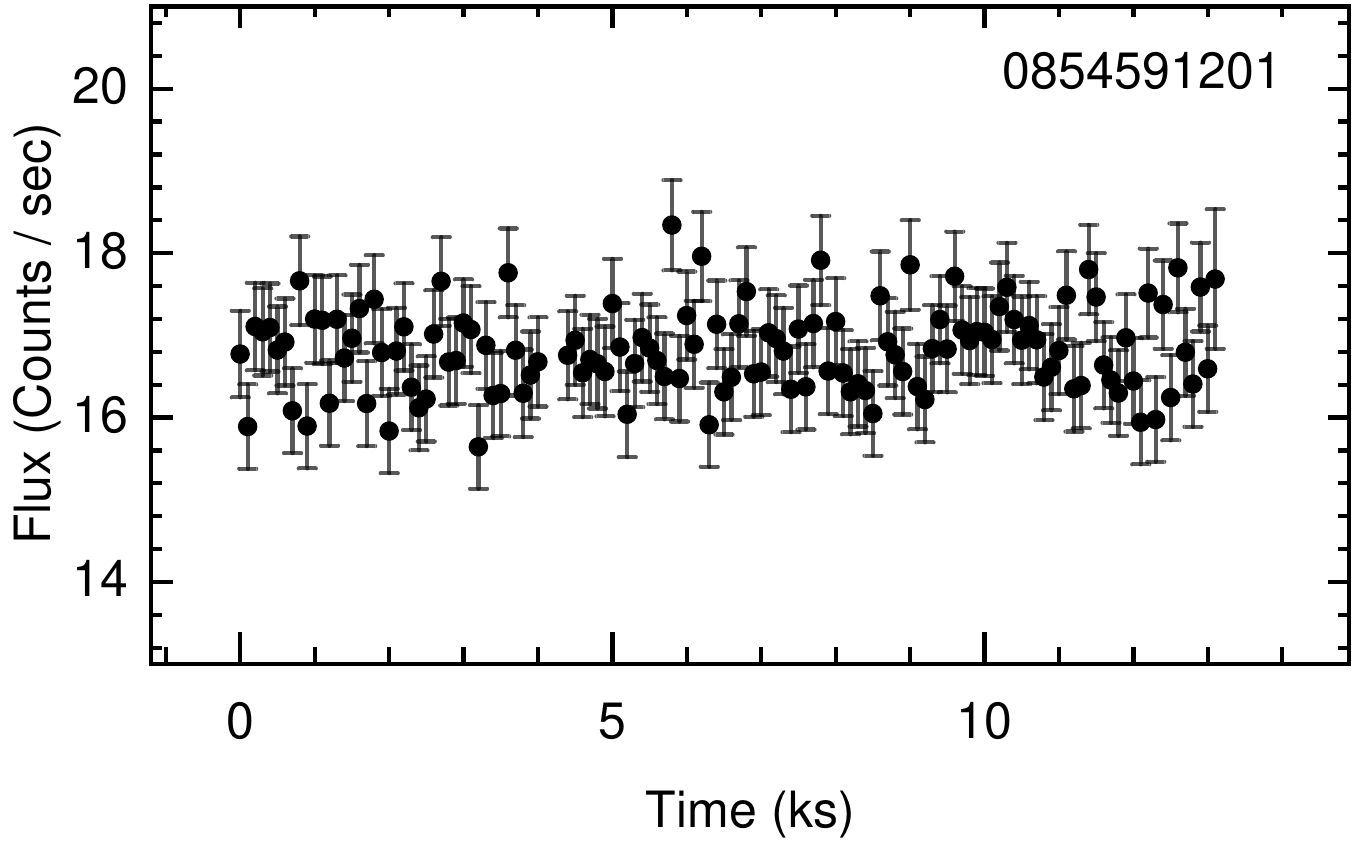}
    \includegraphics[scale=0.35]{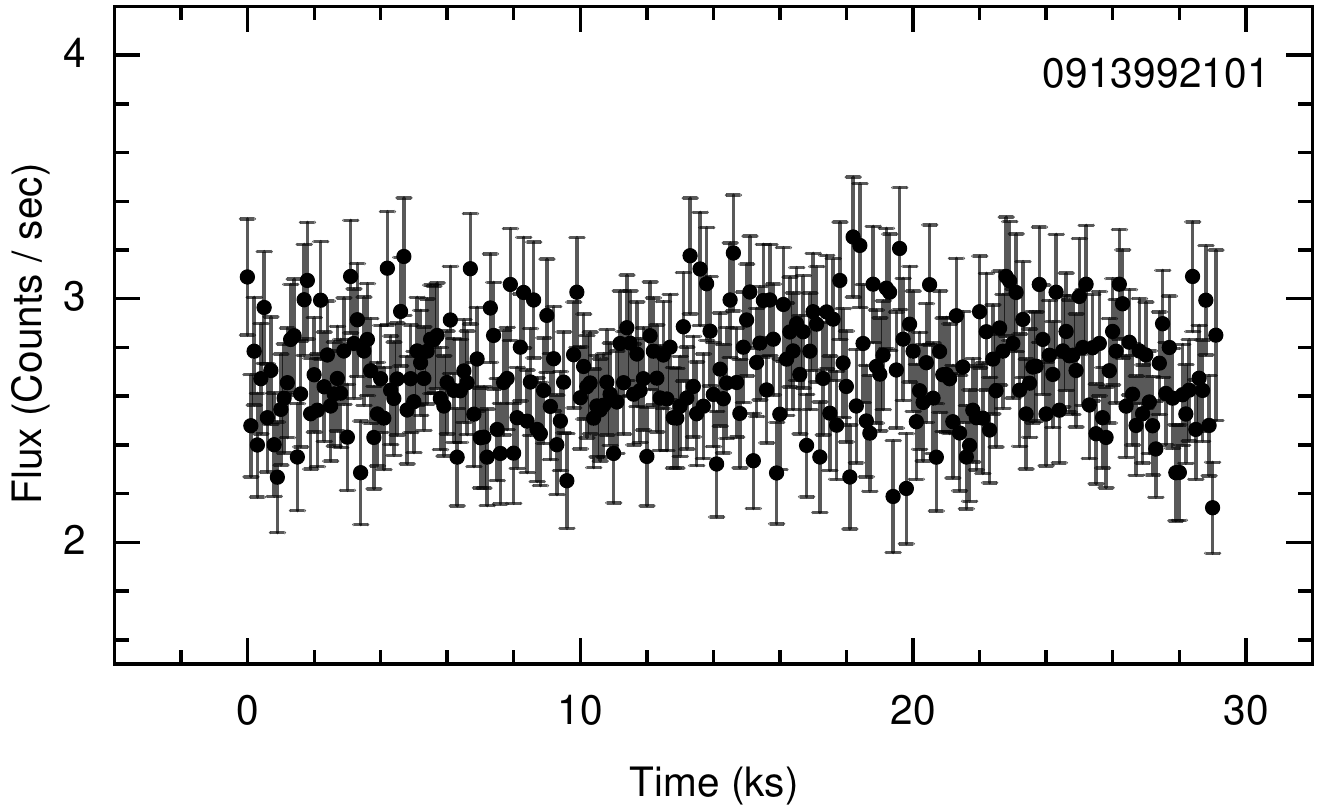}
    
    \caption{XMM-Newton observations of OJ 287: LCs for all observation IDs in the total band (0.2 -- 10 keV).}
    \label{total-LCs}
\end{figure*}

\begin{figure*}
    \centering

    \includegraphics[scale=0.35]{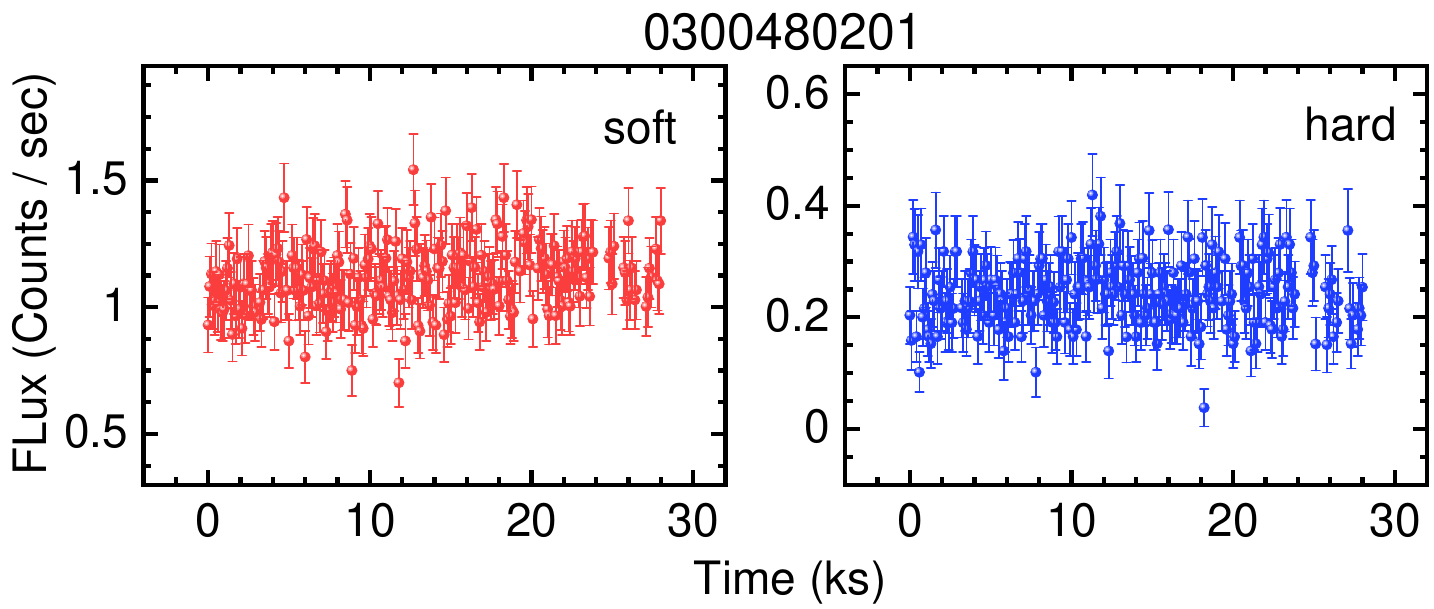}
    \includegraphics[scale=0.35]{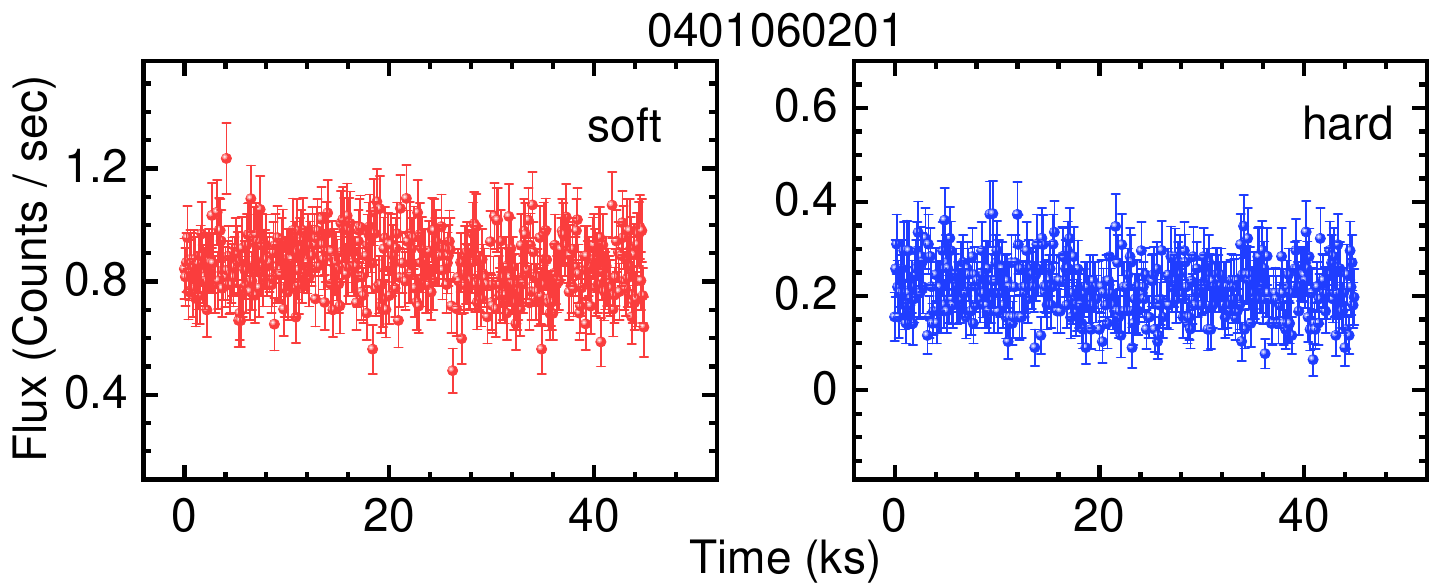}

    \vspace*{0.1in}
    \includegraphics[scale=0.35]{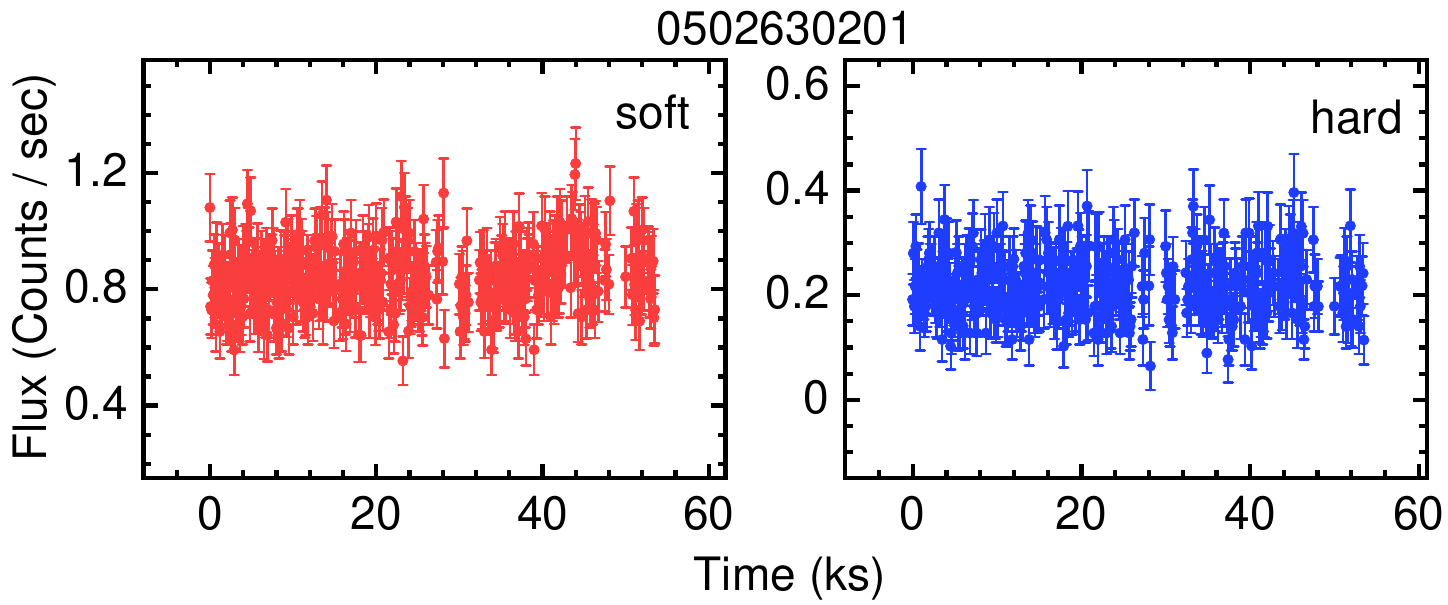}
    \includegraphics[scale=0.35]{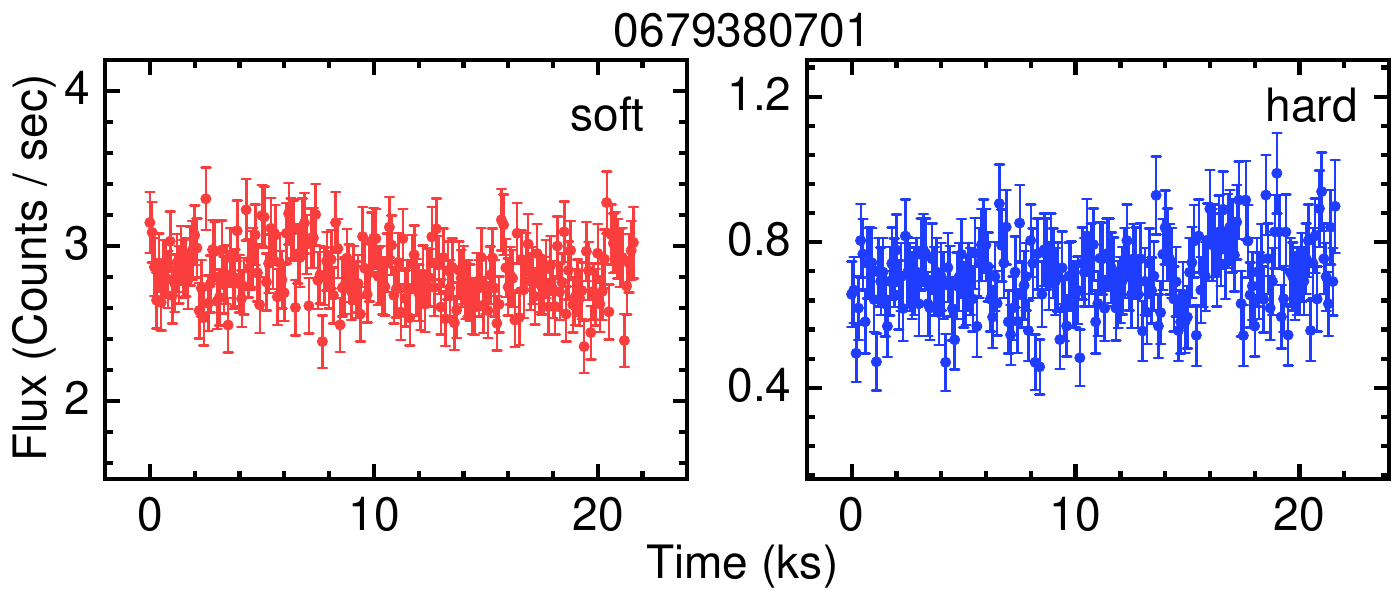}
    
    \vspace*{0.1in}
    \includegraphics[scale=0.35]{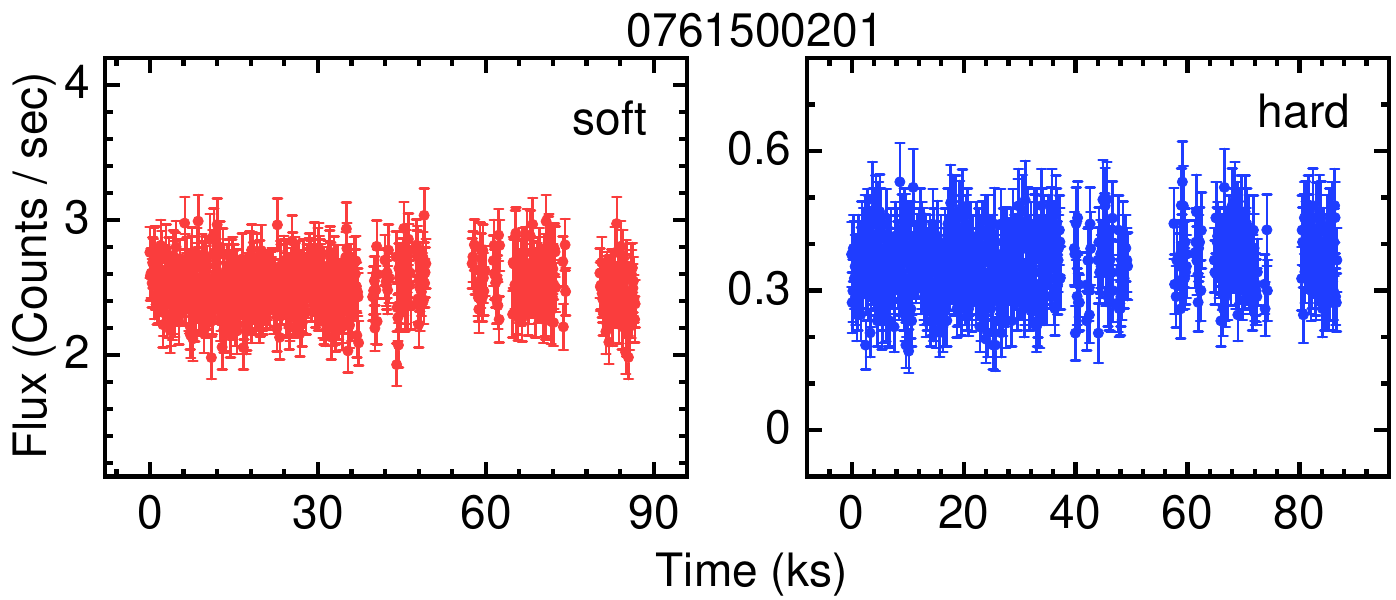}
    \includegraphics[scale=0.35]{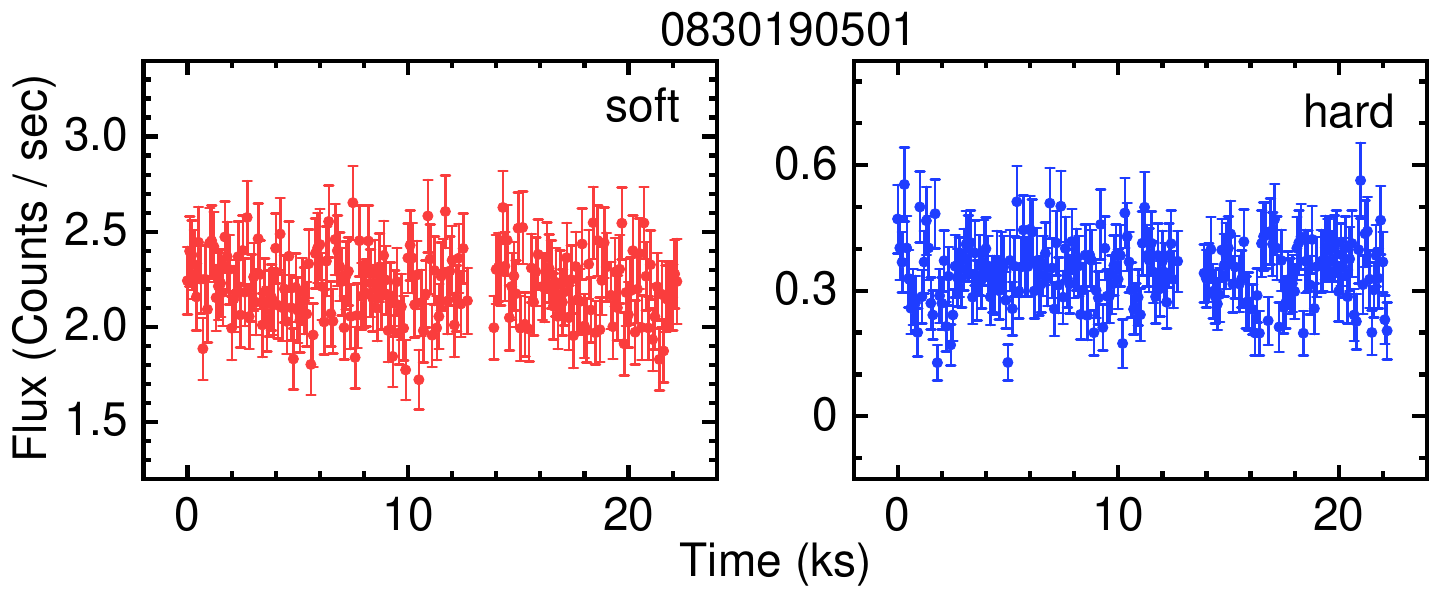}

    \vspace*{0.1in}
    \includegraphics[scale=0.35]{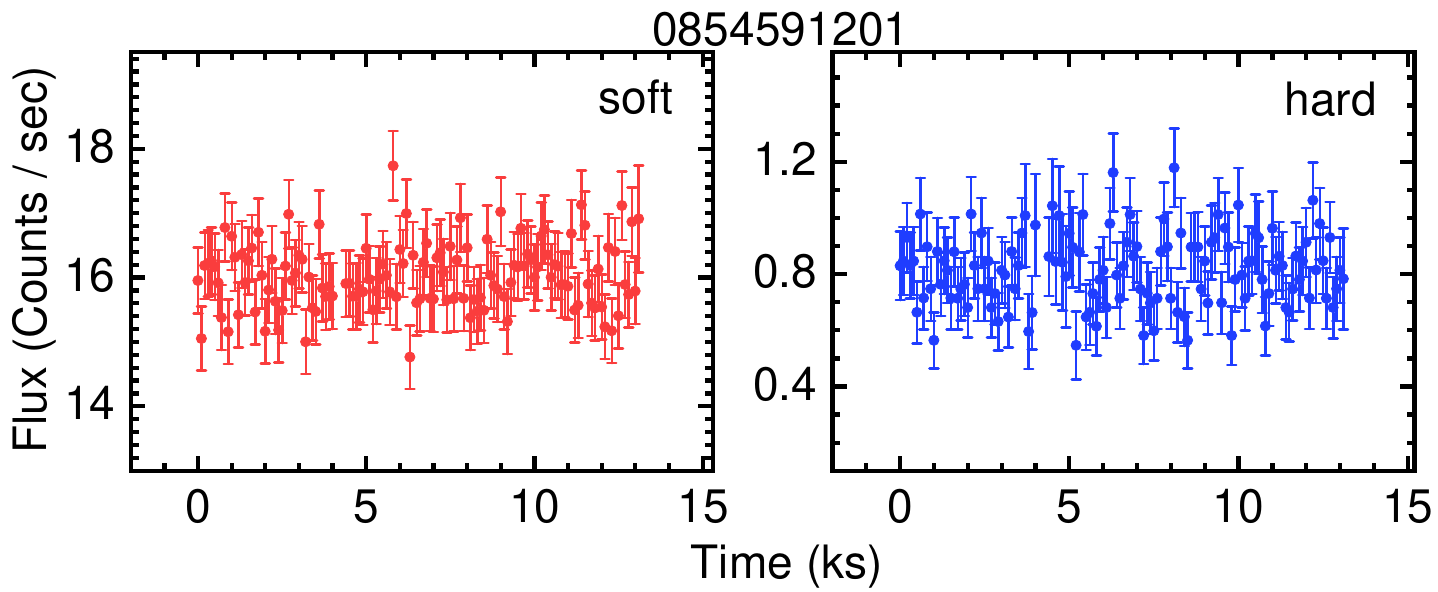}
    \includegraphics[scale=0.35]{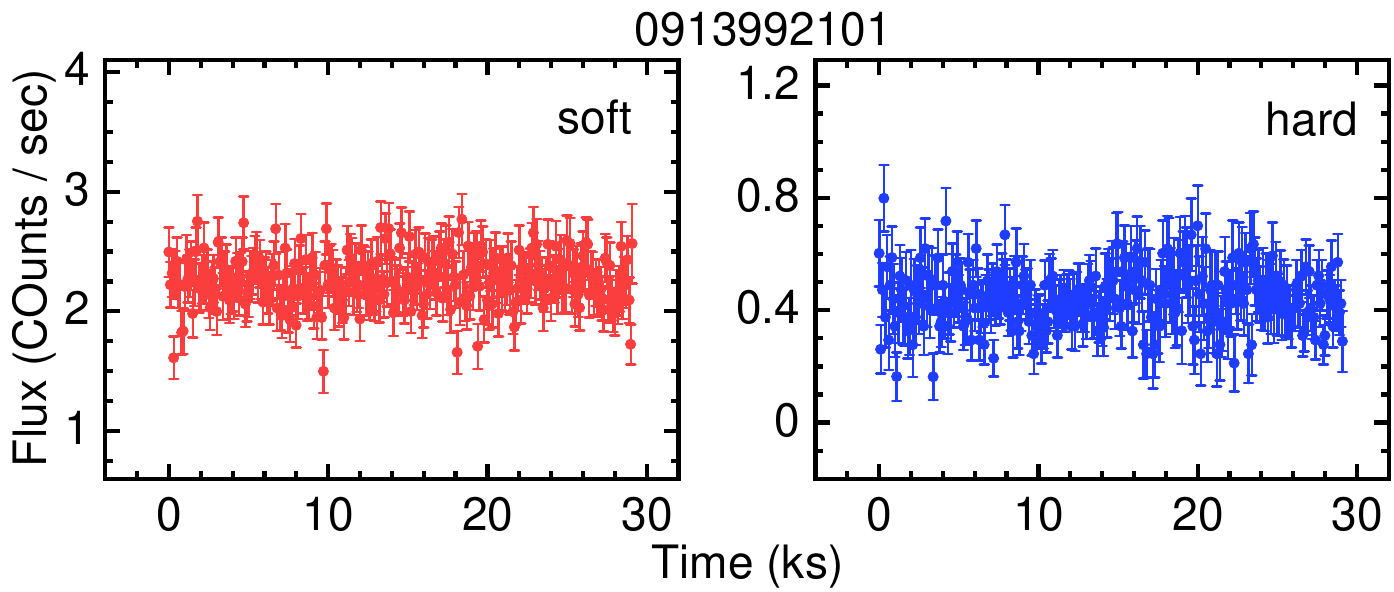}

\caption{Light curves (LCs) from 8 pointed XMM-Newton observations are shown in the soft (red points) and hard (blue points) bands. The corresponding observation IDs are indicated above each plot.}
\label{LCs:soft and hard}
\end{figure*}

\section{Results} \label{sec:results}
\noindent
In this section, we analyze the observational data from Table \ref{tab:oj287_obslog} using these methods introduced in the previous section and present corresponding results.

\subsection{Intraday Variation In X-ray Flux} 
\noindent
We generated LCs for the soft (0.2 - 2.0 keV), hard (2.0 - 10.0 keV) and total (0.2 - 10.0 keV) energy bands using data from 8 pointed observations made with XMM-Newton. Figure \ref{total-LCs} presents the total energy band LCs for all 8 observation IDs, while Figure \ref{LCs:soft and hard} shows the corresponding soft (red points) and hard (blue points) energy band decompositions. \\
\\
We adopted the excess variance approach (Section \ref{subsec:excess variance}) to quantify the strength of flux variability in OJ 287 on intraday timescales. For each observation, we calculated $F_{var}$ and its associated uncertainty in three energy bands (Table \ref{tab:variability}). We regard a LC as significant variability if its fractional variability satisfies the condition $F_{var} > 3\times(F_{var})_{err}$ (see \cite{2021MNRAS.506.1198D}). In observation IDs 0502630201 and 0401060201, IDV is detected across all three energy bands and other observations exhibit IDV in only one or two bands. IDV is identified only in the hard band and the flux is much greater than other observations. For observation ID 0913992101, variability is detected solely in the total band, where the ratio $F_{var} / (F_{var})_{err}$ just exceeds 3. Among the 8 LCs analysed in each energy band, those that meet the above criterion are considered to exhibit significant IDV. Based on this criterion, IDV is detected in 4 LCs in the soft band, 5 in hard band, and 6 in total band. The corresponding variability timescales, as defined in Section \ref{subsec:timescale}, are also provided in Table \ref{tab:variability}.
\noindent

\begin{figure*}
    \centering
    \includegraphics[scale=0.35]{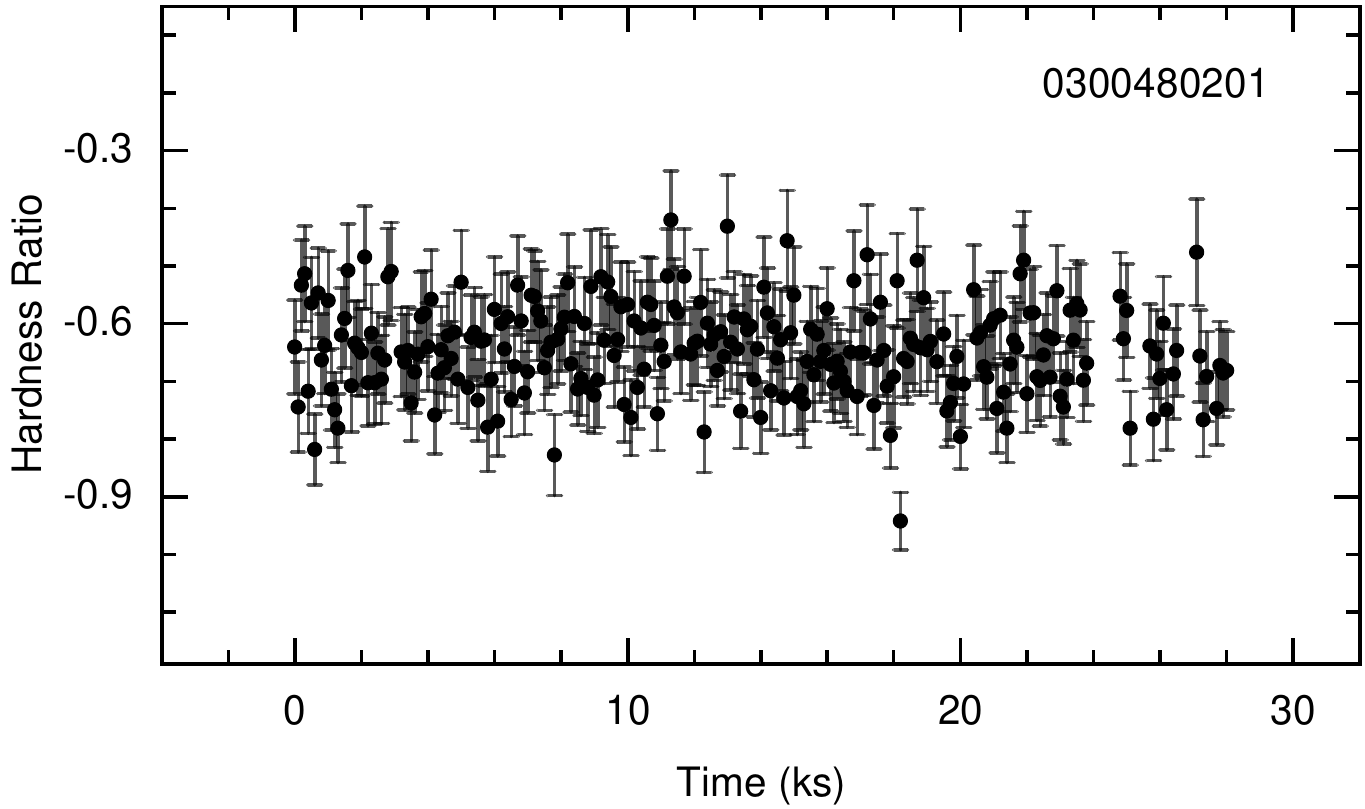}
    \includegraphics[scale=0.35]{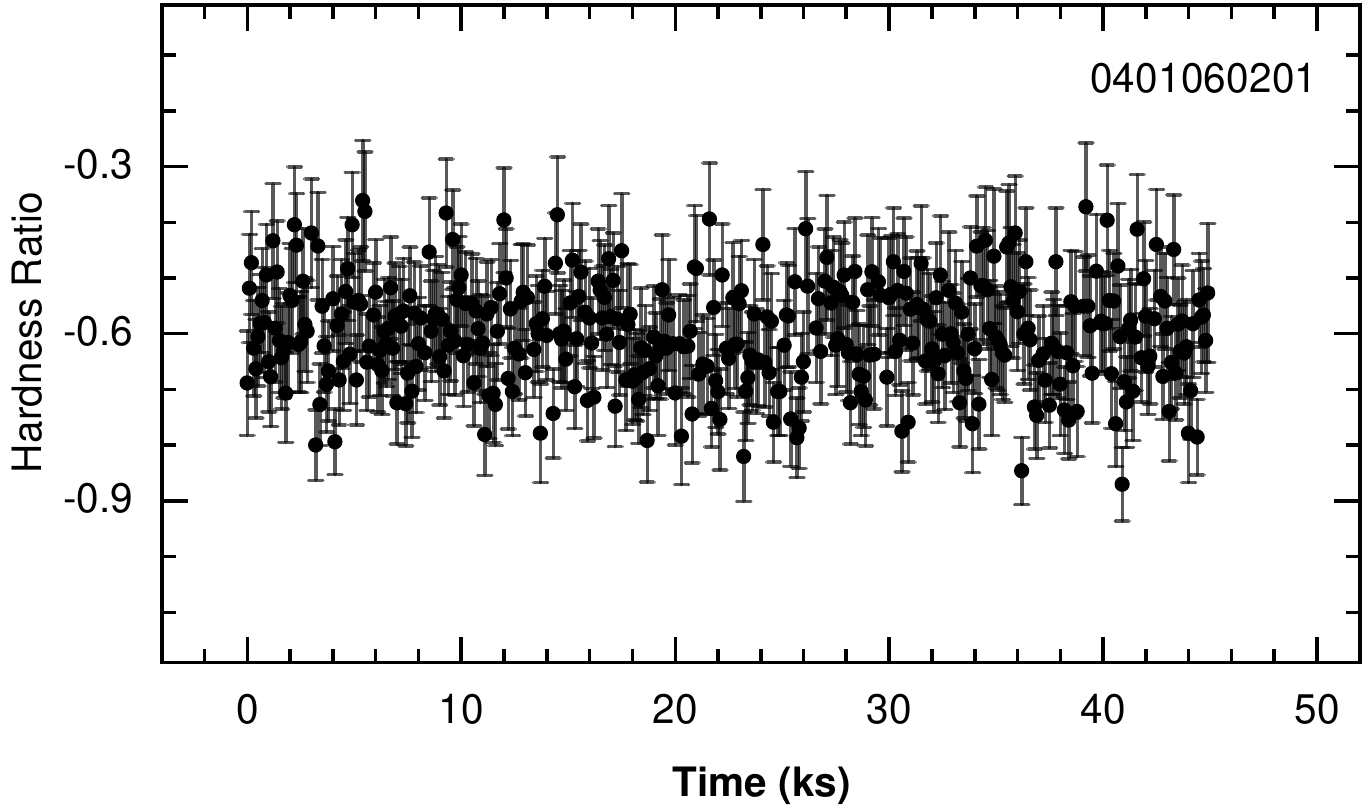}

    \vspace*{0.1in}
    \includegraphics[scale=0.35]{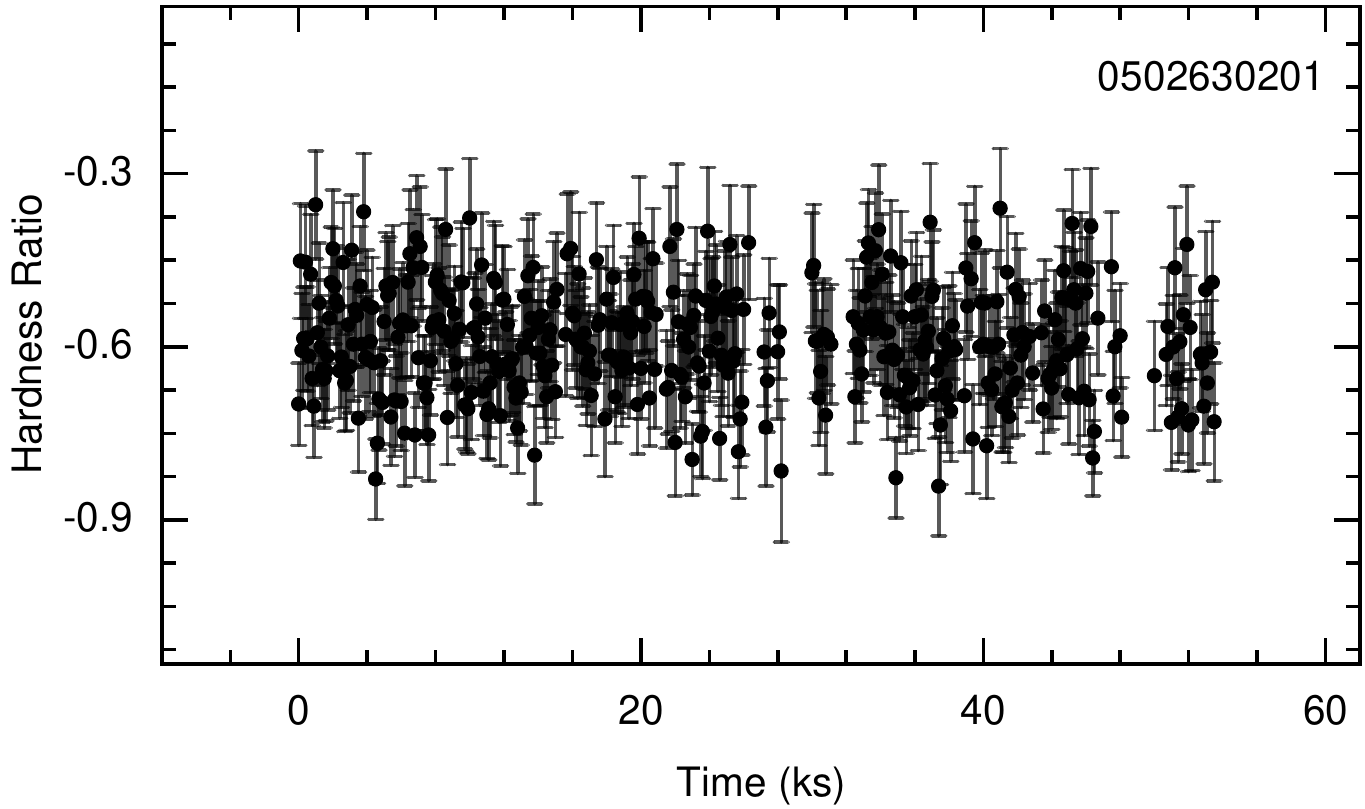}
    \includegraphics[scale=0.35]{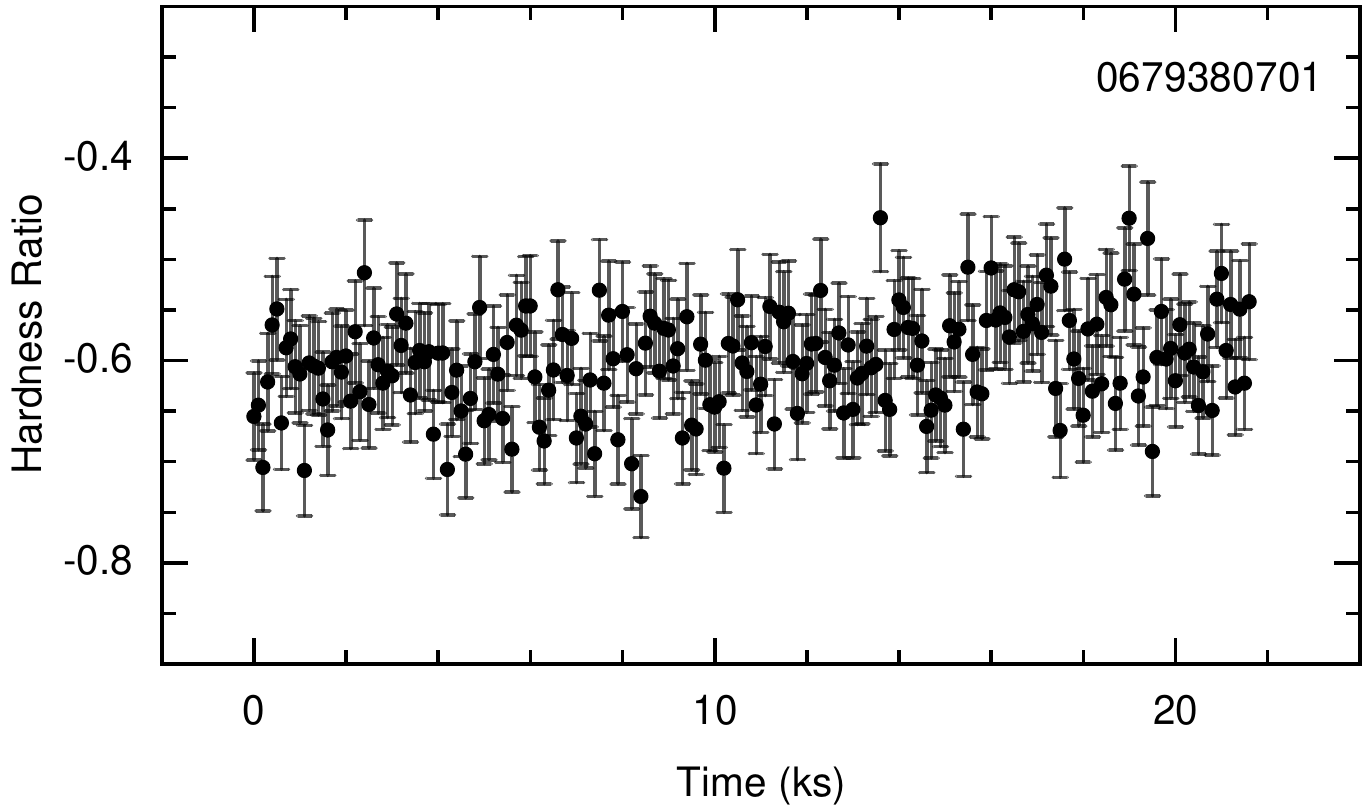}

    \vspace*{0.1in}
    \includegraphics[scale=0.35]{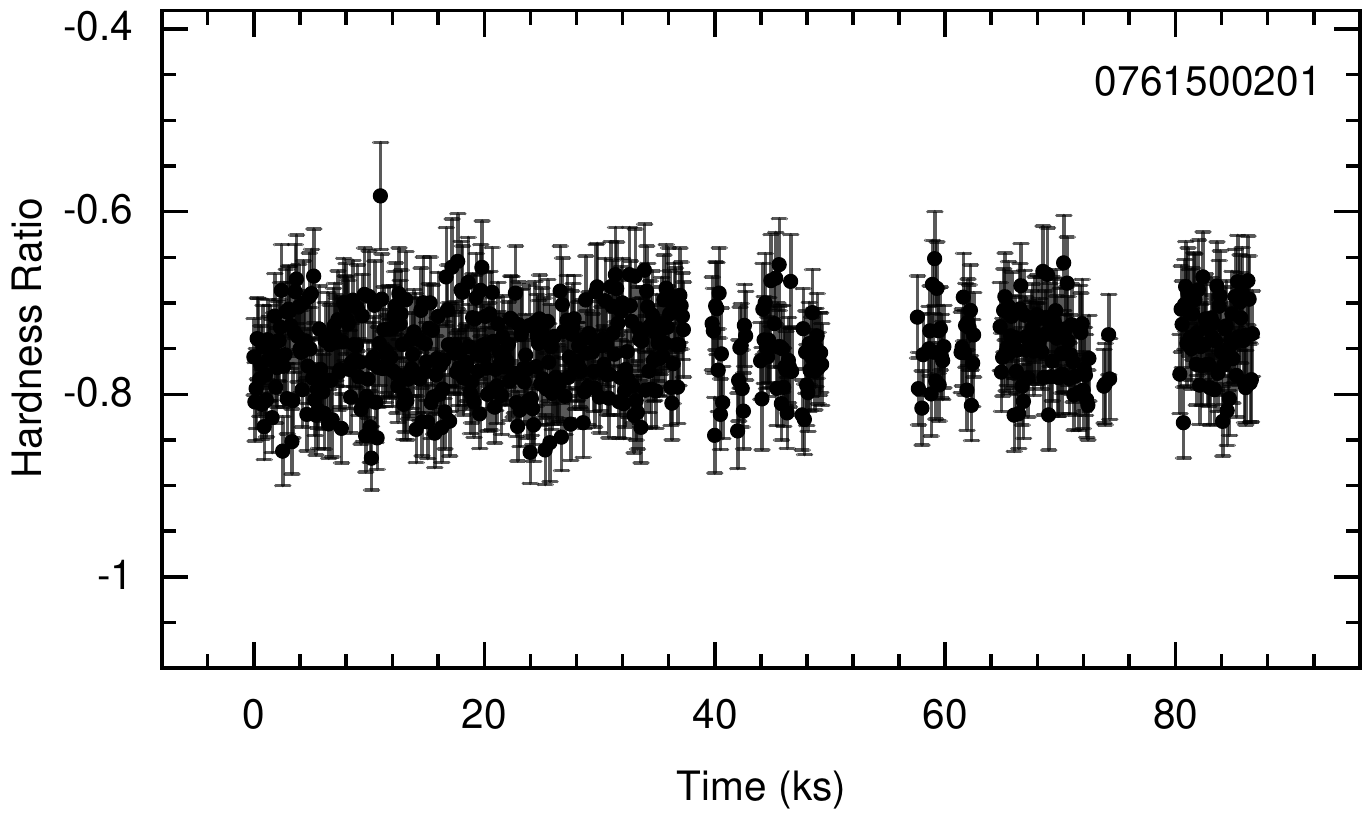}
    \includegraphics[scale=0.35]{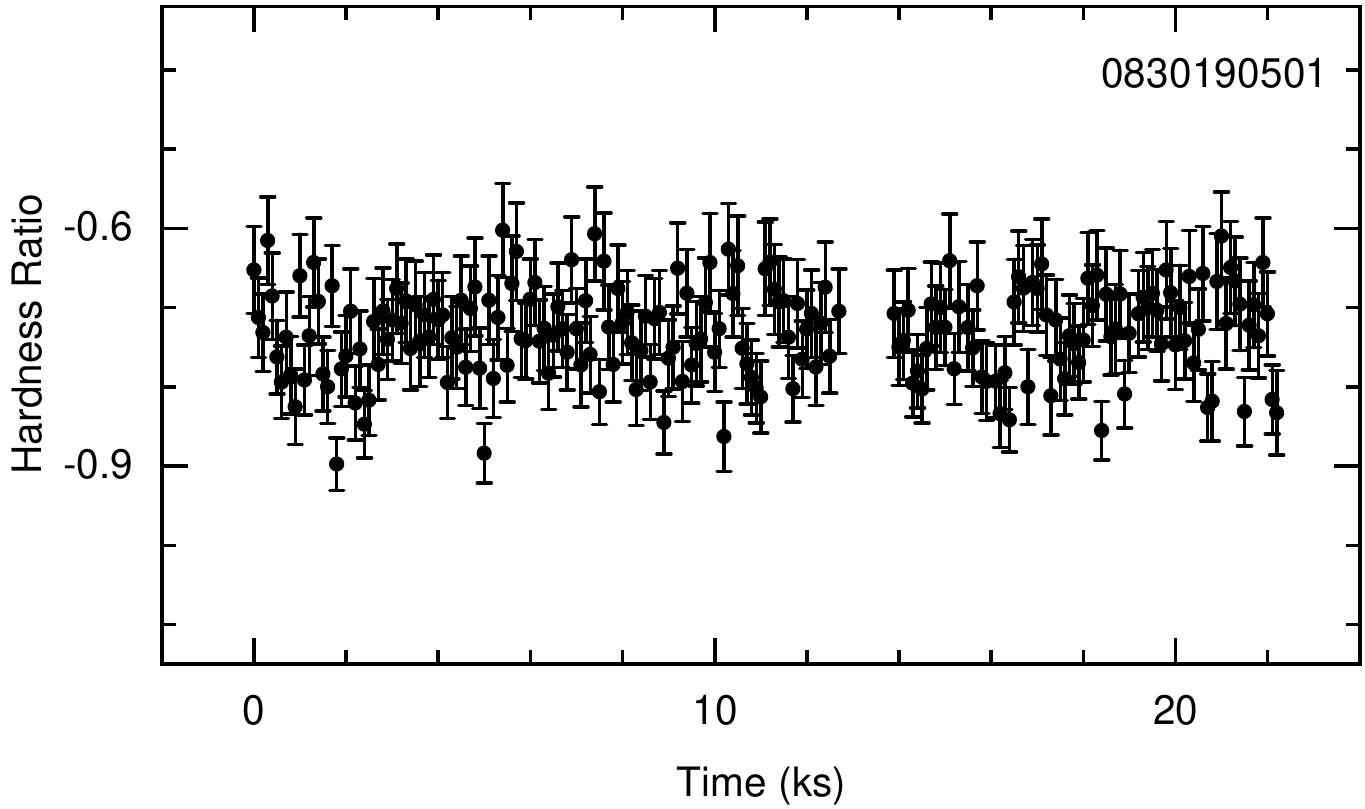}

    \vspace*{0.1in}
    \includegraphics[scale=0.35]{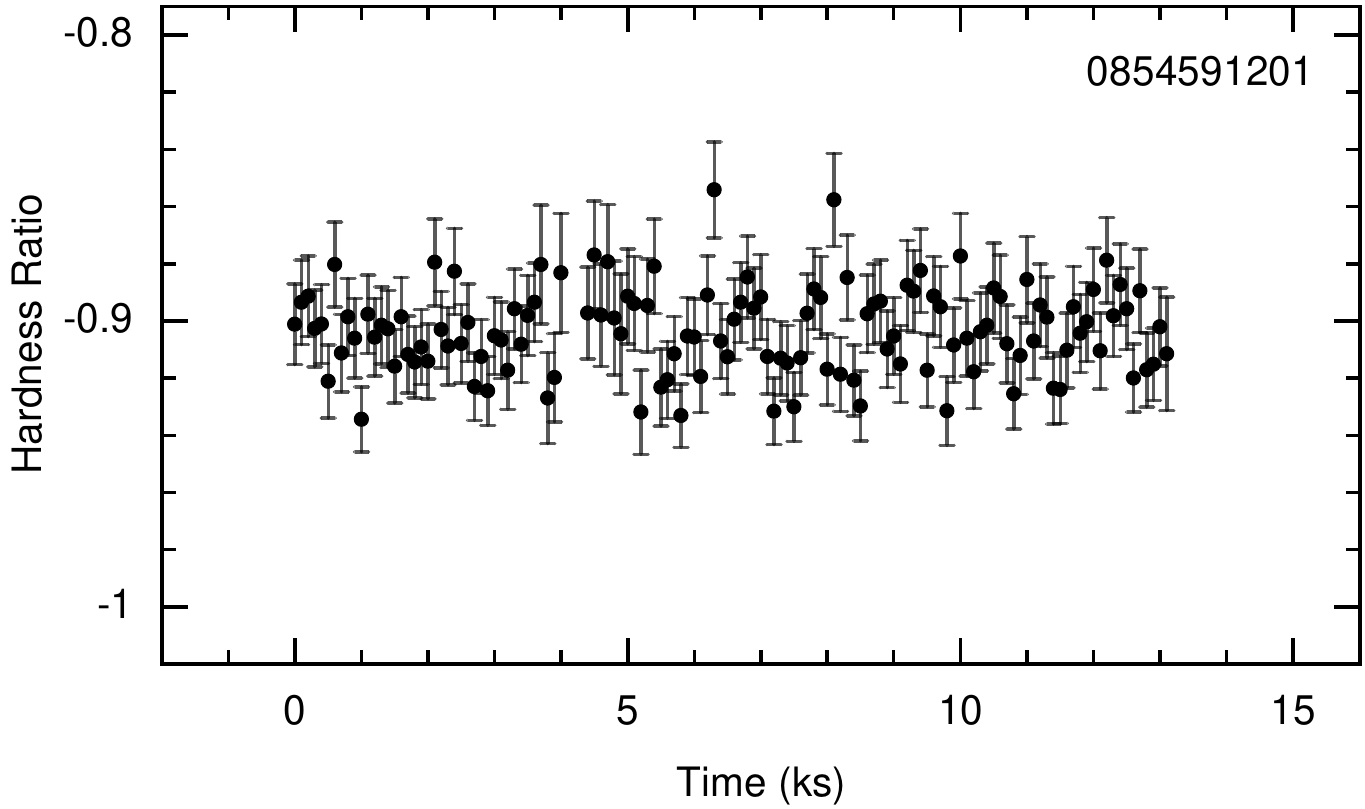}
    \includegraphics[scale=0.35]{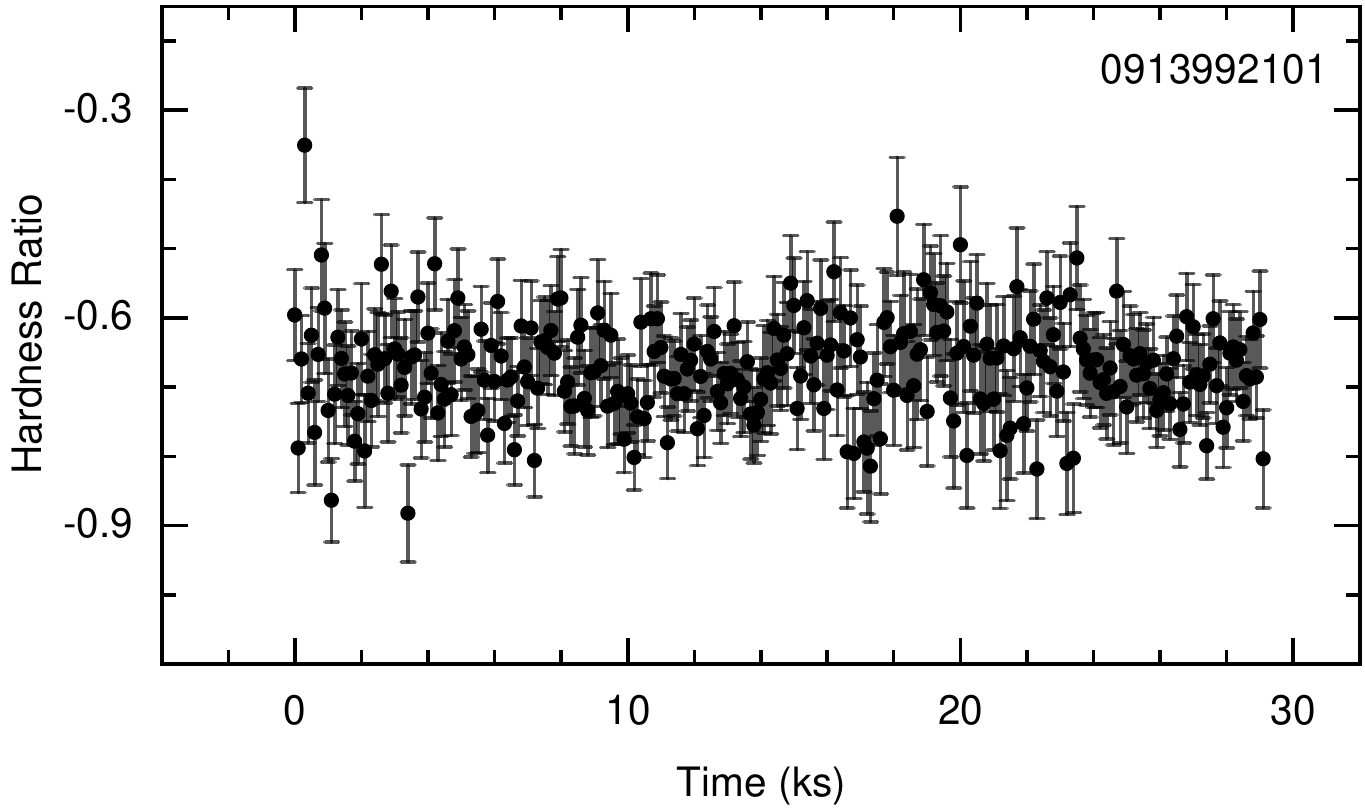}
    
    \caption{HR plots for all observations by XMM-Newton.}
    \label{fig:HR}
\end{figure*}

\begin{figure}
    \centering
    \includegraphics[scale=0.18]{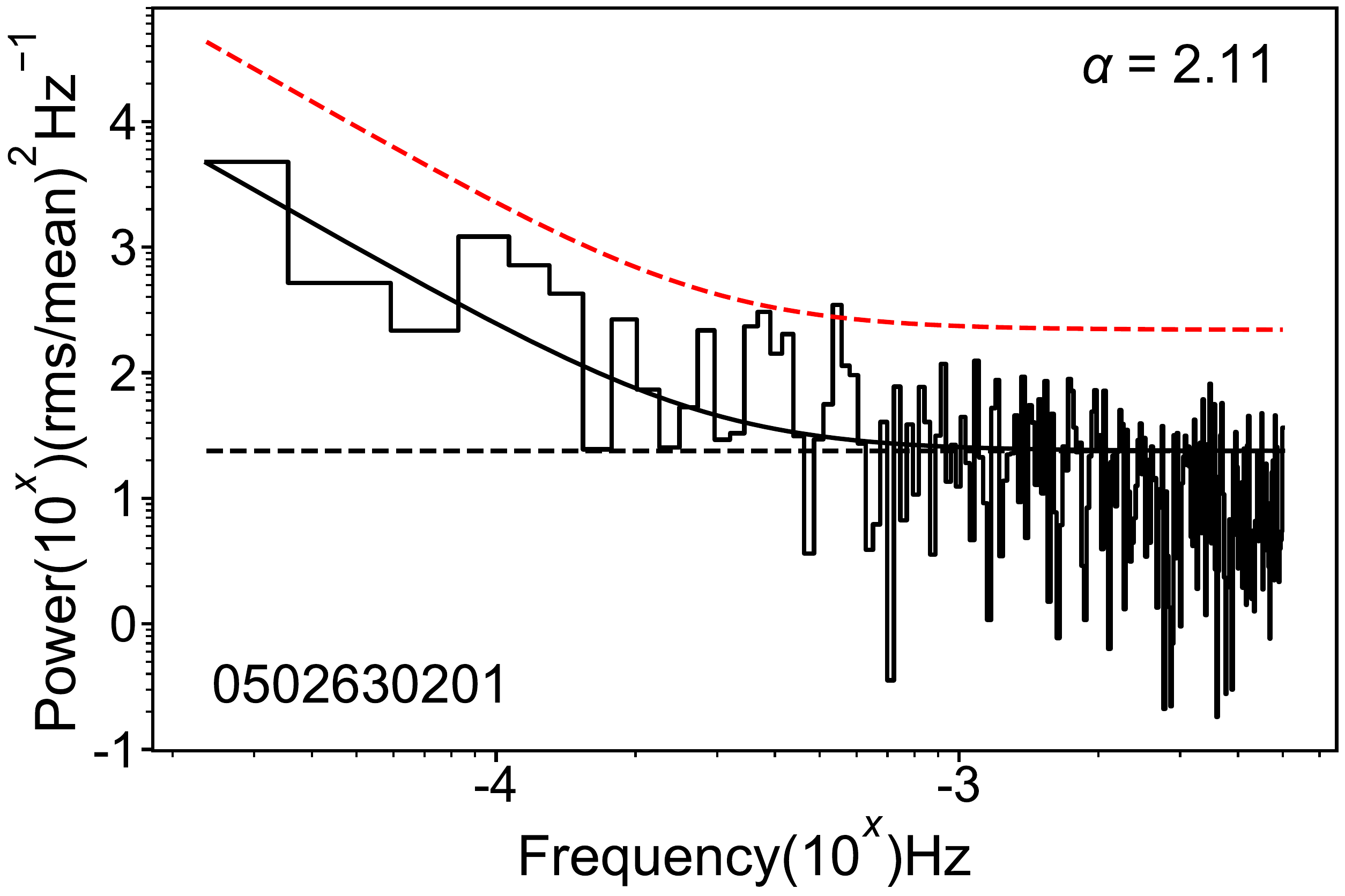}
    \includegraphics[scale=0.18]{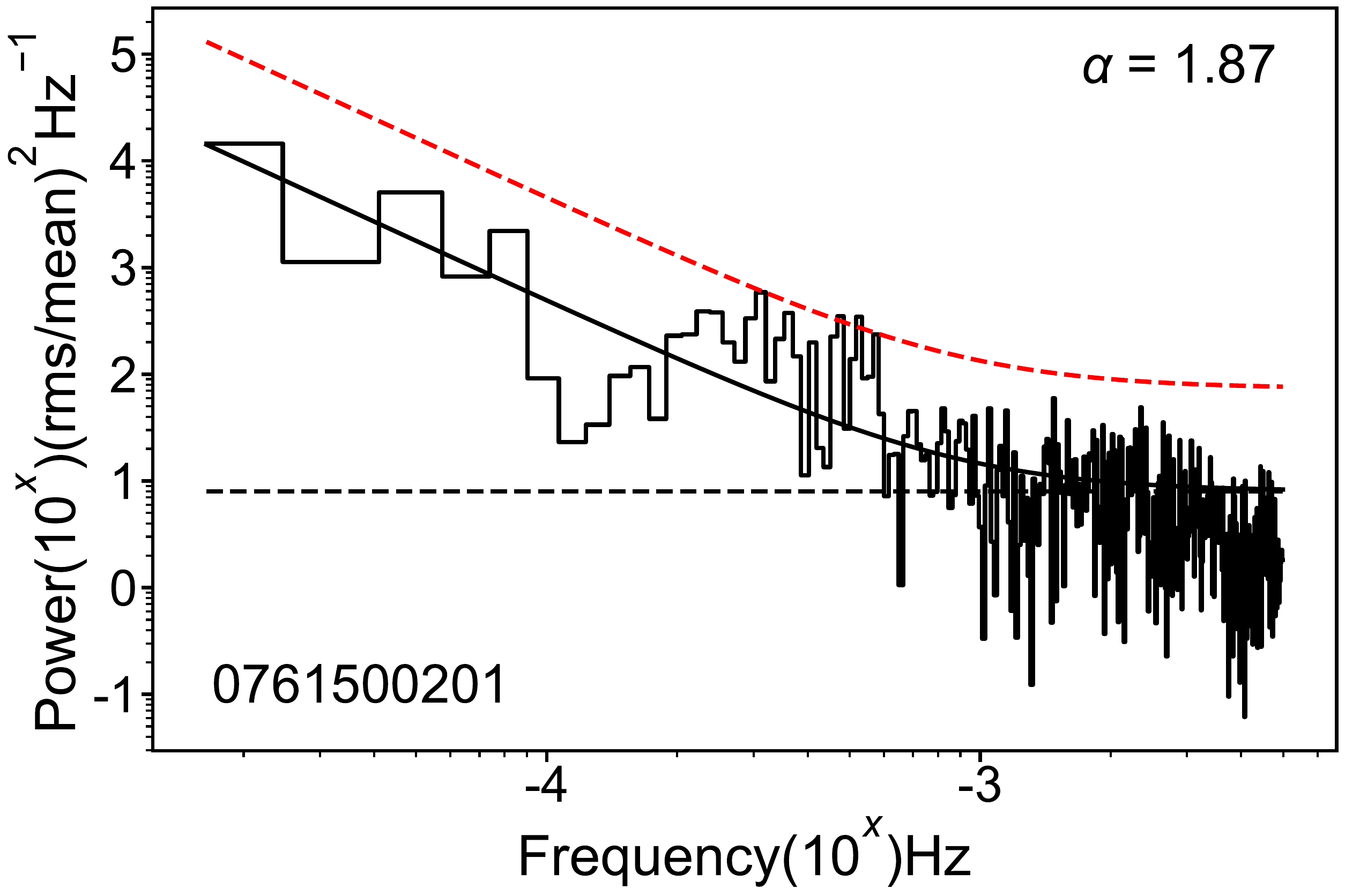}
    \caption{The PSD of all LCs (variable) in the soft energy band from XMM-Newton
    observations. The Obs ID and spectral index are labeled. {{The solid line represents the power-law fit to the red noise. The red dashed curve represents the 3$\sigma$ level of the red noise, and the dashed horizontal line indicates the Poisson noise level.} }}
    \label{fig:PSD_soft}
\end{figure}

\begin{figure}
    \centering
    \includegraphics[scale=0.18]{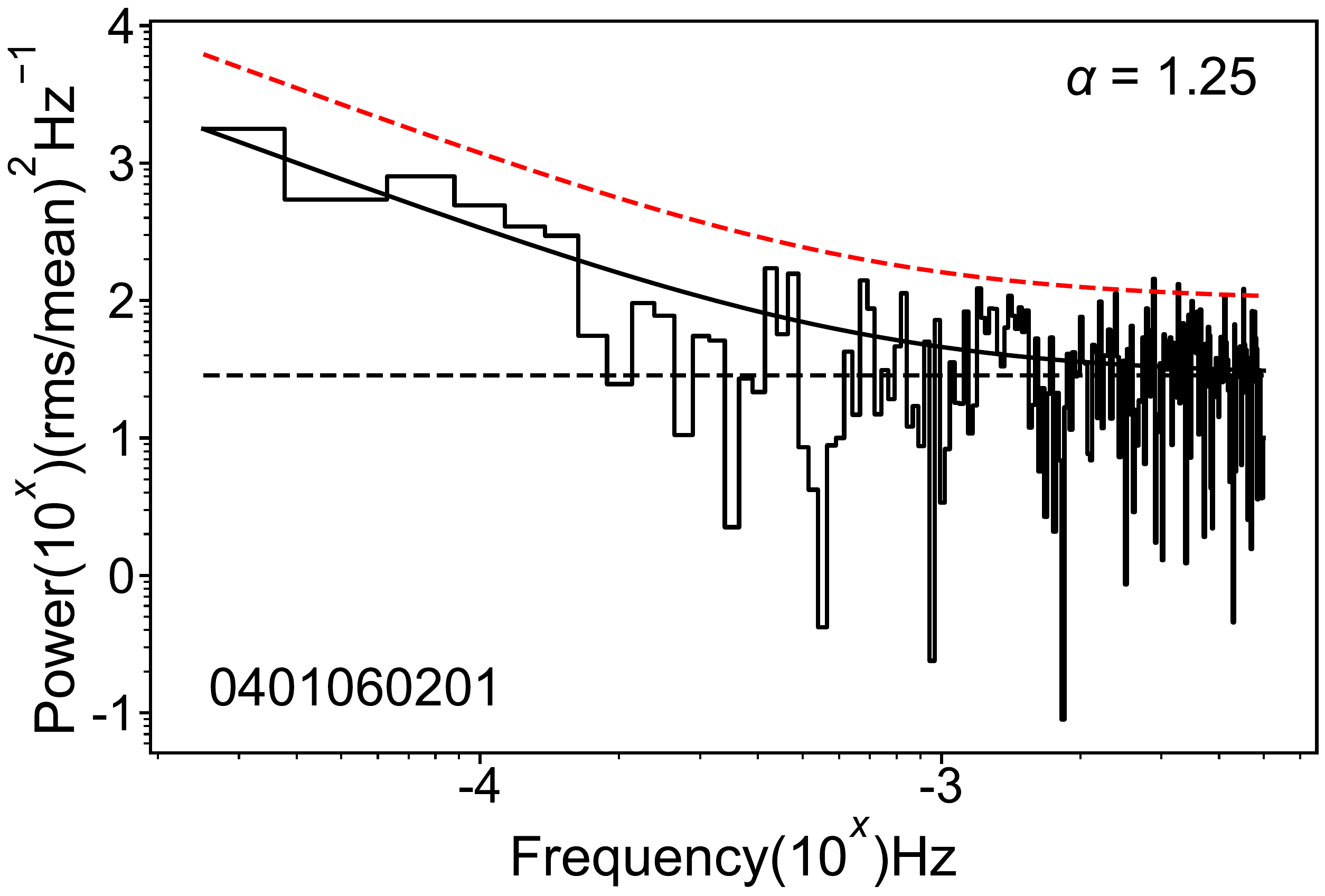}
    \includegraphics[scale=0.18]{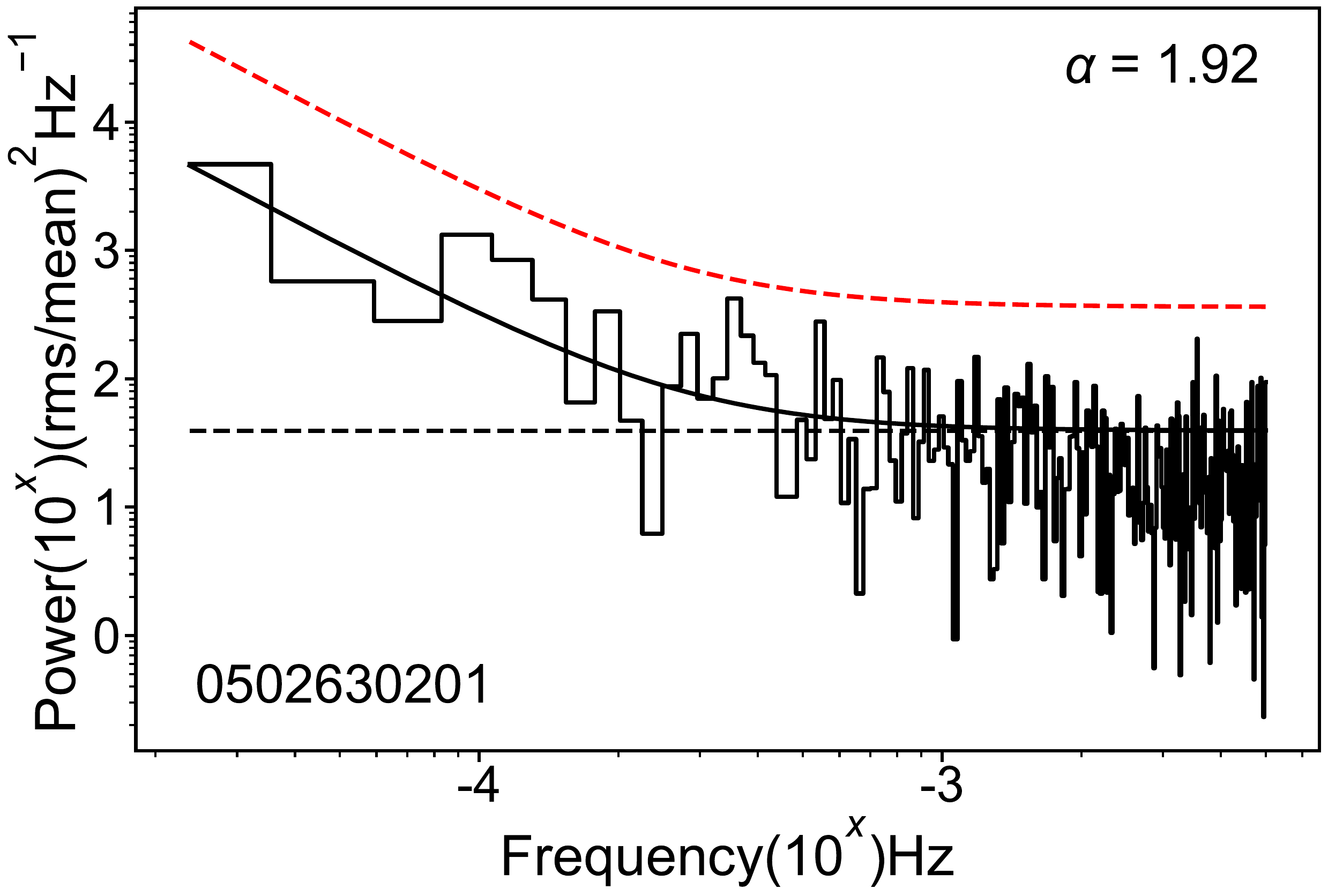}
    \caption{The PSD of all LCs (variable) in the hard energy band from XMM-Newton
    observations. The Obs ID and spectral index are labeled. {{The solid line represents the power-law fit to the red noise. The red dashed curve represents the 3$\sigma$ level of the red noise, and the dashed horizontal line indicates the Poisson noise level.} }}
    \label{fig:PSD_hard}
\end{figure}

\begin{figure*}
    \centering
    \includegraphics[scale=0.2]{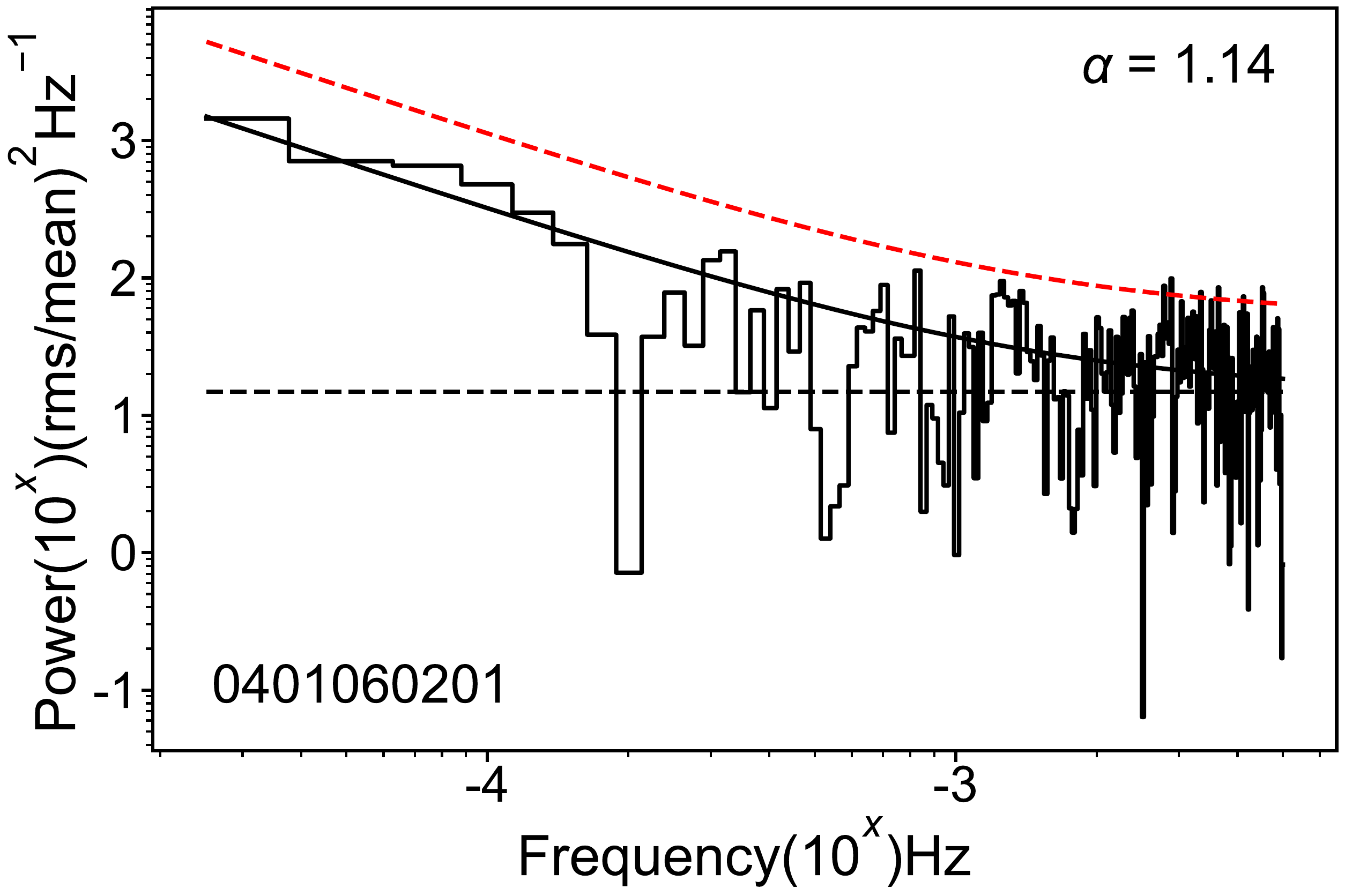}
    \includegraphics[scale=0.2]{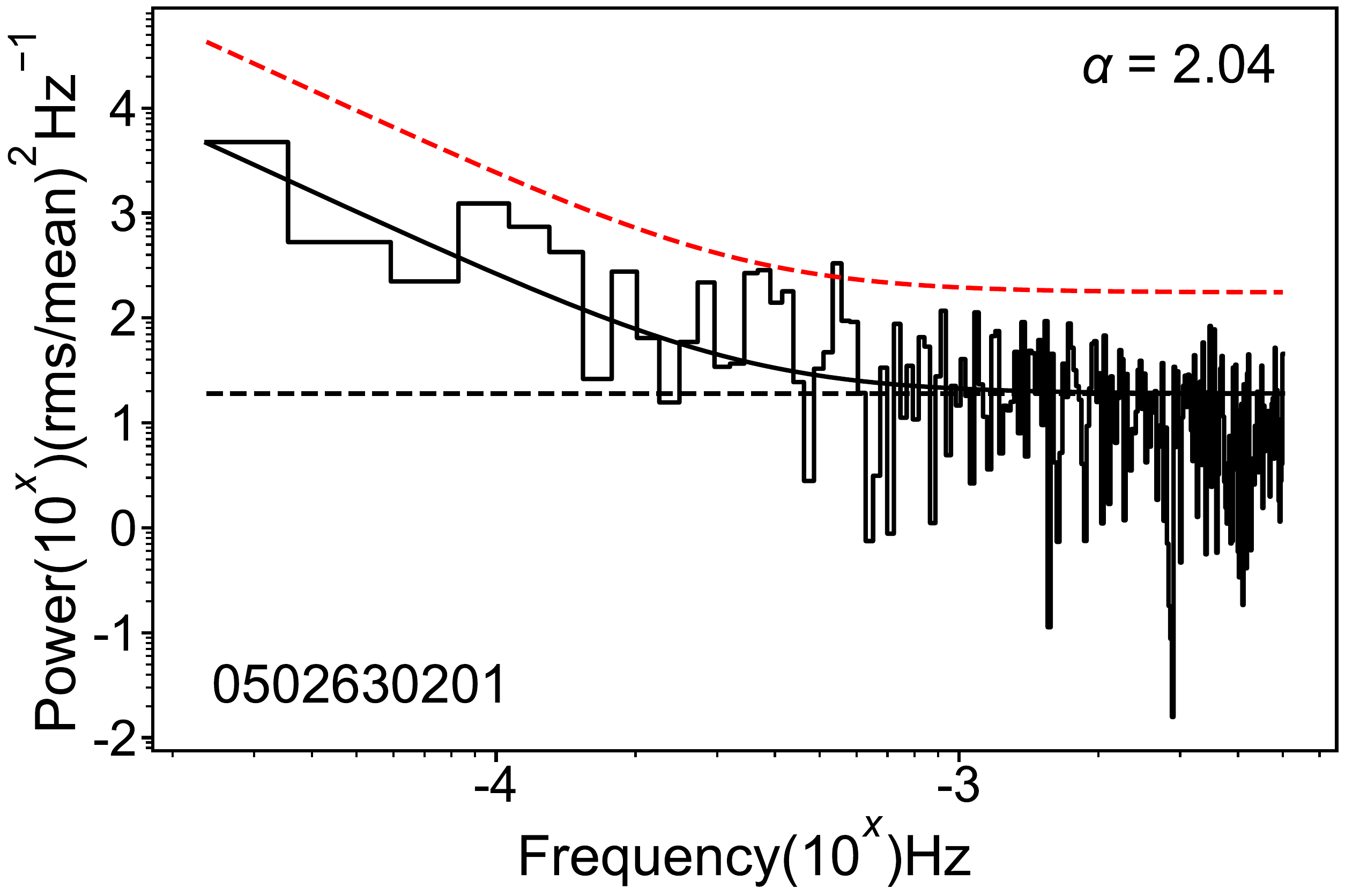}
    \includegraphics[scale=0.2]{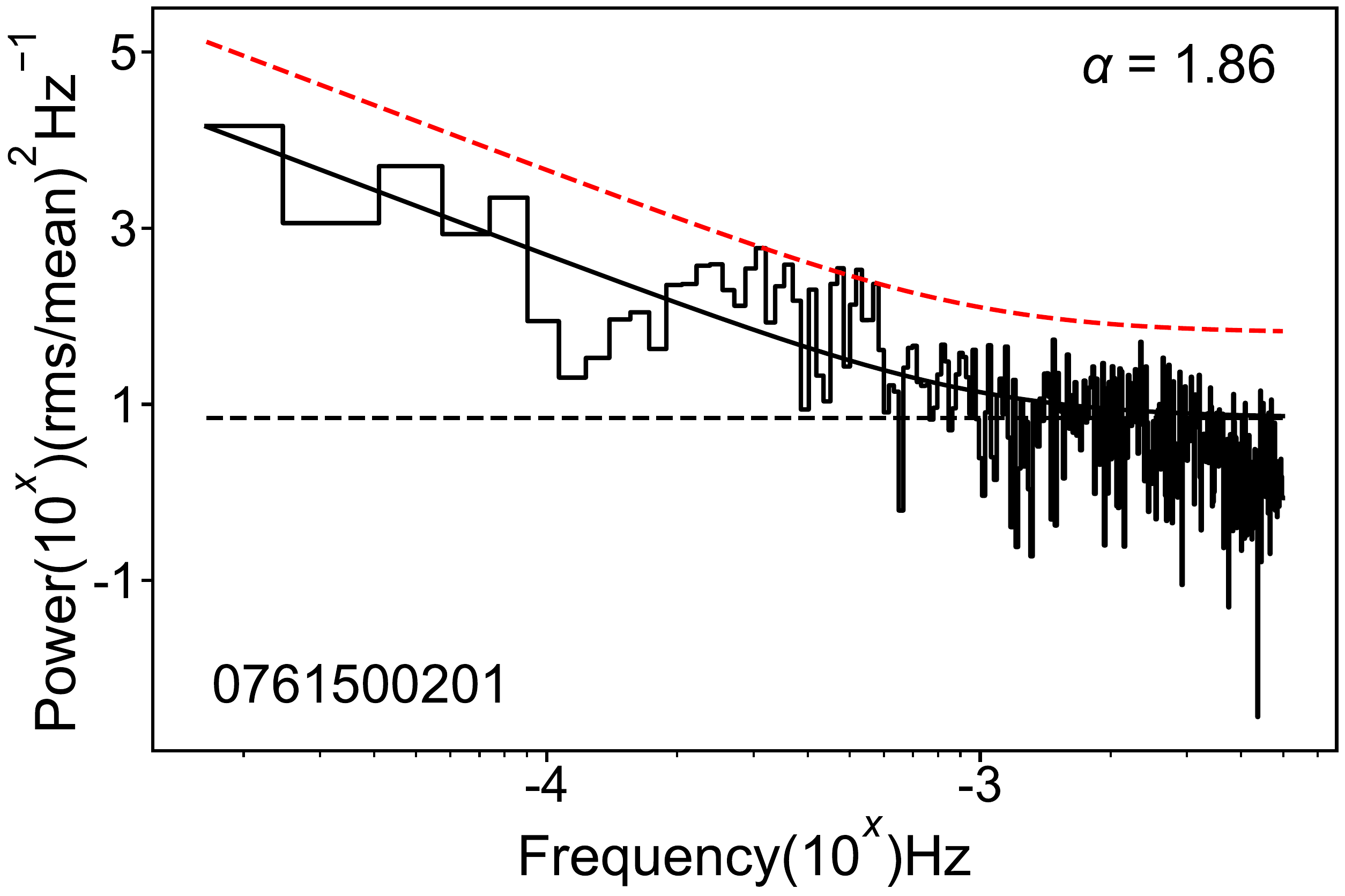}
    \includegraphics[scale=0.2]{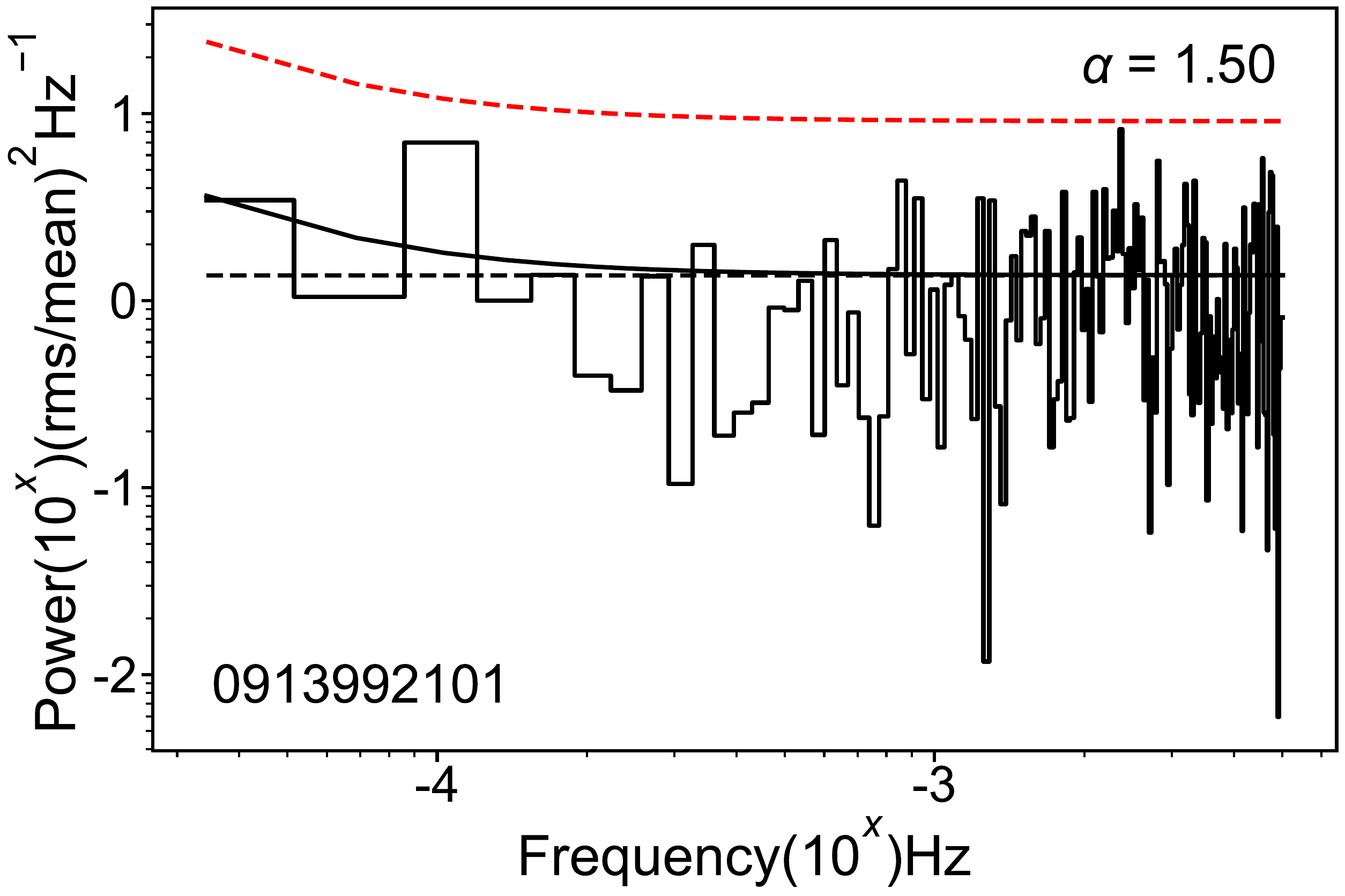}
    \caption{The PSD of all LCs (variable) in the total energy band from XMM-Newton
    observations. The Obs ID and spectral index are labeled. {{The solid line represents the power-law fit to the red noise. The red dashed curve represents the 3$\sigma$ level of the red noise, and the dashed horizontal line indicates the Poisson noise level.} }}
    \label{fig:PSD_total}
\end{figure*}

\subsection{Intraday Spectral Variability}
\noindent
The HR is a fundamental and model-independent metric commonly used to characterize spectral variations in X-ray emission across different energy bands (soft and hard bands). Figure \ref{fig:HR} presents the HR versus time for all observation IDs.  The mean HR values we obtained for each LC are reported in the last column of \autoref{tab:oj287_obslog}. Assuming an absorbed power law spectral model, the observed HR values in the range -0.59 to -0.90 correspond to effective photon indices roughly in the range of $\Gamma_{eff}\simeq1.71-2.70$ as shown in \autoref{tab:oj287_obslog}.
Our analysis shows that none of the 8 LCs exhibit significant HR evolution over time. This result is consistent with expectations, as most LCs show only small variations that are insufficient to cause significant spectral variations. 

\subsection{Power Spectral Density Analysis}\label{subsec:PSD analysis}
\begin{figure}
    \centering
    \includegraphics[width=\linewidth]{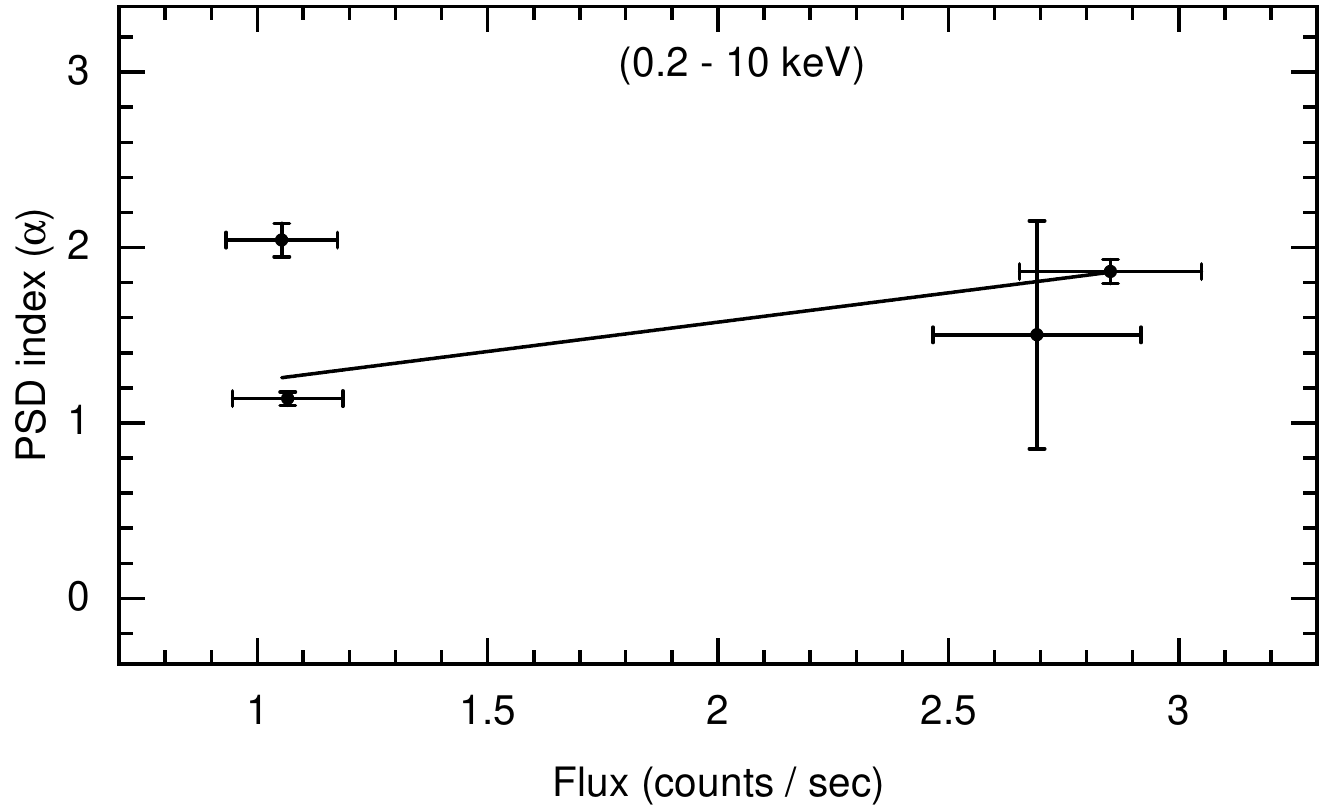}
    \caption{The PSD spectral index versus flux for the four observation IDs that show IDV in the total X-ray energy band.}
    \label{fig: PSD and flux}
\end{figure}

\noindent
A comprehensive study analyzing the X-ray LCs of different AGN subclasses has revealed that their power PSD is predominantly characterized by red noise, i.e., the PSD decreases steeply with frequency ($\nu$) in a power-law form $P(\nu) \sim \nu^{-\alpha}$, with a typical spectral exponent $\alpha \approx 2$ \citep{2012A&A...544A..80G}. Below a specific bending frequency ($\nu_b$), the PSD flattens out, and the $\nu_b$ value is approximately inversely proportional to central (AGN) SMBH mass \citep{2012A&A...544A..80G}. \\
\\
Based on this, we performed a PSD analysis of the blazar OJ 287. We used data obtained by the XMM-Newton throughout its entire observation period to perform PSD analysis on four observational data sets (i.e., variable LCs) that detected soft, hard, or total bands IDV, aiming to characterise the type of variability noise and search for QPOs. The analysis results provided the slope and normalisation parameters characterising the PSD power-law red noise component (see Table \ref{tab:psd_params}). The power-law fitting results of the PSD for the soft energy band are presented in Fig \ref{fig:PSD_soft}, while those for the hard energy band are presented in Fig \ref{fig:PSD_hard}, and the total energy band results are displayed in Figure \ref{fig:PSD_total}. The observation IDs and corresponding power-law indices are labeled in each plot. The analysis indicates that no evidence of QPO was found in these PSDs. \\
\\
We analyzed the PSD of LCs showing variability across different energy bands. The PSD slope ($\alpha$) ranged from 1.14 $\pm$ 0.04 (close to 1/f flicker noise) to 2.11 $\pm$ 0.10 (red noise). This slope range is frequently observed in AGN X-ray variability. The $log_{10}N$ values for the different energy bands fell between -6.35 $\pm$ 2.91 and -2.07 $\pm$ 0.17. \\
\\
In Fig \ref{fig: PSD and flux}, we analyzed the relationship between the PSD index ($\alpha$) and total X-ray flux. The results show a slope of 0.33, corresponding to a correlation coefficient $r = 0.44$ and a p-value = 0.34. A higher p-value indicates does not support a statistically significant between the PSD index and flux. Due to the limited data points and the lack of data in the soft and hard bands, we could not draw reliable conclusions about trends.

\begin{deluxetable*}{ccccccc}
\tablenum{3}
\tablecaption{Power-law fitting parameters of the PSD for OJ 287 in the soft, hard, and total energy bands}
\tablewidth{0pt}
\tablehead{
\colhead{Obs ID} & 
\multicolumn{2}{c}{Soft (0.2 -- 2 keV)} & 
\multicolumn{2}{c}{Hard (2 -- 10 keV)} & 
\multicolumn{2}{c}{Total (0.2 -- 10 keV)} \\
\colhead{} & 
\colhead{$\log_{10} N$} & 
\colhead{$\alpha$} & 
\colhead{$\log_{10} N$} & 
\colhead{$\alpha$} & 
\colhead{$\log_{10} N$} & 
\colhead{$\alpha$}
}
\decimalcolnumbers
\startdata
0300480301 & -- & -- & {{NV}} & {{NV}} & -- & -- \\
0401060201 & {{--}} & {{--}} & $-2.51\pm0.22$ & $1.25\pm0.05$ & $-2.07\pm0.17$ & $1.14\pm0.04$ \\
0502630201 & $-6.10\pm0.46$ & $2.11\pm0.10$ & $-5.23\pm0.42$ & $1.92\pm0.09$ & $-5.78\pm0.44$ & $2.04\pm0.10$ \\
0679380701 & {{NV}} & {{NV}} & {{--}} & {{--}} & {{--}} & {{--}} \\
0761500201 & $-4.79\pm0.33$ & $1.87\pm0.06$ & {{NV}} & {{NV}} & $-4.76\pm0.32$ & $1.86 \pm 0.07$ \\
0830190501 & {{NV}} & {{NV}} & -- & -- & {{NV}} & {{NV}} \\
0854591201 & {{NV}} & {{NV}} & -- & -- & {{NV}} & {{NV}} \\
0913992101 & {{NV}} & {{NV}} & {{NV}} & {{NV}} & $-6.35\pm2.91$ & $1.50\pm0.65$ \\
\enddata
\tablenotetext{}{Note. $N$ represents the normalization factor, $\alpha$ denotes spectral index.}
\tablenotetext{}{{{ ``--'' denotes the variations were so small that the PSD would be dominated by Poisson noise with no red noise component.}}}
\tablenotetext{}{{{``NV'' denotes the observation is non-variable.}}}
\label{tab:psd_params}
\end{deluxetable*}

\begin{figure}
    \centering
    \includegraphics[width=\linewidth]{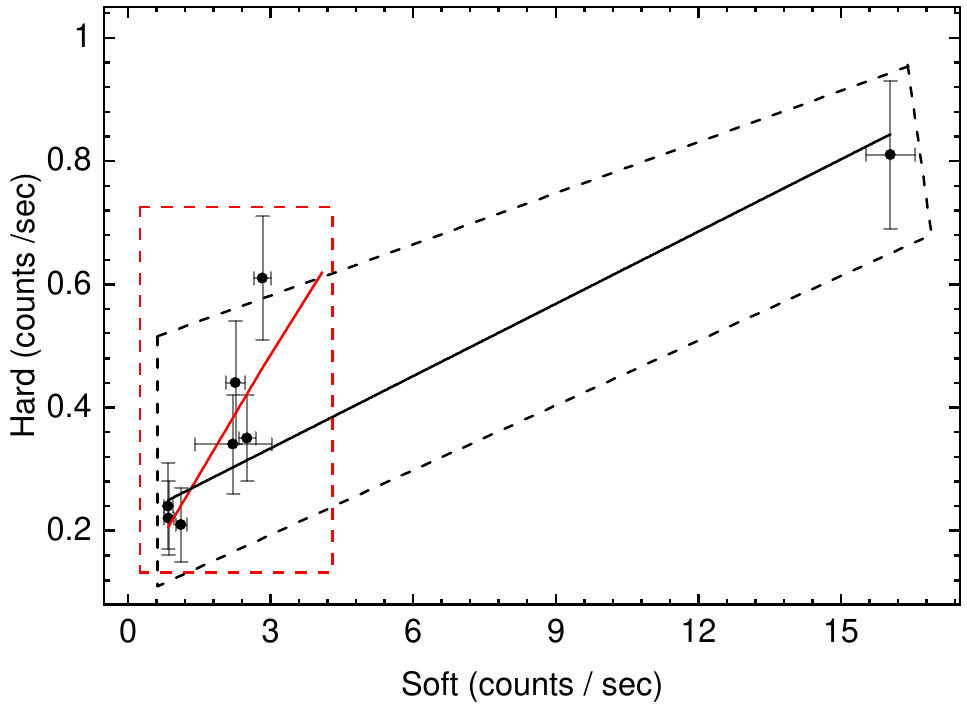}
    \caption{Inter-observation correlation of soft and hard X-ray count rates. {{The red dashed box represents the data set utilized for the linear fitting procedure after the highest flux data point was excluded. The black dashed box represents the data set utilized for the linear fitting procedure after the second-highest flux data point was excluded.}}}
    \label{fig:hard and soft flux}
\end{figure}

\subsection{Intraday cross-correlated variability}
\label{cross-correlated}
\noindent
Fig.\ref{fig:hard and soft flux} shows the relationship between the soft and hard X-ray fluxes across different epochs. A linear fit to the data within the black dashed box (black line) yields $r = 0.955$,  p-value = $8.11 \times 10^{-4}$. After excluding a high flux data point and performing a separate fit (red line), we get $r=0.869$, p-value = 0.011. Both fitting results indicate a statistically significant positive correlation between the two energy bands, suggesting that the HR remains relatively stable across different observations. In other words, when the soft flux increases, the hard flux also increases proportionally. Such synchronous variability implies that these photons are likely produced within the same emission region by a single population of leptons \citep[e.g.,][]{2022MNRAS.511.3101P}. \\
\\
{{We performed a DCF analysis to investigate the potential correlation between the two energy bands. However, the overall lack of variability in the LCs precludes a more sophisticated DCF analysis from detecting such a correlation at high significance.}}

\begin{figure}
    \centering
    \includegraphics[width=\linewidth]{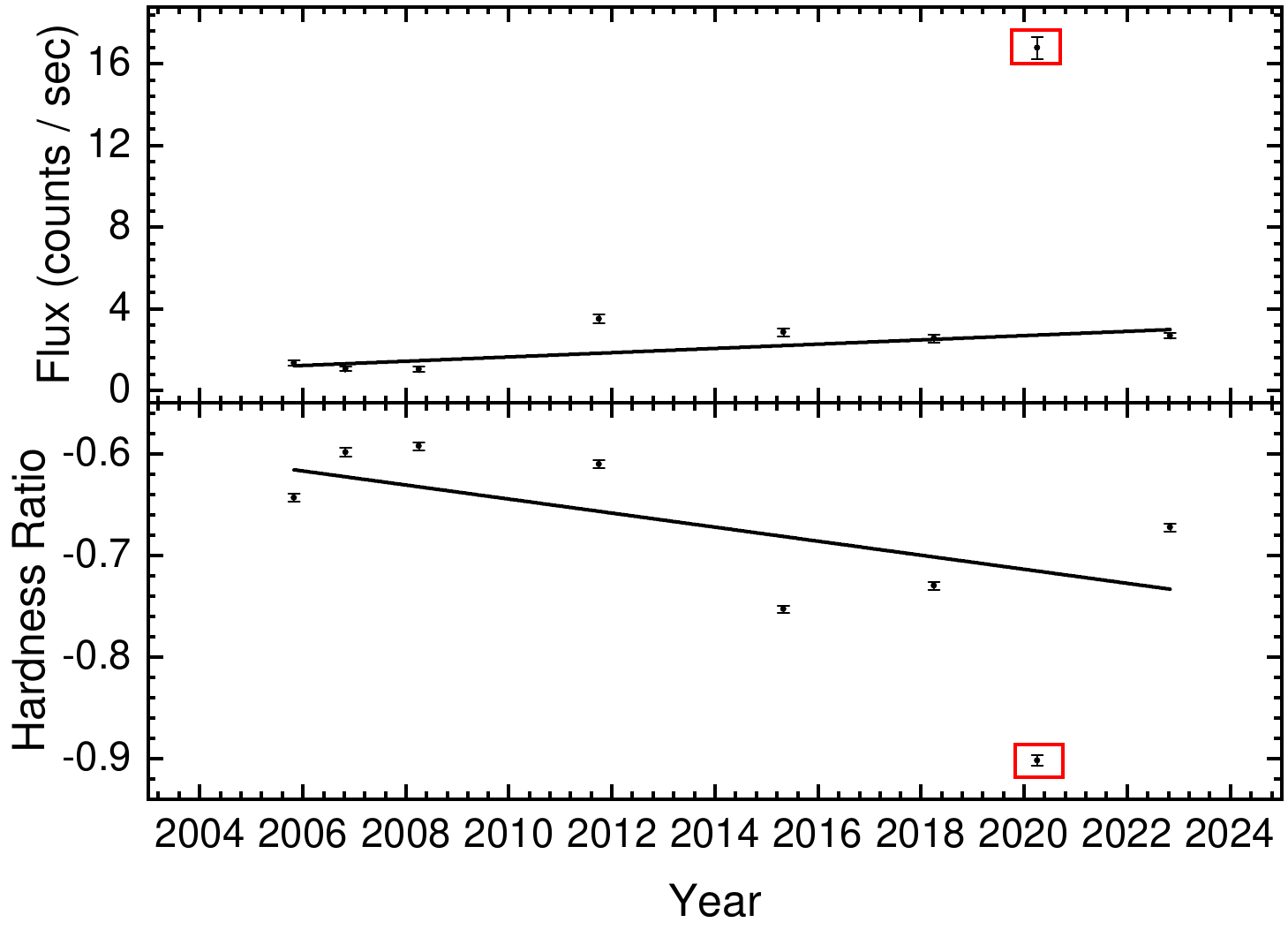}
    \caption{Top panel: Long-term variation of X-ray flux of OJ 287 from 2005 to 2022 based on XMM-Newton observations; Bottom panel: spectral variability of OJ 287. The points within the red box are flux outliers.}
    \label{fig:LTV}
\end{figure}

\begin{figure}
    \centering
    \includegraphics[width=\linewidth]{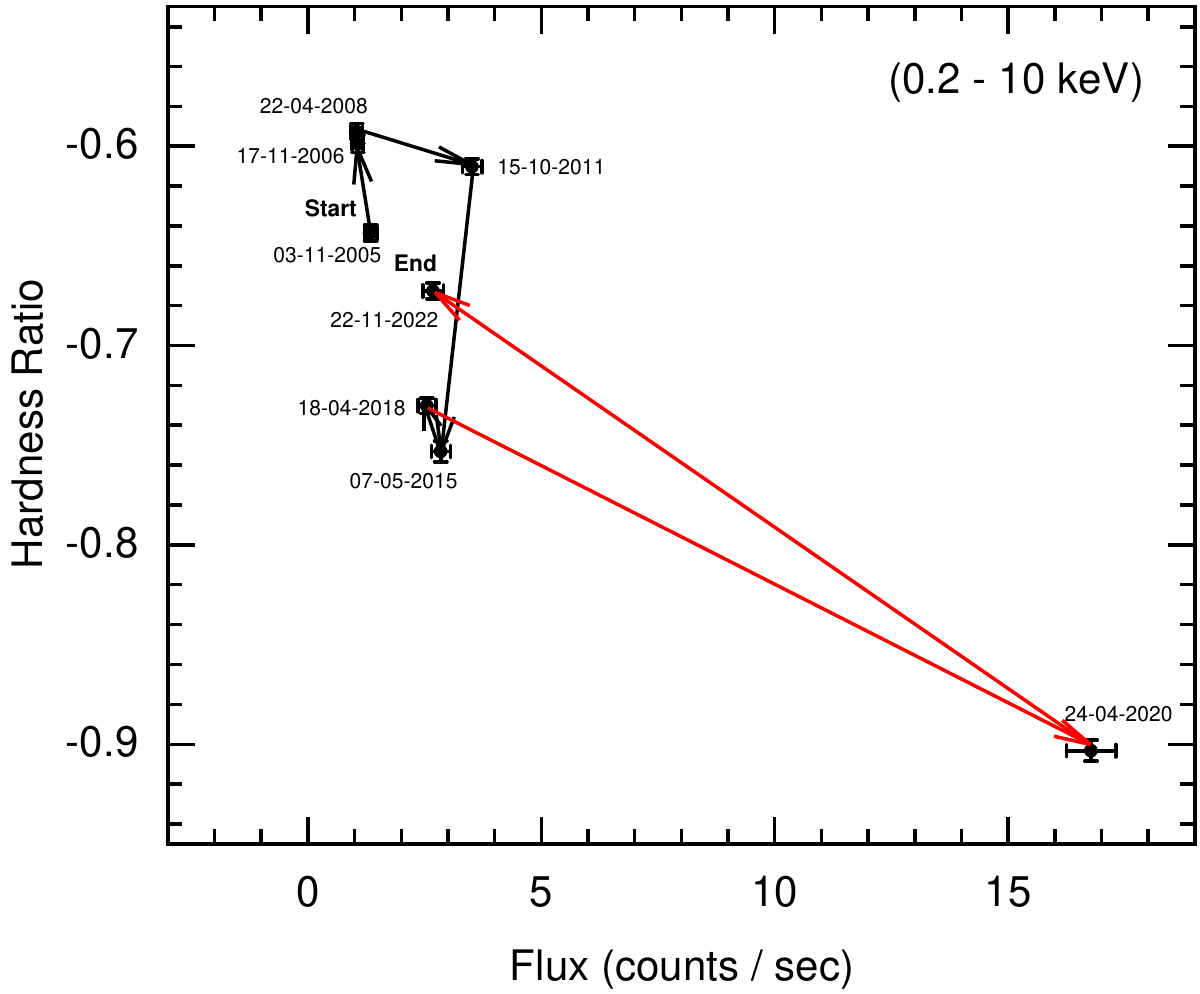}
    \caption{Hardness ratio versus flux relationship at different times}
    \label{fig:epoch}
\end{figure}

\subsection{Long-term variations in flux and spectrum}
\noindent
XMM-Newton has monitored OJ 287 over a span of 17 years, providing a valuable opportunity to study the LTV of its X-ray flux and spectral properties. We analyzed observational data collected between 2005 and 2022 to investigate how OJ 287's X-ray flux and HR vary with time. \\
\\
In the upper section of Fig. \ref{fig:LTV}, we performed a least-squares linear regression to the flux-versus-time data after excluding one apparent outlier (extremely high flux, marked by a red square), which significantly deviates from the general trend. {{This observation was excluded based on the physical phenomenon shown in Fig. \ref{fig:epoch}, which indicates that the source transitioned to a different emission state during this period (high soft state or flare state). It implies a change in the physical mechanism, causing this period to differ from others.}} {{The resulting fit reveals a moderate upward trend in flux from 2005 to 2022, with $r=0.755$, p-value = 0.049, providing tentative evidence for a potential brightening trend.}} \\
\\
The bottom panel of Fig. \ref{fig:LTV} displays HR relative to observational time. The linear fit yields a slope of $-8.1 \times 10^{-3} \text{yr}^{-1}$, corresponding to a total HR decrease of $\sim$ 0.13 over 17 years. {{The fit gives $r = -0.742$, p-value = 0.044, providing tentative evidence of a negative correlation.}} The negative slope indicates a possible gradual softening of OJ 287’s X-ray spectrum over time, consistent with the findings in \cite{2021MNRAS.507.5690G}. \\
\\
The prevailing view holds that shock propagation within the relativistic jets of blazars acts as an efficient particle accelerator, propelling particles to extremely high relativistic energies. These highly relativistic particles subsequently generate intense synchrotron radiation within the distorted magnetic field. This radiation not only spans the X-ray wavelength range but also constitutes the dominant cooling mechanism within the radiation zone. This is because, due to the rapid energy loss of high-energy particles, particles must be repeatedly or continuously accelerated to sustain continuous X-ray emission \citep{1987PhR...154....1B}.
We next examine the relationship between hardness ratio and flux in the total energy band across different epochs. Eight observations spanning from 2005 to 2022 were analyzed. Notably, the observations on 2006 November 17 and 2008 April 22 exhibit similar HR and flux values, as shown in Figure \ref{fig:epoch}, where the data points from different epochs are connected in chronological order with arrows, forming distinct loops. The black arrows correspond to clockwise evolution, while the red arrows indicate anticlockwise evolution. The evolution of OJ 287 from 2005 to 2022 exhibits two different looping behaviors. From 2005 November 3 to 2018 April 18 (black arrows), the source follows a clockwise loop, corresponding to a soft lag, which is indicative of synchrotron cooling dominated processes. From 2018 April 18 to 2022 November 22 (red arrows), the source exhibits a counterclockwise loop, corresponding to a hard lag, suggesting that particle acceleration  dominates during this period. 

\section{Discussion} \label{sec:discussion}
\noindent
We investigated the eight pointed observations of the blazar OJ 287, made by the XMM-Newton satellite. The purpose of our study was to analyze the flux and spectral variability (both intraday and long-term); flux variability, PSD analysis, and cross-correlated variability on intraday time scales. Multi-year X-ray observation of OJ 287 with XMM-Newton revealed that the IDV X-ray LCs exhibit small-scale variations, while large-scale variations occur over longer timescales. On IDV timescales, soft (0.2 - 2 keV) and hard (2 - 10 keV) energy band (0.2 - 10 keV) flux show similar variability patterns and are well correlated on all variable LCs. During an X-ray outburst at one pointed observation, the HR value changes significantly compared to other pointed observations. PSD slopes are found in the range of 1.14 to 2.11. The findings are consistent with a similar study of another LSP blazar \citep{2010ApJ...716...30A}, 3C 273 \citep{2022MNRAS.511.3101P}. \\
\\
\citet{2012A&A...544A..80G} utilized XMM-Newton data for examining the PSD characteristics of 104 AGN, including OJ 287. The 2 observational IDs considered in that study are also included in this study, and the PSD measurements from these datasets are consistent with our findings. \citet{2016MNRAS.462.1508G} considered a sample of 50 LCs of XMM-Newton in the energy range of 0.3 - 10 keV of 12 LSP blazars on IDV timescales, which also included 5 LCs of OJ 287, and found that out of 50 LCs, genuine IDV is present in only two of the LCs, i.e., 4 percent. The finding is consistent with the results of the present study. Using three long to extremely long pointed observations of about 225 ks (1st epoch), 209 ks (2nd epoch), and 751 ks (3rd epoch) of the source by \emph{Suzaku} during its whole operational period, small amplitude (3 - 12\%) X-ray IDV were found in all three observations in the energy range 0.5 - 10 keV; in the 3rd epoch (May 2015) of observation, soft X-ray excess was also seen, which is rare evidence for a blazar \citep{2024MNRAS.532.3285Z}. This source was also observed by XMM-Newton in May 2015 (3rd epoch of Suzaku observation) and also shows strong evidence of soft X-ray excess, which was well explained by two accretion disk-based models: the cool Comptonization component in the accretion disk and the blurred reflection from the partially ionized accretion disk \citep{2020ApJ...890...47P}. Using multi-wavelength observations of Swift/XRT and Swift/UVOT, which cover X-ray, ultraviolet, and optical bands.  During an outburst in 2020, optical and UV bands were correlated. X-ray brightness changes exhibit a certain degree of anticorrelation with spectral changes in the optical-ultraviolet band: hardening in optical-ultraviolet band is typically accompanied by softening in X-ray band; the X-ray spectrum shows a ``softer-when-brighter" trend \citep{2021ApJ...921...18K}. \\
\\
The X-ray variations observed in OJ 287 at these timescales may originate from the jet, the accretion disk corona, or an assemble of both. The characteristics of these variations, especially the PSD, can help distinguish between these possibilities. Accretion disk models often generate PSD slopes within the realm of $1.3\leq\alpha\leq2.1$, though slightly steeper PSDs may occur under certain geometric configurations. In contrast, jet models exhibit steeper PSD slopes, typically within the range $1.7\leq\alpha\leq2.9$, though such simulations are primarily focused on timescales realming from weeks to years \citep[e.g.,][]{2019ApJ...877..151W,2022MNRAS.511.3101P}. The PSD slopes obtained from the X-ray analysis of OJ 287 in this paper range in $1.14\leq\alpha\leq2.11$, with a small portion approaching the flicker region ($\alpha\approx1$), making it difficult to distinguish the origin. However, most results appear to align more closely with jet model predictions. However, given the limited range of PSD values in this study's sample and the current scarcity of theoretical calculations for PSD, we can only offer preliminary inferences here, suggesting that these fluctuations are more likely related to jets. \\
\\
Through these 8 observations with XMM-Newton, the shortest variability timescale $\tau_{var}$ is 0.14 ks. Using basic causality constraints, we can use $\tau_{var}$ to calculate the upper limit of radiation area size R, with the relationship expressed as
\begin{equation}
    R\leq\frac{c\delta\tau_{var}}{1+z}.
\end{equation}
Here, the Doppler factor $\delta$ for OJ 287 ranges from 3.4 to 18.6 \citep[e.g.,][]{1969ApJ...155L..71K,2024MNRAS.532.3285Z}. Using a minimum variance timescale of 140 s and $\delta = (3.4-18.6)$, the size of emission region is $(1.09 - 5.98)\times10^{13}$ cm. The obtained range agrees well with that reported by \citet{2017ApJ...835..275R} derived from their analysis of optical and polarization data for OJ 287. This further supports the notion that the fluctuations primarily originate from the jet, as mentioned in the PSD analysis. \\
\\
According to \cite{2019ApJ...884..125Z,2021MNRAS.506.1198D,2024MNRAS.532.3285Z}, we estimated the relativistic electron synchronization cooling timescale $t_{cool}(\gamma)$ in the observational framework, where the electron energy is $E=\gamma m_ec^2$, with the following formula
\begin{equation} \label{t_cool}
    t_{cool}(\gamma) \simeq 7.74 \times 10^{8} (1+z)\delta^{-1} \frac{1}{B^2\gamma} \ \mathrm{s}.
\end{equation}
Here, $B$ is the magnetic field strength in Gauss. The synchrotron emission frequency detected is
\begin{equation}
    \nu \simeq 4.2 \times 10^{6} \frac{\delta B \gamma^{2}}{1+z} \ \mathrm{Hz} \simeq 10^{18} \nu_{18} \ \mathrm{Hz}.
\end{equation}
Here $0.05<\nu_{18}<2.42$ (in XMM-Newton), and the calculation method adopted by other researchers \citep{2022ApJ...939...80D,2024MNRAS.532.3285Z} can be used to obtain
\begin{equation} \label{B}
    B \geq 5.51 \, \delta^{-1/3} \nu_{18}^{-1/3} \ \mathrm{G}.
\end{equation}
Substituting $\delta (3.4 - 18.6)$ into equation \ref{B} , we estimate a lower limit for the magnetic field strength of OJ 287 as $ B > (2.76-3.66) \, \nu_{18}^{-1/3} \ \mathrm{G}$. Previous studies have estimated the magnetic field strength of OJ 287 to range from 0.7 to 11.5 G under different observational periods and flux intensity conditions \citep[e.g.,][]{2018MNRAS.473.1145K,2021A&A...654A..38P}. Given that the frequency range of $\nu_{18}$ is between 0.05 and 2.42, the magnetic field range we calculated is consistent with the estimated values given in the literature. Using Equations \ref{t_cool} and \ref{B}, constraints can also be set on the electronic Lorentz factor $\gamma$ of OJ 287:
\begin{equation}
    \gamma\leq0.24\times10^{6}\delta^{-\frac{1}{3}}\nu_{18}^{\frac{2}{3}}.
\end{equation}
By substituting the same range of $\delta$ values, the upper limit of $\gamma$ can be estimated to lie between $(0.9-1.6)\times10^{5}\nu_{18}^{\frac{2}{3}}$. Our findings are consistent with the conclusions reported by other reasearchers \citep{2018MNRAS.473.1145K,2021A&A...654A..38P,2022MNRAS.509.2696S}. \\
\\
The sudden increase in flux observed in OJ 287 in 2020 indicates that it is undergoing a significant X-ray burst. There are currently two main hypotheses regarding the cause of this burst: one is the aftereffect of black hole accretion disk interaction, the other is a tidal disruption event (TDE). Through analysis of multi-wavelength data, it was found that the temporal evolution of the 2020 X-ray surge aligns with the characteristics of a TDE, particularly in terms of HR changes (showing no significant temporal evolution), and spectral index \citep{2022MNRAS.515.2778H}. In our study, the HR of the 2020 burst also showed no significant temporal evolution. The tidal disruption event provides a more reasonable explanation for the 2020 X-ray burst.

\section{Conclusions} \label{sec:conclusion}
\noindent
We investigated eight observations of the blazar OJ 287 made by XMM-Newton during its operation from 2005 to 2022. Our study focused on investigating IDV and its timescales, possible time lags between the soft and hard X-ray bands, spectral changes through HR analysis, and PSD analysis to characterize the type of noise and search potential QPOs. \\
\\
The main conclusions of this study are:
\begin{enumerate}
    \item Among the total X-ray energy band light curves, six showed small-amplitude fractional variability ranging from $1.59 \pm 0.49\%$ to $1.98 \pm 0.27\%$, with a maximum variability amplitude of only about $3\%$. The remaining two LCs did not exhibit significant variations. No strong flares were observed in any of the datasets.
    \item We estimated the minimum variability timescale $\tau_{var,min}$, and used it to derive key physical parameters, including the size of the emission region, the magnetic field strength, and the electron Lorentz factor $\gamma$ responsible for X-ray emission.
    \item HR analysis showed no notable spectral variation during all the observations.
    \item A strong positive correlation was found between soft and hard X-ray fluxes, with r = 0.852 and p-value = 0.007, indicating statistical significance. {{However, the overall lack of variability in the LCs precludes a more sophisticated DCF analysis from detecting such a correlation at high significance.}}
    \item We studied PSD analysis on the variable LCs in the soft, hard, and total X-ray bands. Each PSD was adequately described by a power-law model. The PSD slopes ranged from 1.14 to 2.11, mostly falling between 1.7 and 2.8, indicating dominant red noise. QPOs were not found in all PSDs.
    \item A negative correlation was found between flux and HR in the long-term observation, suggesting a softer-when-brighter trend for OJ 287.
    \item The combined results from flux and spectral analysis imply that both particle acceleration and synchrotron cooling processes play important roles in the emission mechanism of OJ 287.
\end{enumerate}

\section*{Acknowledgements}
\noindent
This work was supported by the Tianshan Talent Training Program (grant No. 2023TSYCCX0099) and the National Key R\&D Program of China (grant No. 2024YFA1611500 and 2021YFA0718500). ACG is partially supported by the CAS  `President’s International Fellowship Initiative (PIFI)' with Grant No. 2026PVA0040. AT and YFH acknowledge the support from the Xinjiang Tianchi Talent Program.

\bibliography{sample701}{}

\begin{thebibliography}{}
\expandafter\ifx\csname natexlab\endcsname\relax\def\natexlab#1{#1}\fi
\providecommand{\url}[1]{\href{#1}{#1}}
\providecommand{\dodoi}[1]{doi:~\href{http://doi.org/#1}{\nolinkurl{#1}}}
\providecommand{\doeprint}[1]{\href{http://ascl.net/#1}{\nolinkurl{http://ascl.net/#1}}}
\providecommand{\doarXiv}[1]{\href{https://arxiv.org/abs/#1}{\nolinkurl{https://arxiv.org/abs/#1}}}

\bibitem[{A.~A. {Abdo} {et~al.}(2010){Abdo}, {Ackermann}, {Agudo}, {Ajello}, {Aller}, {Aller}, {Angelakis}, {Arkharov}, {Axelsson}, {Bach}, {Baldini}, {Ballet}, {Barbiellini}, {Bastieri}, {Baughman}, {Bechtol}, {Bellazzini}, {Benitez}, {Berdyugin}, {Berenji}, {Blandford}, {Bloom}, {Boettcher}, {Bonamente}, {Borgland}, {Bregeon}, {Brez}, {Brigida}, {Bruel}, {Burnett}, {Burrows}, {Buson}, {Caliandro}, {Calzoletti}, {Cameron}, {Capalbi}, {Caraveo}, {Carosati}, {Casandjian}, {Cavazzuti}, {Cecchi}, {{\c{C}}elik}, {Charles}, {Chaty}, {Chekhtman}, {Chen}, {Chiang}, {Chincarini}, {Ciprini}, {Claus}, {Cohen-Tanugi}, {Colafrancesco}, {Cominsky}, {Conrad}, {Costamante}, {Cutini}, {D'ammando}, {Deitrick}, {D'Elia}, {Dermer}, {de Angelis}, {de Palma}, {Digel}, {Donnarumma}, {Silva}, {Drell}, {Dubois}, {Dultzin}, {Dumora}, {Falcone}, {Farnier}, {Favuzzi}, {Fegan}, {Focke}, {Forn{\'e}}, {Fortin}, {Frailis}, {Fuhrmann}, {Fukazawa}, {Funk}, {Fusco}, {G{\'o}mez}, {Gargano}, {Gasparrini}, {Gehrels}, {Germani}, {Giebels},
  {Giglietto}, {Giommi}, {Giordano}, {Giuliani}, {Glanzman}, {Godfrey}, {Grenier}, {Gronwall}, {Grove}, {Guillemot}, {Guiriec}, {Gurwell}, {Hadasch}, {Hanabata}, {Harding}, {Hayashida}, {Hays}, {Healey}, {Heidt}, {Hiriart}, {Horan}, {Hoversten}, {Hughes}, {Itoh}, {Jackson}, {J{\'o}hannesson}, {Johnson}, {Johnson}, {Jorstad}, {Kadler}, {Kamae}, {Katagiri}, {Kataoka}, {Kawai}, {Kennea}, {Kerr}, {Kimeridze}, {Kn{\"o}dlseder}, {Kocian}, {Kopatskaya}, {Koptelova}, {Konstantinova}, {Kovalev}, {Kovalev}, {Kurtanidze}, {Kuss}, {Lande}, {Larionov}, {Latronico}, {Leto}, {Lindfors}, {Longo}, {Loparco}, {Lott}, {Lovellette}, {Lubrano}, {Madejski}, {Makeev}, {Marchegiani}, {Marscher}, {Marshall}, {Max-Moerbeck}, {Mazziotta}, {McConville}, {McEnery}, {Meurer}, {Michelson}, {Mitthumsiri}, {Mizuno}, {Moiseev}, {Monte}, {Monzani}, {Morselli}, {Moskalenko}, {Murgia}, {Nestoras}, {Nilsson}, {Nizhelsky}, {Nolan}, {Norris}, {Nuss}, {Ohsugi}, {Ojha}, {Omodei}, {Orlando}, {Ormes}, {Osborne}, {Ozaki}, {Pacciani}, {Padovani},
  {Pagani}, {Page}, {Paneque}, {Panetta}, {Parent}, {Pasanen}, {Pavlidou}, {Pelassa}, {Pepe}, {Perri}, {Pesce-Rollins}, {Piranomonte}, {Piron}, {Pittori}, {Porter}, {Puccetti}, {Rahoui}, {Rain{\`o}}, {Raiteri}, {Rando}, {Razzano}, {Reimer}, \& {Reimer}}]{2010ApJ...716...30A}
{Abdo}, A.~A., {Ackermann}, M., {Agudo}, I., {et~al.} 2010, \bibinfo{title}{{The Spectral Energy Distribution of Fermi Bright Blazars},} \apj, 716, 30, \dodoi{10.1088/0004-637X/716/1/30}

\bibitem[{K.~A. {Arnaud}(1996){Arnaud}}]{Arn96}
{Arnaud}, K.~A. 1996, \bibinfo{title}{{XSPEC: The First Ten Years},} in Astronomical Society of the Pacific Conference Series, Vol. 101, Astronomical Data Analysis Software and Systems V, ed. G.~H. {Jacoby} \& J.~{Barnes}, 17

\bibitem[{J. {Bhagwan} {et~al.}(2014){Bhagwan}, {Gupta}, {Papadakis}, \& {Wiita}}]{2014MNRAS.444.3647B}
{Bhagwan}, J., {Gupta}, A.~C., {Papadakis}, I.~E., \& {Wiita}, P.~J. 2014, \bibinfo{title}{{Spectral energy distributions of the BL Lac PKS 2155 - 304 from XMM-Newton},} \mnras, 444, 3647, \dodoi{10.1093/mnras/stu1703}

\bibitem[{J. {Bhagwan} {et~al.}(2016){Bhagwan}, {Gupta}, {Papadakis}, \& {Wiita}}]{2016NewA...44...21B}
{Bhagwan}, J., {Gupta}, A.~C., {Papadakis}, I.~E., \& {Wiita}, P.~J. 2016, \bibinfo{title}{{Flux and spectral variability of the blazar PKS 2155 -304 with XMM-Newton: Evidence of particle acceleration and synchrotron cooling},} \na, 44, 21, \dodoi{10.1016/j.newast.2015.08.005}

\bibitem[{R. {Blandford} \& D. {Eichler}(1987){Blandford} \& {Eichler}}]{1987PhR...154....1B}
{Blandford}, R., \& {Eichler}, D. 1987, \bibinfo{title}{{Particle acceleration at astrophysical shocks: A theory of cosmic ray origin},} \physrep, 154, 1, \dodoi{10.1016/0370-1573(87)90134-7}

\bibitem[{G.~R. {Burbidge} {et~al.}(1974){Burbidge}, {Jones}, \& {O'Dell}}]{1974ApJ...193...43B}
{Burbidge}, G.~R., {Jones}, T.~W., \& {O'Dell}, S.~L. 1974, \bibinfo{title}{{Physics of compact nonthermal sources. III. Energetic considerations.},} \apj, 193, 43, \dodoi{10.1086/153125}

\bibitem[{L. {Carrasco} {et~al.}(1985){Carrasco}, {Dultzin-Hacyan}, \& {Cruz-Gonzalez}}]{1985Natur.314..146C}
{Carrasco}, L., {Dultzin-Hacyan}, D., \& {Cruz-Gonzalez}, I. 1985, \bibinfo{title}{{Periodicity in the BL Lac object OJ287},} \nat, 314, 146, \dodoi{10.1038/314146a0}

\bibitem[{S. {Ciprini} {et~al.}(2007){Ciprini}, {Raiteri}, {Rizzi}, {Agudo}, {Foschini}, {Fiorucci}, {Takalo}, {Villata}, {Ostorero}, {Sillanp{\"a}{\"a}}, {Valtonen}, {Tosti}, {Wagner}, {Aller}, {Aller}, {Arai}, {Arkharov}, {Bakis}, {Bagaglia}, {B{\"o}ttcher}, {Buemi}, {Carosati}, {Chen}, {Efimov}, {Emmanoulopoulos}, {Erdem}, {Fuhrmann}, {Frasca}, {Fullhart}, {Goyal}, {Heidt}, {Hovatta}, {Hroch}, {Ibrahimov}, {Jilkov{\'a}}, {Joshi}, {Kamada}, {Katsuura}, {Kinoshita}, {Kostov}, {Kotaka}, {Kovalev}, {Krejcov{\'a}}, {Krichbaum}, {Gopal-Krishna}, {Kurosaki}, {Kurtanidze}, {Lahteenmaki}, {Lanteri}, {Larionov}, {Lee}, {Letho}, {Leto}, {Li}, {Lindfors}, {M{\"u}nz}, {Marilli}, {Matsubara}, {Mizoguchi}, {Mondal}, {Nakamura}, {Nieppola}, {Nilsson}, {Nishiyama}, {Nucciarelli}, {Ogino}, {Ohlert}, {Oksanen}, {Ovcharov}, {Pak}, {Pasanen}, {Pullen}, {Pursimo}, {Ros}, {Sadakane}, {Sadun}, {Sagar}, {Sohnk}, {Sumitomo}, {Tanaka}, {Trigilio}, {Torniainenk I.}, {Tornikoski}, {Umana}, {Ungerechts}, {Valtaoja}, {Volvach}, {Webb},
  {Wu}, {Yim}, \& {Zhang}}]{2007MmSAI..78..741C}
{Ciprini}, S., {Raiteri}, C.~M., {Rizzi}, N., {et~al.} 2007, \bibinfo{title}{{Prominent activity of the blazar OJ 287 in 2005. XMM-Newton and multiwavelength observations},} \memsai, 78, 741

\bibitem[{P.~U. {Devanand} {et~al.}(2022){Devanand}, {Gupta}, {Jithesh}, \& {Wiita}}]{2022ApJ...939...80D}
{Devanand}, P.~U., {Gupta}, A.~C., {Jithesh}, V., \& {Wiita}, P.~J. 2022, \bibinfo{title}{{Study of X-Ray Intraday Variability of HBL Blazars Based on Observations Obtained with XMM-Newton},} \apj, 939, 80, \dodoi{10.3847/1538-4357/ac9064}

\bibitem[{P.~U. {Devanand} {et~al.}(2025){Devanand}, {Gupta}, {Jithesh}, {Wiita}, \& {Gupta}}]{2025ApJS..278...20D}
{Devanand}, P.~U., {Gupta}, A.~C., {Jithesh}, V., {Wiita}, P.~J., \& {Gupta}, A. 2025, \bibinfo{title}{{X-Ray Spectral Variability of 13 TeV High-energy-peaked Blazars with XMM-Newton},} \apjs, 278, 20, \dodoi{10.3847/1538-4365/adc10d}

\bibitem[{V. {Dhiman} {et~al.}(2021){Dhiman}, {Gupta}, {Gaur}, \& {Wiita}}]{2021MNRAS.506.1198D}
{Dhiman}, V., {Gupta}, A.~C., {Gaur}, H., \& {Wiita}, P.~J. 2021, \bibinfo{title}{{Multiband variability of the TeV blazar PG 1553+113 with XMM-Newton},} \mnras, 506, 1198, \dodoi{10.1093/mnras/stab1743}

\bibitem[{J.~R. {Dickel} {et~al.}(1967){Dickel}, {Yang}, {McVittie}, \& {Swenson}}]{1967AJ.....72..757D}
{Dickel}, J.~R., {Yang}, K.~S., {McVittie}, G.~C., \& {Swenson}, Jr., G.~W. 1967, \bibinfo{title}{{A survey of the sky at 610.5 MHz. II. The region between declinations +15 and +22 degrees.},} \aj, 72, 757, \dodoi{10.1086/110305}

\bibitem[{R. {Edelson} {et~al.}(2002){Edelson}, {Turner}, {Pounds}, {Vaughan}, {Markowitz}, {Marshall}, {Dobbie}, \& {Warwick}}]{2002ApJ...568..610E}
{Edelson}, R., {Turner}, T.~J., {Pounds}, K., {et~al.} 2002, \bibinfo{title}{{X-Ray Spectral Variability and Rapid Variability of the Soft X-Ray Spectrum Seyfert 1 Galaxies Arakelian 564 and Ton S180},} \apj, 568, 610, \dodoi{10.1086/323779}

\bibitem[{R.~A. {Edelson} \& J.~H. {Krolik}(1988){Edelson} \& {Krolik}}]{1988ApJ...333..646E}
{Edelson}, R.~A., \& {Krolik}, J.~H. 1988, \bibinfo{title}{{The Discrete Correlation Function: A New Method for Analyzing Unevenly Sampled Variability Data},} \apj, 333, 646, \dodoi{10.1086/166773}

\bibitem[{G. {Fossati} {et~al.}(1998){Fossati}, {Maraschi}, {Celotti}, {Comastri}, \& {Ghisellini}}]{1998MNRAS.299..433F}
{Fossati}, G., {Maraschi}, L., {Celotti}, A., {Comastri}, A., \& {Ghisellini}, G. 1998, \bibinfo{title}{{A unifying view of the spectral energy distributions of blazars},} \mnras, 299, 433, \dodoi{10.1046/j.1365-8711.1998.01828.x}

\bibitem[{H. {Gaur} {et~al.}(2010){Gaur}, {Gupta}, {Lachowicz}, \& {Wiita}}]{2010ApJ...718..279G}
{Gaur}, H., {Gupta}, A.~C., {Lachowicz}, P., \& {Wiita}, P.~J. 2010, \bibinfo{title}{{Detection of Intra-day Variability Timescales of Four High-energy Peaked Blazars with XMM-Newton},} \apj, 718, 279, \dodoi{10.1088/0004-637X/718/1/279}

\bibitem[{P. {Giommi} {et~al.}(2021){Giommi}, {Perri}, {Capalbi}, {D'Elia}, {Barres de Almeida}, {Brandt}, {Pollock}, {Arneodo}, {Di Giovanni}, {Chang}, {Civitarese}, {De Angelis}, {Leto}, {Verrecchia}, {Ricard}, {Di Pippo}, {Middei}, {Penacchioni}, {Ruffini}, {Sahakyan}, {Israyelyan}, \& {Turriziani}}]{2021MNRAS.507.5690G}
{Giommi}, P., {Perri}, M., {Capalbi}, M., {et~al.} 2021, \bibinfo{title}{{X-ray spectra, light curves and SEDs of blazars frequently observed by Swift},} \mnras, 507, 5690, \dodoi{10.1093/mnras/stab2425}

\bibitem[{J.~L. {G{\'o}mez} {et~al.}(2022){G{\'o}mez}, {Traianou}, {Krichbaum}, {Lobanov}, {Fuentes}, {Lico}, {Zhao}, {Bruni}, {Kovalev}, {L{\"a}hteenm{\"a}ki}, {Voitsik}, {Lisakov}, {Angelakis}, {Bach}, {Casadio}, {Cho}, {Dey}, {Gopakumar}, {Gurvits}, {Jorstad}, {Kovalev}, {Lister}, {Marscher}, {Myserlis}, {Pushkarev}, {Ros}, {Savolainen}, {Tornikoski}, {Valtonen}, \& {Zensus}}]{2022ApJ...924..122G}
{G{\'o}mez}, J.~L., {Traianou}, E., {Krichbaum}, T.~P., {et~al.} 2022, \bibinfo{title}{{Probing the Innermost Regions of AGN Jets and Their Magnetic Fields with RadioAstron. V. Space and Ground Millimeter-VLBI Imaging of OJ 287},} \apj, 924, 122, \dodoi{10.3847/1538-4357/ac3bcc}

\bibitem[{O. {Gonz{\'a}lez-Mart{\'\i}n} \& S. {Vaughan}(2012){Gonz{\'a}lez-Mart{\'\i}n} \& {Vaughan}}]{2012A&A...544A..80G}
{Gonz{\'a}lez-Mart{\'\i}n}, O., \& {Vaughan}, S. 2012, \bibinfo{title}{{X-ray variability of 104 active galactic nuclei. XMM-Newton power-spectrum density profiles},} \aap, 544, A80, \dodoi{10.1051/0004-6361/201219008}

\bibitem[{A. {Goyal} {et~al.}(2018){Goyal}, {Stawarz}, {Zola}, {Marchenko}, {Soida}, {Nilsson}, {Ciprini}, {Baran}, {Ostrowski}, {Wiita}, {Gopal-Krishna}, {Siemiginowska}, {Sobolewska}, {Jorstad}, {Marscher}, {Aller}, {Aller}, {Hovatta}, {Caton}, {Reichart}, {Matsumoto}, {Sadakane}, {Gazeas}, {Kidger}, {Piirola}, {Jermak}, {Alicavus}, {Baliyan}, {Baransky}, {Berdyugin}, {Blay}, {Boumis}, {Boyd}, {Bufan}, {Campas Torrent}, {Campos}, {Carrillo G{\'o}mez}, {Dalessio}, {Debski}, {Dimitrov}, {Drozdz}, {Er}, {Erdem}, {Escartin P{\'e}rez}, {Fallah Ramazani}, {Filippenko}, {Gafton}, {Garcia}, {Godunova}, {G{\'o}mez Pinilla}, {Gopinathan}, {Haislip}, {Haque}, {Harmanen}, {Hudec}, {Hurst}, {Ivarsen}, {Joshi}, {Kagitani}, {Karaman}, {Karjalainen}, {Kaur}, {Kozie{\l}-Wierzbowska}, {Kuligowska}, {Kundera}, {Kurowski}, {Kvammen}, {LaCluyze}, {Lee}, {Liakos}, {Lozano de Haro}, {Moore}, {Mugrauer}, {Naves Nogues}, {Neely}, {Ogloza}, {Okano}, {Pajdosz}, {Pandey}, {Perri}, {Poyner}, {Provencal}, {Pursimo}, {Raj}, {Rajkumar},
  {Reinthal}, {Reynolds}, {Saario}, {Sadegi}, {Sakanoi}, {Salto Gonz{\'a}lez}, {Sameer}, {Simon}, {Siwak}, {Schweyer}, {Sold{\'a}n Alfaro}, {Sonbas}, {Strobl}, {Takalo}, {Tremosa Espasa}, {Valdes}, {Vasylenko}, {Verrecchia}, {Webb}, {Yoneda}, {Zejmo}, {Zheng}, {Zielinski}, {Janik}, {Chavushyan}, {Mohammed}, {Cheung}, \& {Giroletti}}]{2018ApJ...863..175G}
{Goyal}, A., {Stawarz}, {\L}., {Zola}, S., {et~al.} 2018, \bibinfo{title}{{Stochastic Modeling of Multiwavelength Variability of the Classical BL Lac Object OJ 287 on Timescales Ranging from Decades to Hours},} \apj, 863, 175, \dodoi{10.3847/1538-4357/aad2de}

\bibitem[{A.~C. {Gupta} {et~al.}(2004){Gupta}, {Banerjee}, {Ashok}, \& {Joshi}}]{2004A&A...422..505G}
{Gupta}, A.~C., {Banerjee}, D.~P.~K., {Ashok}, N.~M., \& {Joshi}, U.~C. 2004, \bibinfo{title}{{Near infrared intraday variability of Mrk 421},} \aap, 422, 505, \dodoi{10.1051/0004-6361:20040306}

\bibitem[{A.~C. {Gupta} {et~al.}(2016){Gupta}, {Kalita}, {Gaur}, \& {Duorah}}]{2016MNRAS.462.1508G}
{Gupta}, A.~C., {Kalita}, N., {Gaur}, H., \& {Duorah}, K. 2016, \bibinfo{title}{{Peak of spectral energy distribution plays an important role in intra-day variability of blazars?},} \mnras, 462, 1508, \dodoi{10.1093/mnras/stw1667}

\bibitem[{A.~C. {Gupta} {et~al.}(2022){Gupta}, {Kushwaha}, {Carrasco}, {Xu}, {Wiita}, {Escobedo}, {Porras}, {Recillas}, {Mayya}, {Chavushyan}, {Villarroel}, \& {Zhang}}]{2022ApJS..260...39G}
{Gupta}, A.~C., {Kushwaha}, P., {Carrasco}, L., {et~al.} 2022, \bibinfo{title}{{Long-term Multiband Near-infrared Variability of the Blazar OJ 287 during 2007-2021},} \apjs, 260, 39, \dodoi{10.3847/1538-4365/ac6c2c}

\bibitem[{A.~C. {Gupta} {et~al.}(2023){Gupta}, {Kushwaha}, {Valtonen}, {Savchenko}, {Jorstad}, {Imazawa}, {Wiita}, {Gu}, {Marscher}, {Zhang}, {Bachev}, {Borman}, {Gaur}, {Grishina}, {Hagen-Thorn}, {Kopatskaya}, {Larionov}, {Larionova}, {Larionova}, {Morozova}, {Nakaoka}, {Strigachev}, {Troitskaya}, {Troitsky}, {Uemura}, {Vasilyev}, {Weaver}, \& {Zhovtan}}]{2023ApJ...957L..11G}
{Gupta}, A.~C., {Kushwaha}, P., {Valtonen}, M.~J., {et~al.} 2023, \bibinfo{title}{{Quasi-simultaneous Optical Flux and Polarization Variability of the Binary Super Massive Black Hole Blazar OJ 287 from 2015 to 2023: Detection of an Anticorrelation in Flux and Polarization Variability},} \apjl, 957, L11, \dodoi{10.3847/2041-8213/acfd2e}

\bibitem[{V.~A. {Hagen-Thorn} {et~al.}(2008){Hagen-Thorn}, {Larionov}, {Jorstad}, {Arkharov}, {Hagen-Thorn}, {Efimova}, {Larionova}, \& {Marscher}}]{2008ApJ...672...40H}
{Hagen-Thorn}, V.~A., {Larionov}, V.~M., {Jorstad}, S.~G., {et~al.} 2008, \bibinfo{title}{{The Outburst of the Blazar AO 0235+164 in 2006 December: Shock-in-Jet Interpretation},} \apj, 672, 40, \dodoi{10.1086/523841}

\bibitem[{ {HI4PI\ Collaboration} {et~al.}(2016){HI4PI\ Collaboration}, {Ben Bekhti}, {Fl{\"o}er}, {Keller}, {Kerp}, {Lenz}, {Winkel}, {Bailin}, {Calabretta}, {Dedes}, {Ford}, {Gibson}, {Haud}, {Janowiecki}, {Kalberla}, {Lockman}, {McClure-Griffiths}, {Murphy}, {Nakanishi}, {Pisano}, \& {Staveley-Smith}}]{HI4PI16}
{HI4PI\ Collaboration}, {Ben Bekhti}, N., {Fl{\"o}er}, L., {et~al.} 2016, \bibinfo{title}{{HI4PI: A full-sky H I survey based on EBHIS and GASS},} \aap, 594, A116, \dodoi{10.1051/0004-6361/201629178}

\bibitem[{S. {Huang} {et~al.}(2022){Huang}, {Hu}, {Yin}, {Chen}, {Alexeeva}, \& {Jiang}}]{2022MNRAS.515.2778H}
{Huang}, S., {Hu}, S., {Yin}, H., {et~al.} 2022, \bibinfo{title}{{Exploration of the origin of the 2020 X-ray outburst in OJ 287},} \mnras, 515, 2778, \dodoi{10.1093/mnras/stac2022}

\bibitem[{N. {Kalita} {et~al.}(2017){Kalita}, {Gupta}, {Wiita}, {Dewangan}, \& {Duorah}}]{2017MNRAS.469.3824K}
{Kalita}, N., {Gupta}, A.~C., {Wiita}, P.~J., {Dewangan}, G.~C., \& {Duorah}, K. 2017, \bibinfo{title}{{Origin of X-rays in the low state of the FSRQ 3C 273: evidence of inverse Compton emission},} \mnras, 469, 3824, \dodoi{10.1093/mnras/stx1108}

\bibitem[{K.~I. {Kellermann} \& I.~I.~K. {Pauliny-Toth}(1969){Kellermann} \& {Pauliny-Toth}}]{1969ApJ...155L..71K}
{Kellermann}, K.~I., \& {Pauliny-Toth}, I.~I.~K. 1969, \bibinfo{title}{{The Spectra of Opaque Radio Sources},} \apjl, 155, L71, \dodoi{10.1086/180305}

\bibitem[{S. {Komossa} {et~al.}(2021){Komossa}, {Grupe}, {Gallo}, {Gonzalez}, {Yao}, {Hollett}, {Parker}, \& {Ciprini}}]{2021ApJ...923...51K}
{Komossa}, S., {Grupe}, D., {Gallo}, L.~C., {et~al.} 2021, \bibinfo{title}{{MOMO. IV. The Complete Swift X-Ray and UV/Optical Light Curve and Characteristic Variability of the Blazar OJ 287 during the Last Two Decades},} \apj, 923, 51, \dodoi{10.3847/1538-4357/ac1442}

\bibitem[{S. {Komossa} {et~al.}(2020){Komossa}, {Grupe}, {Parker}, {Valtonen}, {G{\'o}mez}, {Gopakumar}, \& {Dey}}]{2020MNRAS.498L..35K}
{Komossa}, S., {Grupe}, D., {Parker}, M.~L., {et~al.} 2020, \bibinfo{title}{{The 2020 April-June super-outburst of OJ 287 and its long-term multiwavelength light curve with Swift: binary supermassive black hole and jet activity},} \mnras, 498, L35, \dodoi{10.1093/mnrasl/slaa125}

\bibitem[{P. {Kushwaha} {et~al.}(2021){Kushwaha}, {Pal}, {Kalita}, {Kumari}, {Naik}, {Gupta}, {de Gouveia Dal Pino}, \& {Gu}}]{2021ApJ...921...18K}
{Kushwaha}, P., {Pal}, M., {Kalita}, N., {et~al.} 2021, \bibinfo{title}{{Blazar OJ 287 after First VHE Activity: Tracking the Reemergence of the HBL-like Component in 2020},} \apj, 921, 18, \dodoi{10.3847/1538-4357/ac19b8}

\bibitem[{P. {Kushwaha} {et~al.}(2020){Kushwaha}, {Sarkar}, {Gupta}, {Tripathi}, \& {Wiita}}]{2020MNRAS.499..653K}
{Kushwaha}, P., {Sarkar}, A., {Gupta}, A.~C., {Tripathi}, A., \& {Wiita}, P.~J. 2020, \bibinfo{title}{{A possible {\ensuremath{\gamma}}-ray quasi-periodic oscillation of {\ensuremath{\sim}}314 days in the blazar OJ 287},} \mnras, 499, 653, \dodoi{10.1093/mnras/staa2899}

\bibitem[{P. {Kushwaha} {et~al.}(2018){Kushwaha}, {Gupta}, {Wiita}, {Gaur}, {de Gouveia Dal Pino}, {Bhagwan}, {Kurtanidze}, {Larionov}, {Damljanovic}, {Uemura}, {Semkov}, {Strigachev}, {Bachev}, {Vince}, {Gu}, {Zhang}, {Abe}, {Agarwal}, {Borman}, {Fan}, {Grishina}, {Hirochi}, {Itoh}, {Kawabata}, {Kopatskaya}, {Kurtanidze}, {Larionova}, {Larionova}, {Mishra}, {Morozova}, {Nakaoka}, {Nikolashvili}, {Savchenko}, {Troitskaya}, {Troitsky}, \& {Vasilyev}}]{2018MNRAS.473.1145K}
{Kushwaha}, P., {Gupta}, A.~C., {Wiita}, P.~J., {et~al.} 2018, \bibinfo{title}{{Multiwavelength temporal and spectral variability of the blazar OJ 287 during and after the 2015 December flare: a major accretion disc contribution},} \mnras, 473, 1145, \dodoi{10.1093/mnras/stx2394}

\bibitem[{P. {Lachowicz} {et~al.}(2009){Lachowicz}, {Gupta}, {Gaur}, \& {Wiita}}]{2009A&A...506L..17L}
{Lachowicz}, P., {Gupta}, A.~C., {Gaur}, H., \& {Wiita}, P.~J. 2009, \bibinfo{title}{{A \raisebox{-0.5ex}\textasciitilde4.6 h quasi-periodic oscillation in the BL Lacertae PKS 2155-304?},} \aap, 506, L17, \dodoi{10.1051/0004-6361/200913161}

\bibitem[{S. {Laine} {et~al.}(2020){Laine}, {Dey}, {Valtonen}, {Gopakumar}, {Zola}, {Komossa}, {Kidger}, {Pihajoki}, {G{\'o}mez}, {Caton}, {Ciprini}, {Drozdz}, {Gazeas}, {Godunova}, {Haque}, {Hildebrandt}, {Hudec}, {Jermak}, {Kong}, {Lehto}, {Liakos}, {Matsumoto}, {Mugrauer}, {Pursimo}, {Reichart}, {Simon}, {Siwak}, \& {Sonbas}}]{2020ApJ...894L...1L}
{Laine}, S., {Dey}, L., {Valtonen}, M., {et~al.} 2020, \bibinfo{title}{{Spitzer Observations of the Predicted Eddington Flare from Blazar OJ 287},} \apjl, 894, L1, \dodoi{10.3847/2041-8213/ab79a4}

\bibitem[{G.~M. {Madejski} \& D.~A. {Schwartz}(1988){Madejski} \& {Schwartz}}]{1988ApJ...330..776M}
{Madejski}, G.~M., \& {Schwartz}, D.~A. 1988, \bibinfo{title}{{Studies of BL Lacertae Objects with the Einstein Observatory: The Soft X-Ray Spectra of OJ 287 and PKS 0735+178},} \apj, 330, 776, \dodoi{10.1086/166511}

\bibitem[{A.~P. {Marscher}(1983){Marscher}}]{1983ApJ...264..296M}
{Marscher}, A.~P. 1983, \bibinfo{title}{{Accurate formula for the self-Compton X-ray flux density from a uniform, spherical, compact radio source.},} \apj, 264, 296, \dodoi{10.1086/160597}

\bibitem[{P. {Mohan} \& A. {Mangalam}(2015){Mohan} \& {Mangalam}}]{2015ApJ...805...91M}
{Mohan}, P., \& {Mangalam}, A. 2015, \bibinfo{title}{{Kinematics of and Emission from Helically Orbiting Blobs in a Relativistic Magnetized Jet},} \apj, 805, 91, \dodoi{10.1088/0004-637X/805/2/91}

\bibitem[{A.~P. {Noel} {et~al.}(2022){Noel}, {Gaur}, {Gupta}, {Wierzcholska}, {Ostrowski}, {Dhiman}, \& {Bhatta}}]{2022ApJS..262....4N}
{Noel}, A.~P., {Gaur}, H., {Gupta}, A.~C., {et~al.} 2022, \bibinfo{title}{{X-Ray Intraday Variability of the TeV Blazar Markarian 421 with XMM-Newton},} \apjs, 262, 4, \dodoi{10.3847/1538-4365/ac7799}

\bibitem[{M. {Pal} {et~al.}(2020){Pal}, {Kushwaha}, {Dewangan}, \& {Pawar}}]{2020ApJ...890...47P}
{Pal}, M., {Kushwaha}, P., {Dewangan}, G.~C., \& {Pawar}, P.~K. 2020, \bibinfo{title}{{Strong Soft X-Ray Excess in 2015 XMM-Newton Observations of BL Lac OJ 287},} \apj, 890, 47, \dodoi{10.3847/1538-4357/ab65ee}

\bibitem[{A. {Pandey} {et~al.}(2017){Pandey}, {Gupta}, \& {Wiita}}]{2017ApJ...841..123P}
{Pandey}, A., {Gupta}, A.~C., \& {Wiita}, P.~J. 2017, \bibinfo{title}{{X-Ray Intraday Variability of Five TeV Blazars with NuSTAR},} \apj, 841, 123, \dodoi{10.3847/1538-4357/aa705e}

\bibitem[{A. {Pandey} {et~al.}(2018){Pandey}, {Gupta}, \& {Wiita}}]{2018ApJ...859...49P}
{Pandey}, A., {Gupta}, A.~C., \& {Wiita}, P.~J. 2018, \bibinfo{title}{{X-Ray Flux and Spectral Variability of Six TeV Blazars with NuSTAR},} \apj, 859, 49, \dodoi{10.3847/1538-4357/aabc5b}

\bibitem[{G.~S. {Pavana Gowtami} {et~al.}(2022){Pavana Gowtami}, {Gaur}, {Gupta}, {Wiita}, {Liao}, \& {Ward}}]{2022MNRAS.511.3101P}
{Pavana Gowtami}, G.~S., {Gaur}, H., {Gupta}, A.~C., {et~al.} 2022, \bibinfo{title}{{X-ray intraday variability and power spectral density profiles of the blazar 3C 273 with XMM-Newton during 2000-2021},} \mnras, 511, 3101, \dodoi{10.1093/mnras/stac286}

\bibitem[{R. {Prince} {et~al.}(2021){Prince}, {Agarwal}, {Gupta}, {Majumdar}, {Czerny}, {Cellone}, \& {Andruchow}}]{2021A&A...654A..38P}
{Prince}, R., {Agarwal}, A., {Gupta}, N., {et~al.} 2021, \bibinfo{title}{{Multiwavelength analysis and modeling of OJ 287 during 2017-2020},} \aap, 654, A38, \dodoi{10.1051/0004-6361/202140708}

\bibitem[{S. {Rakshit} {et~al.}(2017){Rakshit}, {Stalin}, {Muneer}, {Neha}, \& {Paliya}}]{2017ApJ...835..275R}
{Rakshit}, S., {Stalin}, C.~S., {Muneer}, S., {Neha}, S., \& {Paliya}, V.~S. 2017, \bibinfo{title}{{Flux and Polarization Variability of OJ 287 during the Early 2016 Outburst},} \apj, 835, 275, \dodoi{10.3847/1538-4357/835/2/275}

\bibitem[{ {Rees}(1984){Rees}}]{1984ARA&A..22..471R}
{Rees}. 1984, \bibinfo{title}{{Black Hole Models for Active Galactic Nuclei},} \araa, 22, 471, \dodoi{10.1146/annurev.aa.22.090184.002351}

\bibitem[{J.~C. {Rodr{\'\i}guez-Ram{\'\i}rez} {et~al.}(2020){Rodr{\'\i}guez-Ram{\'\i}rez}, {Kushwaha}, {de Gouveia Dal Pino}, \& {Santos-Lima}}]{2020MNRAS.498.5424R}
{Rodr{\'\i}guez-Ram{\'\i}rez}, J.~C., {Kushwaha}, P., {de Gouveia Dal Pino}, E.~M., \& {Santos-Lima}, R. 2020, \bibinfo{title}{{A hadronic emission model for black hole-disc impacts in the blazar OJ 287},} \mnras, 498, 5424, \dodoi{10.1093/mnras/staa2664}

\bibitem[{R.~M. {Sambruna} {et~al.}(1994){Sambruna}, {Barr}, {Giommi}, {Maraschi}, {Tagliaferri}, \& {Treves}}]{1994ApJS...95..371S}
{Sambruna}, R.~M., {Barr}, P., {Giommi}, P., {et~al.} 1994, \bibinfo{title}{{The X-Ray Spectra of Blazars: Analysis of the Complete EXOSAT Archive},} \apjs, 95, 371, \dodoi{10.1086/192102}

\bibitem[{A. {Sillanpaa} {et~al.}(1988){Sillanpaa}, {Haarala}, {Valtonen}, {Sundelius}, \& {Byrd}}]{1988ApJ...325..628S}
{Sillanpaa}, A., {Haarala}, S., {Valtonen}, M.~J., {Sundelius}, B., \& {Byrd}, G.~G. 1988, \bibinfo{title}{{OJ 287: Binary Pair of Supermassive Black Holes},} \apj, 325, 628, \dodoi{10.1086/166033}

\bibitem[{A. {Sillanpaa} {et~al.}(1996){Sillanpaa}, {Takalo}, {Pursimo}, {Lehto}, {Nilsson}, {Teerikorpi}, {Heinaemaeki}, {Kidger}, {de Diego}, {Gonzalez-Perez}, {Rodriguez-Espinosa}, {Mahoney}, {Boltwood}, {Dultzin-Hacyan}, {Benitez}, {Turner}, {Robertson}, {Honeycut}, {Efimov}, {Shakhovskoy}, {Charles}, {Schramm}, {Borgeest}, {Linde}, {Weneit}, {Kuehl}, {Schramm}, {Sadun}, {Grashuis}, {Heidt}, {Wagner}, {Bock}, {Kuemmel}, {Heines}, {Fiorucci}, {Tosti}, {Ghisellini}, {Raiteri}, {Villata}, {de Francesco}, {Bosio}, \& {Latini}}]{1996A&A...305L..17S}
{Sillanpaa}, A., {Takalo}, L.~O., {Pursimo}, T., {et~al.} 1996, \bibinfo{title}{{Confirmation of the 12-year optical outburst cycle in blazar OJ 287.},} \aap, 305, L17

\bibitem[{K.~P. {Singh} {et~al.}(2022){Singh}, {Kushwaha}, {Sinha}, {Pal}, {Agarwal}, \& {Dewangan}}]{2022MNRAS.509.2696S}
{Singh}, K.~P., {Kushwaha}, P., {Sinha}, A., {et~al.} 2022, \bibinfo{title}{{Spectral States of OJ 287 blazar from Multiwavelength Observations with AstroSat},} \mnras, 509, 2696, \dodoi{10.1093/mnras/stab3161}

\bibitem[{E. {Traianou} {et~al.}(2025){Traianou}, {G{\'o}mez}, {Cho}, {Chael}, {Fuentes}, {Myserlis}, {Wielgus}, {Zhao}, {Lico}, {Moriyama}, {Dey}, {Bruni}, {Dahale}, {Toscano}, {Gurvits}, {Lisakov}, {Kovalev}, {Lobanov}, {Pushkarev}, \& {Sokolovsky}}]{2025A&A...700A..16T}
{Traianou}, E., {G{\'o}mez}, J.~L., {Cho}, I., {et~al.} 2025, \bibinfo{title}{{Revealing a ribbon-like jet in OJ 287 with RadioAstron},} \aap, 700, A16, \dodoi{10.1051/0004-6361/202554929}

\bibitem[{M.~J.~L. {Turner} {et~al.}(2001){Turner}, {Abbey}, {Arnaud}, {Balasini}, {Barbera}, {Belsole}, {Bennie}, {Bernard}, {Bignami}, {Boer}, {Briel}, {Butler}, {Cara}, {Chabaud}, {Cole}, {Collura}, {Conte}, {Cros}, {Denby}, {Dhez}, {Di Coco}, {Dowson}, {Ferrando}, {Ghizzardi}, {Gianotti}, {Goodall}, {Gretton}, {Griffiths}, {Hainaut}, {Hochedez}, {Holland}, {Jourdain}, {Kendziorra}, {Lagostina}, {Laine}, {La Palombara}, {Lortholary}, {Lumb}, {Marty}, {Molendi}, {Pigot}, {Poindron}, {Pounds}, {Reeves}, {Reppin}, {Rothenflug}, {Salvetat}, {Sauvageot}, {Schmitt}, {Sembay}, {Short}, {Spragg}, {Stephen}, {Str{\"u}der}, {Tiengo}, {Trifoglio}, {Tr{\"u}mper}, {Vercellone}, {Vigroux}, {Villa}, {Ward}, {Whitehead}, \& {Zonca}}]{2001A&A...365L..27T}
{Turner}, M.~J.~L., {Abbey}, A., {Arnaud}, M., {et~al.} 2001, \bibinfo{title}{{The European Photon Imaging Camera on XMM-Newton: The MOS cameras},} \aap, 365, L27, \dodoi{10.1051/0004-6361:20000087}

\bibitem[{C.~M. {Urry} \& P. {Padovani}(1995){Urry} \& {Padovani}}]{1995PASP..107..803U}
{Urry}, C.~M., \& {Padovani}, P. 1995, \bibinfo{title}{{Unified Schemes for Radio-Loud Active Galactic Nuclei},} \pasp, 107, 803, \dodoi{10.1086/133630}

\bibitem[{E. {Valtaoja} {et~al.}(1985){Valtaoja}, {Lehto}, {Teerikorpi}, {Korhonen}, {Valtonen}, {Ter{\"a}sranta}, {Salonen}, {Urpo}, {Tiuri}, {Piirola}, \& {Saslaw}}]{1985Natur.314..148V}
{Valtaoja}, E., {Lehto}, H., {Teerikorpi}, P., {et~al.} 1985, \bibinfo{title}{{A 15.7-min periodicity in OJ287},} \nat, 314, 148, \dodoi{10.1038/314148a0}

\bibitem[{M.~J. {Valtonen} {et~al.}(2009){Valtonen}, {Nilsson}, {Villforth}, {Lehto}, {Takalo}, {Lindfors}, {Sillanp{\"a}{\"a}}, {Hentunen}, {Mikkola}, {Zola}, {Drozdz}, {Koziel}, {Ogloza}, {Kurpinska-Winiarska}, {Siwak}, {Winiarski}, {Heidt}, {Kidger}, {Pursimo}, {Wu}, {Zhou}, {Sadakane}, {Marchev}, {Nissinen}, {Niarchos}, {Liakos}, {Gazeas}, {Dogru}, {Poyner}, {Dietrich}, {Assef}, {Atlee}, {Bird}, {DePoy}, {Eastman}, {Peeples}, {Prieto}, {Watson}, {Yee}, {Mattingly}, \& {Ohlert}}]{2009ApJ...698..781V}
{Valtonen}, M.~J., {Nilsson}, K., {Villforth}, C., {et~al.} 2009, \bibinfo{title}{{Tidally Induced Outbursts in OJ 287 during 2005-2008},} \apj, 698, 781, \dodoi{10.1088/0004-637X/698/1/781}

\bibitem[{M.~J. {Valtonen} {et~al.}(2016){Valtonen}, {Zola}, {Ciprini}, {Gopakumar}, {Matsumoto}, {Sadakane}, {Kidger}, {Gazeas}, {Nilsson}, {Berdyugin}, {Piirola}, {Jermak}, {Baliyan}, {Alicavus}, {Boyd}, {Campas Torrent}, {Campos}, {Carrillo G{\'o}mez}, {Caton}, {Chavushyan}, {Dalessio}, {Debski}, {Dimitrov}, {Drozdz}, {Er}, {Erdem}, {Escartin P{\'e}rez}, {Fallah Ramazani}, {Filippenko}, {Ganesh}, {Garcia}, {G{\'o}mez Pinilla}, {Gopinathan}, {Haislip}, {Hudec}, {Hurst}, {Ivarsen}, {Jelinek}, {Joshi}, {Kagitani}, {Kaur}, {Keel}, {LaCluyze}, {Lee}, {Lindfors}, {Lozano de Haro}, {Moore}, {Mugrauer}, {Naves Nogues}, {Neely}, {Nelson}, {Ogloza}, {Okano}, {Pandey}, {Perri}, {Pihajoki}, {Poyner}, {Provencal}, {Pursimo}, {Raj}, {Reichart}, {Reinthal}, {Sadegi}, {Sakanoi}, {Salto Gonz{\'a}lez}, {Sameer}, {Schweyer}, {Siwak}, {Sold{\'a}n Alfaro}, {Sonbas}, {Steele}, {Stocke}, {Strobl}, {Takalo}, {Tomov}, {Tremosa Espasa}, {Valdes}, {Valero P{\'e}rez}, {Verrecchia}, {Webb}, {Yoneda}, {Zejmo}, {Zheng}, {Telting},
  {Saario}, {Reynolds}, {Kvammen}, {Gafton}, {Karjalainen}, {Harmanen}, \& {Blay}}]{2016ApJ...819L..37V}
{Valtonen}, M.~J., {Zola}, S., {Ciprini}, S., {et~al.} 2016, \bibinfo{title}{{Primary Black Hole Spin in OJ 287 as Determined by the General Relativity Centenary Flare},} \apjl, 819, L37, \dodoi{10.3847/2041-8205/819/2/L37}

\bibitem[{S. {Vaughan}(2010){Vaughan}}]{2010MNRAS.402..307V}
{Vaughan}, S. 2010, \bibinfo{title}{{A Bayesian test for periodic signals in red noise},} \mnras, 402, 307, \dodoi{10.1111/j.1365-2966.2009.15868.x}

\bibitem[{S. {Vaughan} {et~al.}(2003){Vaughan}, {F}, {Warwick}, \& {Uttley}}]{2003MNRAS.345.1271V}
{Vaughan}, S., {F}, R., {Warwick}, R.~S., \& {Uttley}, P. 2003, \bibinfo{title}{{On characterizing the variability properties of X-ray light curves from active galaxies},} \mnras, 345, 1271, \dodoi{10.1046/j.1365-2966.2003.07042.x}

\bibitem[{S.~J. {Wagner} \& A. {Witzel}(1995){Wagner} \& {Witzel}}]{1995ARA&A..33..163W}
{Wagner}, S.~J., \& {Witzel}, A. 1995, \bibinfo{title}{{Intraday Variability In Quasars and BL Lac Objects},} \araa, 33, 163, \dodoi{10.1146/annurev.aa.33.090195.001115}

\bibitem[{A.~E. {Wehrle} {et~al.}(2019){Wehrle}, {Carini}, \& {Wiita}}]{2019ApJ...877..151W}
{Wehrle}, A.~E., {Carini}, M., \& {Wiita}, P.~J. 2019, \bibinfo{title}{{Measuring the Variability in K2 Optical Light Curves of the Binary Black Hole Candidate OJ 287 and Other Fermi Active Galactic Nuclei in 2014-2015},} \apj, 877, 151, \dodoi{10.3847/1538-4357/ab1b2d}

\bibitem[{J.-H. {Woo} \& C.~M. {Urry}(2002){Woo} \& {Urry}}]{2002ApJ...579..530W}
{Woo}, J.-H., \& {Urry}, C.~M. 2002, \bibinfo{title}{{Active Galactic Nucleus Black Hole Masses and Bolometric Luminosities},} \apj, 579, 530, \dodoi{10.1086/342878}

\bibitem[{B.-K. {Zhang} {et~al.}(2025){Zhang}, {Jin}, {Zong}, {Wang}, {Zhu}, \& {Dai}}]{2025MNRAS.541.3008Z}
{Zhang}, B.-K., {Jin}, M., {Zong}, P., {et~al.} 2025, \bibinfo{title}{{The 39-day quasi-periodic oscillation and spectral characteristics of blazar OJ 287 in the X-ray band},} \mnras, 541, 3008, \dodoi{10.1093/mnras/staf1134}

\bibitem[{Z. {Zhang} {et~al.}(2019){Zhang}, {Gupta}, {Gaur}, {Wiita}, {An}, {Gu}, {Hu}, \& {Xu}}]{2019ApJ...884..125Z}
{Zhang}, Z., {Gupta}, A.~C., {Gaur}, H., {et~al.} 2019, \bibinfo{title}{{X-Ray Intraday Variability of the TeV Blazar Mrk 421 with Suzaku},} \apj, 884, 125, \dodoi{10.3847/1538-4357/ab3f3a}

\bibitem[{Z. {Zhang} {et~al.}(2021){Zhang}, {Gupta}, {Gaur}, {Wiita}, {An}, {Lu}, {Fan}, \& {Xu}}]{2021ApJ...909..103Z}
{Zhang}, Z., {Gupta}, A.~C., {Gaur}, H., {et~al.} 2021, \bibinfo{title}{{X-Ray Intraday Variability of the TeV Blazar PKS 2155-304 with Suzaku during 2005-2014},} \apj, 909, 103, \dodoi{10.3847/1538-4357/abdd38}

\bibitem[{D. {Zhou} {et~al.}(2024){Zhou}, {Zhang}, {Gupta}, {Kushwaha}, {Wiita}, {Gu}, \& {Xu}}]{2024MNRAS.532.3285Z}
{Zhou}, D., {Zhang}, Z., {Gupta}, A.~C., {et~al.} 2024, \bibinfo{title}{{X-ray flux and spectral variability of the blazar OJ 287 with Suzaku},} \mnras, 532, 3285, \dodoi{10.1093/mnras/stae1722}

\end{thebibliography}
\bibliographystyle{aasjournalv7}

\clearpage
 \restartappendixnumbering
\clearpage

\end{document}